\documentclass[prd,reprint,onecolumn,superscriptaddress,tightenlines,nofootinbib, eqsecnum]{revtex4-2}

\usepackage{amsmath}
\usepackage{amsfonts}
\usepackage{amssymb}
\usepackage{bm}
\usepackage[colorlinks]{hyperref}
\usepackage{mathrsfs}
\usepackage{graphicx}
\usepackage{empheq}
\usepackage{ulem}
\usepackage{tensor}
\usepackage[usenames]{color}
\definecolor{darkgreen}{rgb}{0,0.6,0}
\hypersetup{urlcolor=darkgreen, citecolor=darkgreen}
\usepackage{cleveref}

\allowdisplaybreaks

\DeclareSymbolFontAlphabet{\mathrsfs}{rsfs}
\DeclareMathAlphabet{\mathcal}{OMS}{cmsy}{m}{n}

\newcommand{\di}{\mathrm{i}}
\newcommand{\e}{\mathrm{e}}
\newcommand\calO{{\mathcal{O}}}
\newcommand{\dd}{\mathrm{d}}

\newcommand{\nn}{\nonumber}
\newcommand{\fid}{\dot{\phi}}
\newcommand{\rd}{\dot{r}}
\newcommand{\Et}{\widetilde{E}}

\newcommand{\kapS}{\chi^{(2)}_{\kappa+}}
\newcommand{\kapA}{\chi^{(2)}_{\kappa-}}

\usepackage{parskip}
\begin{document}
\title{Spin effects in gravitational waveforms and fluxes for binaries on eccentric orbits \\to the third post-Newtonian order}  

\author{Quentin \textsc{Henry}}\email{quentin.henry@aei.mpg.de}
\affiliation{Max Planck Institute for Gravitational Physics (Albert Einstein Institute), Am M\"uhlenberg 1, Potsdam 14476, Germany}

\author{Mohammed \textsc{Khalil}}\email{mkhalil@perimeterinstitute.ca}
\affiliation{Perimeter Institute for Theoretical Physics, 31 Caroline Street North, Waterloo, Ontario N2L 2Y5, Canada}


\begin{abstract}
Compact binaries can have non-negligible orbital eccentricities in the frequency band of ground-based gravitational-wave detectors, depending on their astrophysical formation channels.
To accurately determine the parameters of such systems, waveform models need to incorporate eccentricity effects.
In this paper, we consider an eccentric binary of spinning nonprecessing compact objects, and derive the energy and angular momentum fluxes at infinity, as well as the gravitational waveform modes to the third post-Newtonian order.
The novel results of this paper include the next-to-leading order instantaneous spin-orbit and spin-spin contributions to the waveform modes, in addition to the hereditary (tail and memory) contributions to the modes and fluxes for eccentric orbits.
The instantaneous contributions are derived for generic motion, while the hereditary contributions are computed in a small-eccentricity expansion, but we consider a resummation that makes them valid for large eccentricities.
We employ a quasi-Keplerian parametrization of the motion using harmonic coordinates and the covariant spin-supplementary condition, which complements some results in the literature in other coordinates.
Our results can be useful in improving the accuracy of waveform models for spinning binaries on eccentric orbits.
\end{abstract}

\maketitle

\section{Introduction}\label{sec:intro}

Gravitational-wave (GW) detections by the LIGO-Virgo-KAGRA collaboration~\cite{LIGOScientific:2018mvr,LIGOScientific:2020ibl,LIGOScientific:2021djp} have so far been consistent with compact binaries in quasi-circular inspirals~\cite{Salemi:2019owp,Romero-Shaw:2019itr,Nitz:2019spj,Romero-Shaw:2020thy,Lenon:2020oza,Wu:2020zwr,Yun:2020aow,Pal:2023dyg}, though the short signal of GW190521 could be consistent with either an eccentric nonprecessing-spin, or a circular precessing-spin,  binary~\cite{Romero-Shaw:2020thy,Romero-Shaw:2022fbf,Gayathri:2020coq}.

Some astrophysical binary formation channels, in dense globular clusters and galactic nuclei, lead to a small fraction of eccentric binaries in the frequency band of current-generation ground-based detectors~\cite{Samsing:2017rat,Samsing:2013kua,Samsing:2017xmd,Rodriguez:2017pec,Zevin:2018kzq,Gondan:2020svr,Antonini:2012ad,Antonini:2015zsa,VanLandingham:2016ccd,Tagawa:2020jnc,Sedda:2023qlx,Chattopadhyay:2023pil}.
Furthermore, for the Laser Interferometer Space Antenna (LISA)~\cite{LISA:2017pwj}, most GW sources are expected to have significant eccentricity~\cite{Sesana:2010qb,Breivik:2016ddj,Willems:2007xe,Samsing:2018isx,Kremer:2018cir,Hoang:2019kye,Xuan:2022qkw,Lau:2019wzw,Garg:2023lfg}. 
Therefore, detecting eccentric binaries, or the lack of such a detection, can provide valuable information about the binary formation channels~\cite{Zevin:2021rtf,Nishizawa:2016eza,Breivik:2016ddj}.
To confidently measure the eccentricity of a binary system, it is important to use accurate waveform models valid for eccentric orbits~\cite{Ramos-Buades:2020eju,Lower:2018seu}, otherwise one might encounter systematic biases in parameter estimation~\cite{Bhat:2022amc,Saini:2022igm,Cho:2022cdy,Divyajyoti:2023rht}.

Most waveform models rely on the post-Newtonian (PN) approximation (small-velocity and weak-field expansion, i.e., $v^2/c^2 \sim GM/c^2r \ll 1 $) to describe the binary dynamics.
The PN conservative dynamics is known for generic orbits to 5.5PN order~\cite{Bini:2019nra,Bini:2020wpo,Blumlein:2021txe,Almeida:2023yia,Henry:2023sdy}, though the nonlocal-in-time contributions need to be computed separately for bound and unbound orbits in an eccentricity expansion~\cite{Damour:2014jta,Damour:2015isa}.
The radiative sector (energy and angular momentum fluxes, and the waveform phase and amplitude) is generally more difficult to derive than the conservative dynamics, and currently the PN information is known at a lower order for eccentric than circular orbits.
In the nonspinning case, and for quasi-circular inspirals, the fluxes and GW phase have recently been derived to 4.5PN, with the dominant (2,2) waveform mode to 4PN~\cite{Blanchet:2023sbv,Blanchet:2023bwj}, and the higher modes at 3.5PN~\cite{Blanchet:2008je,Faye:2012we,Faye:2014fra,Henry:2021cek,Henry:2022ccf}. 
However, for eccentric orbits, the radiative dynamics is known to 3PN;
in particular, the energy flux was derived in Refs.~\cite{Arun:2007rg,Arun:2007sg} and angular momentum flux in Ref.~\cite{Arun:2009mc}, while the waveform modes were completed to 3PN in Refs.~\cite{Mishra:2015bqa,Boetzel:2019nfw,Ebersold:2019kdc}.

The spin contributions, for quasi-circular inspirals and nonprecessing spins, are fully known in the waveform modes and energy flux to 3.5PN~\cite{Kidder:1992fr,Kidder:1995zr,Blanchet:2006gy,Blanchet:2011zv,Buonanno:2012rv,Bohe:2013cla,Marsat:2013caa,Marsat:2014xea,Bohe:2015ana,Henry:2022dzx}.
However, for eccentric orbits, the spin part is known in the modes only to 2PN~\cite{Khalil:2021txt,Paul:2022xfy}, which includes the leading order (LO) spin-orbit (SO) and spin-spin (SS) effects.
The instantaneous contributions to the energy and angular momentum fluxes are known to 3PN from Refs.~\cite{Racine:2008kj,Bohe:2015ana,Cho:2021mqw}, which includes next-to-leading (NLO) SO and SS contributions, but the hereditary contributions (tail and memory effects) are known for quasi-circular orbits only~\cite{Blanchet:2011zv,Marsat:2013caa,Cho:2022syn,Mitman:2022kwt}. 
The highest known order in the instantaneous contributions to the energy flux is the next-to-NLO SS at 4PN, which was recently derived in Ref.~\cite{Cho:2022syn} for generic orbits and precessing spins.

The instantaneous contributions to the fluxes and waveform modes can be derived in a closed form for generic orbits, but the hereditary contributions have only been computed in a small-eccentricity expansion.
A useful technique in the calculation of the hereditary contributions is the use of the quasi-Keplerian (QK) parametrization~\cite{damour1985general,Damour:1988mr,schafer1993second}, which was derived for nonspinning binaries to 3PN in harmonic and Arnowitt-Deser-Misner (ADM) coordinates in Ref.~\cite{Memmesheimer:2004cv}, and to 4PN in ADM coordinates in Ref.~\cite{Cho:2021oai}. 
The 3.5PN SO and SS contributions were derived in Refs.~\cite{Tessmer:2010hp,Tessmer:2012xr} in ADM coordinates and using the Newton-Wigner spin-supplementary condition (SSC)~\cite{pryce1948mass,newton1949localized}.

In this paper, we complete the fluxes and waveform modes to 3PN for eccentric orbits and nonprecessing spins. The novel results in this paper are the following:
\begin{enumerate}
\item The spin contributions in the QK parametrization to 3PN in \emph{harmonic} coordinates and using the \emph{covariant} Tulczyjew-Dixon SSC~\cite{Tulczyjew:1959,Dixon:1979}.

\item The spin contributions to the tail part of the energy and angular momentum fluxes at infinity, computed in a small-eccentricity expansion to $\calO(e^8)$. 
We also compute the orbit-averaged fluxes using the QK parametrization, resum the tail contribution to improve its validity for high eccentricities, and obtain the time evolution of the secular orbital elements from the fluxes.

\item The spin contributions to the waveform modes at the 2.5PN and 3PN orders, which includes the instantaneous and tail contributions, in addition to direct-current (DC) and oscillatory memory contributions. 
We compute all hereditary contributions to the modes in a small-eccentricity expansion to $\calO(e^6)$. Note that even for circular orbits, our result for the spin part of the memory beyond 2PN is new. 
\end{enumerate}

This paper is structured as follows: in Sec.~\ref{sec:formalism}, we summarize the approach we use for the calculations, which is the PN multipolar post-Minkowskian (PN-MPM) formalism, and explain some techniques for evaluating the hereditary contributions based on the QK parametrization.
Sections~\ref{sec:fluxes} and~\ref{sec:modes} include our results for the fluxes and waveform modes, respectively.
We conclude in Sec.~\ref{sec:conclusions} and include some expressions for the QK parametrization in Appendix~\ref{app:QK}.
All the results presented in this paper are written in harmonic coordinates using the covariant SSC, and are provided as \textit{Mathematica} files in the Supplemental Material~\cite{ancMaterial}.

\subsection*{Notation and conventions}

The conventions employed throughout this work are as follows: we use a mostly plus signature, the Minkowski metric being $\eta_{\mu\nu} = (-,+,+,+)$; Greek letters denote spacetime indices $\mu,\nu,\ldots = (0,1,2,3)$ and Latin ones denote purely spatial indices $i,j,\ldots = (1,2,3)$; bold font denotes three-dimensional vectors, e.g., $\boldsymbol{L} = L^i$; we use the multi-index notation $A_L = 
A_{i_1i_2\ldots i_\ell}$; brackets denote symmetric and trace-free (STF) operators: $A_{\langle L\rangle} = \underset{L}{\text{STF}}\left[A_L\right]$; the d'Alembertian operator is defined with respect to the flat Minkowski metric $\Box \equiv \eta^{\mu\nu}\partial_{\mu\nu} = \Delta - c^{-2}\partial_t^2$; dots and numbers in parentheses denote time differentiation $A^{(n)} = \dd^nA/\dd t^n$; $\calO(n)$ represents a contribution of order $v^n/c^n$, i.e., $(n/2)$PN order.

For a binary with masses $m_1$ and $m_2$, we assume $m_1 \geq m_2$ and define
\begin{gather}
M= m_1 + m_2, \qquad \mu = \frac{m_1m_2}{M}, \qquad \nu = \frac{\mu}{M}, \qquad
\delta =\frac{m_1 - m_2}{M},
\end{gather}
where $M$ is the total mass, $\mu$ is the reduced mass, $\nu$ is the symmetric mass ratio, and $\delta$ is the anti-symmetric mass ratio. 

For a spinning binary with (constant-magnitude) spins $\bm{S}_1$ and $\bm{S}_2$, the dimensionless spin magnitudes are denoted $\chi_1 \equiv c |\bm{S}_1| / (Gm_1^2)$ and $\chi_2 \equiv c |\bm{S}_2| / (Gm_2^2)$. 
The spin-quadrupole constants are denoted $\kappa_1$ and $\kappa_2$, which equal one for black holes.
To simplify our expressions, and to make it easier to specialize them for black holes, we define the following combinations of the spin magnitudes and quadrupole constants:
\begin{subequations}
\begin{align}
\chi_S & \equiv \frac{1}{2} \left(\chi_1 + \chi_2\right),\\
\chi_A & \equiv \frac{1}{2} \left(\chi_1 - \chi_2\right), \\
\kapS &\equiv \frac{1}{2} \left[\chi _1^2 (\kappa_1 - 1)+\chi _2^2 (\kappa_2 - 1)\right], \\
\kapA &\equiv \frac{1}{2} \left[\chi _1^2 (\kappa_1 - 1) - \chi _2^2 (\kappa_2 - 1)\right].
\end{align}
\end{subequations}

Several quantities are used in the QK parametrization, as explained in Subsection~\ref{sec:QK}, and we summarize them in Table~\ref{tab:QKPsummary} below. For a geometric picture for some of these quantities, see Fig.~2 of Ref.~\cite{Tessmer:2012xr}.
\begin{table}[th]
\caption{Definition of the main quantities used in the QK parametrization.}
\label{tab:QKPsummary}
\begin{ruledtabular}
\begin{tabular}{p{0.2\linewidth} p{0.6\linewidth} p{0.2\linewidth}}
symbol & description & defined in \\
\hline
$a_r$ & semi-major axis & Eq.~\eqref{eq:QKr} \\
$e_r$ & radial eccentricity & Eq.~\eqref{eq:QKr} \\
$e_t \equiv e$ & time eccentricity & Eq.~\eqref{eq:QKl} \\
$e_\phi$ & phase eccentricity & Eq.~\eqref{eq:QKphi} \\
$u$ & eccentric anomaly & Eqs.~\eqref{eq:QKr}, \eqref{eq:QKl} \\
$v$ & true anomaly & Eqs.~\eqref{eq:QKphi}, \eqref{eq:QKv} \\
$n \equiv 2\pi/P$ & mean motion (radial angular frequency), with $P$ being the radial period & Eq.~\eqref{eq:Pintegral} \\
$l$ & mean anomaly & Eq.~\eqref{eq:QKl} \\
$\Phi$ & total phase between two successive periastron passages & Eq.~\eqref{eq:Phiintegral} \\
$K\equiv \Phi/(2\pi)$ & periastron advance & \\
$x \equiv (GM\Omega)^{2/3} / c^2$ & dimensionless variable related to the orbital frequency $\Omega = K n$ & Eq.~\eqref{eq:QKx} \\
$\Et \equiv - (E-Mc^2)/\mu$ & (minus) the reduced binding energy & Eq.~\eqref{eq:tildEhdef} \\
$h\equiv L/(GM\mu)$ & reduced orbital angular momentum & Eq.~\eqref{eq:tildEhdef} \\
\end{tabular}
\end{ruledtabular}
\end{table}

\section{Fluxes and waveform in the PN-MPM formalism}
\label{sec:formalism}

In this section, we briefly recall some aspects of the PN-MPM formalism. Then, we derive the QK parametrization, and finally describe how to evaluate the hereditary integrals that are required for the 3PN waveform and fluxes. 

The PN-MPM formalism is an iterative algorithm that allows computing, to a given PN order, the radiated fluxes and the waveform through a multipolar decomposition of the system~\cite{Blanchet:2013haa}. The relevant quantities are parametrized by a set of symmetric trace-free multipole moments $\{U_L,V_L\}$, called ``radiative'' because they are defined at infinity on an asymptotically flat spacetime. 
The PN-MPM formalism allows linking the radiative multipole moments to the ``canonical'' multipole moments $\{M_L,S_L\}$ parametrizing the source and computed in harmonic coordinates. 

Both sets of multipole moments differ by nonlinear interactions of the GW field. Notably, hereditary effects, such as tail and memory integrals, appear at the considered PN order. 
Fortunately for us, the canonical moments were derived in Refs.~\cite{Blanchet:2008je,Faye:2014fra,Henry:2022dzx} for arbitrary motion up to the 3.5PN order. 
Thus, we can obtain the fluxes and waveform modes from those moments, which is straightforward for the instantaneous contributions, but for the hereditary contributions, we first derive the QK parametrization, then express the canonical moments within this parametrization to evaluate the hereditary integrals.

\subsection{Fluxes and waveform modes in terms of radiative multipole moments}

Let us consider an asymptotically flat spacetime with a background Minkowski metric $\eta^{\mu\nu}$. We define $h^{\mu\nu}\equiv\sqrt{-g}g^{\mu\nu}-\eta^{\mu\nu}$ being the deviation to the gothic metric, where $g$ is the determinant of $g_{\mu\nu}$. Within the harmonic gauge, which imposes $\partial_\nu h^{\mu\nu}=0$, the Einstein field equations read
\begin{equation}
\Box h^{\mu\nu}=\frac{16\pi G}{c^4}\tau^{\mu\nu},
\end{equation}
where $\tau^{\mu\nu}$ is the usual Landau-Lifshitz pseudo stress-energy tensor. There exists a coordinate system $(T,\bm{X})$, called radiative coordinates, in which the solution metric, in the transverse and traceless (TT) gauge, reads~\cite{Thorne:1980ru}
\begin{align}\label{eq:hijTT}
h_{ij}^\text{TT} &= - \frac{4G}{c^2R} \,\mathcal{P}_{ijkl} \sum^{+\infty}_{\ell=2}\frac{1}{c^\ell \ell !} \left\{ N_{L-2} \,U_{klL-2} - \frac{2\ell}{c(\ell+1)} \,N_{aL-2} \,\varepsilon_{ab(k} \,V_{l)bL-2}\right\} + \mathcal{O}\left(\frac{1}{R^2}\right)\,,
\end{align}
where $\varepsilon_{ijk}$ is the Levi-Civita symbol, $\mathcal{P}_{ijkl} = \mathcal{P}_{i(k}\mathcal{P}_{l)j}-\frac{1}{2}\mathcal{P}_{ij}\mathcal{P}_{kl}$ is the usual TT projector, with $\mathcal{P}_{ij}=\delta_{ij}-N_iN_j$ being the projector orthogonal to the unit vector $\bm{N}$, which is the direction to the observer. The relation~\eqref{eq:hijTT} defines the radiative multipole moments $\{U_L,V_L\}$. After introducing the orthonormal triad $(\bm{P},\bm{Q},\bm{N})$, we can define the usual GW polarizations
\begin{subequations}\label{eq:hpluscross}
\begin{align}
h_+ &= \frac{1}{2}(P_i P_j - Q_i Q_j)h_{ij}^\text{TT},\\
h_\times &= \frac{1}{2}(P_i Q_j + Q_i P_j)h_{ij}^\text{TT}.
\end{align}
\end{subequations}
Next, the amplitude of the GW can also be linked to the radiative moments. After a mode decomposition and projecting the multipoles from the STF to a spherical-harmonics basis, the complex amplitude $h$ reads
\begin{equation}
h\equiv h_+ -\di h_\times = \sum_{l=2}^\infty \sum_{m=-\ell}^{\ell} h_{\ell m} Y^{\ell m}_{-2}(\Theta,\Phi),
\end{equation}
where 
\begin{equation}\label{eq:hlm}
h_{\ell m} = -\frac{2 G}{R c^{\ell+2}\ell !}\sqrt{\frac{(\ell+1)(\ell+2)}{\ell(\ell-1)}}\,\alpha_L^{\ell m} \left( U_L+\frac{2\ell}{\ell+1}\frac{\di}{c} V_L \right),
\end{equation}
with $R$ being the distance between the source and the observer, and the STF tensors $\alpha_L^{\ell m}$ are defined by
\begin{equation}
\alpha_L^{\ell m} \equiv \int \dd \Omega\,\hat{N}_L\,\overline{Y}^{\,\ell m}.
\end{equation}
The explicit expression for $\alpha_L^{\ell m}$ is given by, e.g., Eq. (4.7) of Ref.~\cite{Henry:2021cek}. 
Note that in the case of planar orbits, the mass-type and current-type multipoles do not mix within a given $(\ell,m)$ mode. More specifically for even $\ell+m$, only the mass-type moments $U_L$ contribute, while for odd $\ell+m$, only the current-type moments $V_L$ contribute.

Let us now focus on the radiated energy and angular momentum. The radiated energy at future null infinity can be written in terms of the TT metric at the lowest order in $R$ as~\cite{Thorne:1980ru}
\begin{equation}\label{eq:flux_gen}
\mathcal{F} = \frac{R^2 c^3}{32 \pi G} \int \dd \Omega \, \dot{h}^\text{TT}_{ij}\dot{h}^\text{TT}_{ij}\,.
\end{equation}
Using Eq.~\eqref{eq:hijTT}, we can write the energy flux in terms of the radiative multipole moments, and a similar procedure can be done to express the angular momentum flux~\cite{Thorne:1980ru,Blanchet:2018yqa}. In the end, both fluxes read
\begin{subequations}\label{eq:fluxes_moments}
\begin{align}
\mathcal{F} &= \sum_{\ell= 2}^{+\infty} \frac{G}{c^{2\ell+1}} \frac{(\ell+1)(\ell+2)}{(\ell-1)\ell\,\ell !(2\ell+1)!!}\bigg[U_L^{(1)}U_L^{(1)} + \frac{4\ell^2}{c^2(\ell+1)^2}\,V_L^{(1)}V_L^{(1)}\bigg]\,,\\
\mathcal{G}_i &= \varepsilon_{iab}\,\sum_{\ell= 2}^{+\infty} \frac{G}{c^{2\ell+1}} \frac{(\ell+1)(\ell+2)}{(\ell-1)\,\ell !(2\ell+1)!!}\bigg[U^{}_{aL-1}U_{bL-1}^{(1)} + \frac{4\ell^2}{c^2(\ell+1)^2}\,V^{}_{aL-1}V_{bL-1}^{(1)}\bigg]\,.
\end{align}
\end{subequations}
We recall that we are interested in the spin contributions to the fluxes and modes to the 3PN order. The spin contributions to the mass-type multipoles arise at 1.5PN, while they contribute to the current-type multipoles at 0.5PN. Thus, restricting ourselves to the 3PN order for the spin contributions, the energy and angular momentum fluxes are obtained only from the mass and current-type multipoles for $\ell=2$ and $\ell=3$. In the next subsection, we explain how the radiative multipole moments are derived.

\subsection{Radiative multipole moments}

The radiative moments are composed of two types of contributions: instantaneous ones, which depend on the local time and can be computed for arbitrary motion; and hereditary contributions, which depend on the entire history of the binary and require specifying the orbit. Furthermore, as we see later, an eccentricity expansion within the QK parametrization is required to evaluate the hereditary integrals analytically. 
In this subsection, we restrict ourselves to the moments that affect the spin terms at the 3PN order. 
An exhaustive list of nonlinearities coming from the difference between radiative and canonical moments can be found in, e.g., Ref.~\cite{Faye:2014fra}. 

The general structure of the radiative moments in terms of the canonical moments at 3PN order read
\begin{subequations}
\begin{align}
U_L &= U_L^\text{inst} + U_L^\text{tail} + U_L^\text{mem}\,,\\
V_L &= V_L^\text{inst} + V_L^\text{tail}\,,
\end{align}
\end{subequations}
where $U_L^\text{inst}$ and $V_L^\text{inst}$ denote the instantaneous contributions, while $U_L^\text{tail}$ and $V_L^\text{tail}$ are the tail contributions. The memory contribution $U_L^\text{mem}$ only appears in the mass-type moments.

\subsubsection{Instantaneous contributions}

The instantaneous part of the radiative multipole moments are generally given by
\begin{subequations}
\label{eq:instMoments}
\begin{align}
U_{L}^{\rm inst}= M_L^{(\ell)} +\delta U_{L}^{\rm inst} \,,\\
V_{L}^{\rm inst}= S_L^{(\ell)}+\delta V_{L}^{\rm inst}\,,
\end{align}
\end{subequations}
where $\delta U_{L}^{\rm inst}$ and $\delta V_{L}^{\rm inst}$ are local-in-time, nonlinear interactions of the gravitational field. The full expressions of these interactions for the 3.5PN waveform are provided in Ref.~\cite{Faye:2014fra}. At the considered order of this paper, the spin contributions of these interactions are given by~\cite{Henry:2022dzx}
\begin{subequations}
\begin{align}
\left[\delta U_{ij}^{\rm inst} \right]_{S} &= \frac{2G}{c^{5}} \left[ \frac{1}{3} \varepsilon_{ab\langle i} M_{j\rangle a}^{(4)}S_{b}\right]_{S} + \calO(8) \,,\\
\left[\delta V_{ijk}^{\rm inst} \right]_{S} &= - \frac{2G}{c^{3}} \left[ S_{\langle i}M^{(4)}_{jk\rangle } \right]_{S} + \calO(6) \,,
\end{align}
\end{subequations}
where the subscript $S$ refers to the spin part of the interaction. The rest of $\delta U_{L}^{\rm inst}$ and $\delta V_{L}^{\rm inst}$ do not contribute to the spin terms for the waveform at 3PN.
Both the linear and nonlinear parts of the radiative moments can be computed for arbitrary motion, i.e., without the need for specifying the orbit.

\subsubsection{Tail contributions}
The tail integrals contributing to the 3PN spin terms in the GW fluxes and modes come from those of the mass quadrupole, current quadrupole, and current octupole. Their expressions read~\cite{Blanchet:1992br}
\begin{subequations}\label{eq:tailsgen}
\begin{align}
\left[U_{ij}^\text{tail}\right]_S &= \frac{2G \mathcal{M}}{c^3}\int_0^\infty \dd\tau \left[\ln\left(\frac{\tau}{2 b}\right) +\frac{11}{12}\right] \left[M_{ij}^{(4)}(T_R-\tau)\right]_S\,,\\
\left[V_{ij}^\text{tail}\right]_S &= \frac{2G \mathcal{M}}{c^3}\int_0^\infty \dd\tau \left[\ln\left(\frac{\tau}{2 b}\right) +\frac{7}{6}\right] \left[S_{ij}^{(4)}(T_R-\tau)\right]_S\,,\\
\left[V_{ijk}^\text{tail}\right]_S &= \frac{2G \mathcal{M}}{c^3}\int_0^\infty \dd\tau \left[\ln\left(\frac{\tau}{2 b}\right) +\frac{5}{3}\right] \left[S_{ijk}^{(5)}(T_R-\tau)\right]_S\,,
\end{align}
\end{subequations}
where $\mathcal{M}$ is the ADM mass, $T_R$ is the retarded time, and $b$ is an arbitrary gauge parameter that can be eliminated after a phase redefinition, as we discuss in Subsection~\ref{sec:phaseRedef}. Contrary to the instantaneous contributions, these integrals require specifying the binary's orbit to be evaluated analytically. We detail the procedure in Subsection~\ref{subsec:heredEval}.

\subsubsection{Memory contributions}

The memory contributions to the radiative moments can be derived within the PN-MPM formalism. They take the form of integrals over the past of products of multipoles without logarithms. Formally, these integrals are of order $\mathcal{O}(c^{-5})$;
for example, the leading order memory term in the radiative mass-quadrupole moment reads
\begin{equation}
U_{ij}^\text{mem} = \frac{G}{c^5} \int_{-\infty}^{T_R} \left[
-\frac{2}{7} M_{a\langle i}^{(3)}(\tau) M_{j\rangle a}^{(3)}(\tau)
\right] \dd \tau + \calO(7).
\end{equation}
These memory contributions have been derived consistently for each multipole entering the 3.5PN waveform in Ref.~\cite{Faye:2014fra}. 
Furthermore, recent work~\cite{Trestini:2023wwg} tackled the computation of the so-called tail-of-memory effects that formally appear at 4PN. 

However, as shown in Subsection~\ref{subsec:heredEval}, the explicit evaluation of these integrals can lower the PN order of the memory contributions to the modes. Thus, in order to be consistent to a given PN order in the waveform amplitude, one would need to push the MPM procedure to higher orders, which is difficult to do using this method.
Fortunately, one can bypass this difficulty by computing the memory contributions to the waveform modes using another method, allowing us to be consistent in the present case to 3PN. 
Following Refs.~\cite{Blanchet:1992br,Favata:2008yd,Favata:2011qi,Blanchet:2023pce}, one can extract from the flux, Eq.~\eqref{eq:flux_gen}, the memory contributions to the amplitude modes. They read
\begin{equation}\label{eq:hlmmem_generic}
h_{\ell m }^\text{mem} = -\frac{16\pi G}{R c^4}\sqrt{\frac{(\ell-2)!}{(\ell+2)!}}\int^{T_R}_{-\infty} \dd t \int \dd \Omega \,\frac{\dd \mathcal{F}}{\dd\Omega} (\Omega) Y_{\ell m}^*(\Omega)\,.
\end{equation}
With this relation at hand, one can express the flux in terms of the time derivatives of the modes instead of the radiative multipole moments. Then, after performing the angular integration, as explained in detail in Ref.~\cite{Favata:2008yd}, Eq.~\eqref{eq:hlmmem_generic} becomes an integral over time of the following source
\begin{equation}\label{eq:hlmmemdotgen}
\dot{h}_{\ell m }^\text{mem} = -\frac{R}{c}\sqrt{\frac{(\ell-2)!}{(\ell+2)!}}\sum_{\ell'=2}^\infty\sum_{\ell''=2}^\infty \sum_{m'=-\ell'}^{\ell'}\sum_{m''=-\ell''}^{\ell''} (-1)^{m+m''} \langle \dot{h}_{\ell' m'}\dot{h}^*_{\ell'' m''}\rangle G^{2 -2 0}_{\ell' \ell'' m'-m''-m}\,,
\end{equation}
where the value of the angular integral $G^{s s' s''}_{l l' l'' m m' m''}$ can be found in Appendix A of Ref.~\cite{Favata:2008yd}. We then use the modes computed in Subsections~\ref{subsec:modesinst} and~\ref{subsec:tailmodes}, take their time derivative and insert them in Eq.~\eqref{eq:hlmmemdotgen} which gives the source of the memory integrals. In Subsection~\ref{subsec:heredEval}, we explain how to evaluate the elementary integrals generated by those sources.

\subsection{Quasi-Keplerian parametrization}\label{sec:QK}

We parametrize the motion of an eccentric-orbit binary using a Keplerian-like parametrization. The periastron advance implies that we cannot take a purely Keplerian parametrization as the motion is no longer elliptic. Instead, we choose the quasi-Keplerian parametrization introduced in Ref.~\cite{damour1985general}, and extended to spin in Refs.~\cite{Tessmer:2010hp,Tessmer:2012xr,Klein:2010ti}. 

The starting point of this computation is the conserved energy $E$ and angular momentum $\bm{J}$, which were derived in harmonic coordinates to 3PN for spinning binaries in Refs.~\cite{Bohe:2012mr,Bohe:2015ana}. The spin definition in these papers, which is the one we adopt here, is chosen to use the covariant (Tulczyjew-Dixon) SSC. Therefore, we rederive the spin contributions to the QK parametrization using the covariant SSC and harmonic coordinates, as opposed to the Newton-Wigner SSC and ADM coordinates used in Refs.~\cite{Tessmer:2010hp,Tessmer:2012xr}.

We restrict ourselves to the nonprecessing case, which implies that the motion is planar. 
When considering spin effects, it is better to work with the orbital angular momentum $\bm{L} = \bm{J} - (\bm{S}_1 +\bm{S}_2)/c $, which is also conserved for nonprecessing spins.
We use the orthonormal unit vectors $(\hat{\bm{l}}, \hat{\bm{n}}, \hat{\bm{\lambda}})$, where $\hat{\bm{n}}$ is the unit vector in the radial direction, $\hat{\bm{l}}$ is the direction of the orbital angular momentum, and $\hat{\bm{\lambda}} \equiv \hat{\bm{l}} \times \hat{\bm{n}}$.
Then, we express the conserved energy and angular momentum, computed in~\cite{Bohe:2012mr,Bohe:2015ana}, in terms of $\rd$ and $\fid$ and obtain, at leading order as an example,
\begin{subequations}
\begin{align}
\Et &= \frac{G M}{r}-\frac{\rd^2}{2}-\frac{r^2\fid^2}{2} + \calO(2)\,,\\
h &= \frac{r^2\fid}{GM} +\calO(2)\,,
\end{align}
\end{subequations}
where $\Et$ is (minus) the reduced binding energy and $h$ is the reduced angular momentum, i.e.,
\begin{equation}
\label{eq:tildEhdef}
\Et \equiv - \frac{E - Mc^2}{\mu}, \qquad
h \equiv \frac{\vert \bm{L}\vert}{GM\mu}.
\end{equation}

We can invert the above relations for $\Et$ and $h$ to obtain an expression for $\rd$ and $\fid$ in terms of $\Et$, $h$ and $s\equiv 1/r$ by replacing iteratively the values of $\rd$ and $\fid$ appearing at higher PN orders. We find
\begin{subequations}\label{eq:rdfid}
\begin{align}
\dot{r}^2 &= \mathcal{P}(s)\,,\label{eq:rd2}\\
\dot{\phi}&=s^2\mathcal{Q}(s)\label{eq:phid}\,,
\end{align}
\end{subequations}
where $\mathcal{P}$ and $\mathcal{Q}$ are two polynomials of $s$ whose coefficients are explicit expressions of $\Et$ and $h$. Note that, at the 3PN order for nonspinning bodies in harmonic coordinates, $\rd^2$ is not a polynomial of $s$ due to the appearance of $\ln(r)$ in the conserved energy and angular momentum. Thus, one has to change the coordinate system to remove these logarithms in order to apply the following procedure~\cite{Memmesheimer:2004cv}. 
However, to derive the spin contributions in the waveform to the 3PN order, we do not need to include the 3PN nonspinning contributions. Hence, we derive the QK parametrization including 2PN nonspinning and 3PN spinning contributions without changing coordinates.

To obtain the QK parametrization, we followed the well-explained method in Ref.~\cite{Memmesheimer:2004cv} which we briefly summarize here. The first step to integrate Eqs.~\eqref{eq:rdfid} is to remark that for bound orbits, $s$ goes through a maximum and a minimum $s_+$ and $s_-$ corresponding to the minimal and maximal value of the separation $r_-$ and $r_+$, respectively, for quasi-elliptic orbits~\cite{Damour:1983tz}. Using this property, one can write
\begin{equation}\label{eq:Pfac}
\mathcal{P}(s) = (s_+ - s)(s - s_-) \mathcal{R}(s)\,,
\end{equation}
where $\mathcal{R}$ is also a polynomial. The above relation fixes $s_+$ and $s_-$. Furthermore, we impose the radial motion to follow the ansatz
\begin{equation}
r = a_r(1-e_r \cos u)\,,
\end{equation}
where the semi-major axis $a_r$ and radial eccentricity $e_r$ read
\begin{equation}
a_r=\frac{1}{2}\frac{s_++ s_-}{s_+ s_-}\,, \qquad e_r = \frac{s_+-s_-}{s_++s_-}\,.
\end{equation}
Thus factorizing $\mathcal{P}$ as in Eq.~\eqref{eq:Pfac} directly fixes $a_r$, $e_r$ and $\mathcal{R}(s)$. At Newtonian order, we recover the Keplerian parametrization of the orbits which means that $\mathcal{R}(s)$ is a constant.

We next focus on the integration of Eq.~\eqref{eq:rd2}. We define the total radial period $P$ as the time integral over one orbit
\begin{equation}\label{eq:Pintegral}
P = \int \dd t'= 2 \int_{s_-}^{s_+}\dd s \,\frac{ \left[\mathcal{R}(s) \right]^{-1/2}}{s^2\sqrt{(s_+ -s)(s-s_-)}}\,.
\end{equation}
This integral is straightforward to compute; we can then express $P$ in terms of $a_r$ and $e_r$, and then the conserved quantities. Then, we compute the time dependency of the QK parametrization using
\begin{equation}
t-t_0 = \int_{s}^{s_+}\dd s \, \frac{ \left[\mathcal{R}(s) \right]^{-1/2}}{s^2\sqrt{(s_+ -s)(s-s_-)}}\,.
\end{equation}
Finally, we define the mean motion $n\equiv 2\pi/P$ and mean anomaly $l=n(t-t_0)$.
The results of the two integrals above can be expressed in terms of the eccentric and true anomalies $u$ and $v$, defined in Eq.~\eqref{eq:QKgen} below.

We treat Eq.~\eqref{eq:phid} in the same way as Eq.~\eqref{eq:rd2}. We define the advance of periastron angle as the angular integral over one orbit
\begin{equation}\label{eq:Phiintegral}
\Phi = \int \dd \phi' = 2 \int_{s_-}^{s_+}\dd s \,\frac{ \mathcal{Q}(s)\left[\mathcal{R}(s) \right]^{-1/2}}{\sqrt{(s_+ -s)(s-s_-)}}\,,
\end{equation}
which is computed the same way as $P$. Then, the phase variable is computed as a function of $u$ and $v$ through
\begin{equation}
\phi-\phi_0 = \int_{s}^{s_+}\dd s\, \frac{ \mathcal{Q}(s) \left[\mathcal{R}(s) \right]^{-1/2}}{\sqrt{(s_+ -s)(s-s_-)}}\,.
\end{equation}
Finally, we get an expression for $\tfrac{2\pi}{\Phi}(\phi-\phi_0)$ which constitutes the last piece of the QK parametrization that we adopt. 

In harmonic gauge, our result reads
\begin{subequations}\label{eq:QKgen}
\begin{align}
r &= a_r(1-e_r \cos u)\,, \label{eq:QKr}\\
l &= u - e_t \sin u + f_{v-u} (v-u) + f_v \sin v\,,\label{eq:QKl}\\
\frac{2\pi}{\Phi} (\phi-\phi_0) &= v+ g_{2v} \sin(2v)+ g_{3v} \sin(3v)\,,\label{eq:QKphi}\\
v&= 2 \arctan\left[\sqrt{\frac{1+e_\phi}{1-e_\phi}} \tan \frac{u}{2} \right], \label{eq:QKv}
\end{align}
\end{subequations}
where we have introduced the time eccentricity $e_t$ and the phase eccentricity $e_\phi$. 
This choice of parametrization is done so that at Newtonian order, we recover the Keplerian parametrization~\cite{damour1985general}. The phase eccentricity $e_\phi$ is fixed by imposing that $\phi$ does not contain any term proportional to $\sin(v)$.

The explicit expressions for the main quantities are displayed in Appendix~\ref{app:QK}.
In the nonspinning case, we recover the results in Ref.~\cite{Memmesheimer:2004cv} to 2PN, at which the structure of the 3PN spin contribution is similar to the 2PN nonspinning contribution. 
The 3.5PN SO and SS contributions are known from Refs.~\cite{Tessmer:2010hp,Tessmer:2012xr} in ADM coordinates and the NW SSC. Our results differ from those references because of using harmonic coordinates and the covariant SSC, but we checked that the gauge-invariant quantities, such as $n$ and $\Phi$ when expressed in terms of conserved quantities, are in agreement.

With this parametrization at hand, we can express $(r,\dot{r},\phi,\dot{\phi})$ in terms of $(x,e_t,u)$, where $x$ is a gauge-independent PN parameter defined as
\begin{equation}
\label{eq:QKx}
x = \left(\frac{G M \Omega}{c^3}\right)^{2/3},
\end{equation}
where $\Omega = K n$ is the orbital frequency.
First, we express the dynamical quantities in the QK parametrization in terms of $x$ and $e_t$, then make use of the chain rule which gives, e.g., for the separation
\begin{equation}
\frac{\dd r}{\dd t} = \frac{\dd r}{\dd u}\frac{n}{(\dd l/\dd u)}\,,
\end{equation}
and we derive $\dot{r}(x,e_t,u)$ using Eq.~\eqref{eq:QKv} by eliminating $v$ for $u$. The same steps can be used to derive $\dot{\phi}(x,e_t,u)$. 

Note that the above results for the QK parametrization correspond to the conservative part of the dynamics; the radiation-reaction part, coming from the emission of GWs, is derived in Subsection~\ref{subsec:SecOrb} by taking into account the energy and angular momentum radiated by the system. 
For now, the expressions of the conservative part are exact in the sense that no eccentricity expansion has been performed. One can also choose to express the variables of the problem in terms of $(x,e_t,l)$, which requires performing a small eccentricity expansion as we see in the following subsection.

\subsection{Evaluating hereditary contributions using the QK parametrization}\label{subsec:heredEval}

As mentioned, there are two types of hereditary integrals to compute: the tail integrals containing logarithms, and the memory integrals. The latter enter the even $\ell+m$ modes and can be split into two classes of contributions: the DC memory, affecting only the $(\ell,0)$ modes, and the oscillatory memory. 
For both the memory and tail contributions, we compute the integrands for generic orbits, i.e., in terms of $(r,\dot{r},\phi,\dot{\phi})$. Then, we express them in terms of $(x,e_t,l)$ in a small-eccentricity expansion, in order to evaluate the integrals analytically.

\subsubsection{Tail integrals}

To express $(r,\dot{r},\phi,\dot{\phi})$ as functions of $(x,e_t,l)$, we invert Eq.~\eqref{eq:QKl} to obtain $u$ as a function of $l$. This relation reads (see, e.g., Refs.~\cite{Boetzel:2017zza,Boetzel:2019nfw})
\begin{equation}\label{eq:uofl}
u = l + \sum_{k=1}^\infty A_k \sin(k l),
\end{equation}
with 
\begin{equation}
A_k = \frac{2}{k} J_k(k e_t)+\sum_{j=1}^\infty \alpha_j\left[ J_{k+j}(k e_t)-J_{k-j}(k e_t) \right],
\end{equation}
where $J_k$ are Bessel functions and $\alpha_k$ reads
\begin{equation}
\alpha_k = 2 \beta_\phi^k\left(\frac{\sqrt{1-e_\phi^2}}{e_\phi} f_v+\frac{1}{k}f_{v-u}  \right)\,,
\end{equation}
with $\beta_\phi = (1-\sqrt{1-e_\phi^2})/e_\phi$. Note that the value of $\alpha_k$ is specific to this computation and has a more general form for higher PN orders in the QK parametrizations. Once $u(l)$ is known, one can insert it into the expressions for $(r,\dot{r},\phi,\dot{\phi})$ in terms of $(x,e_t,u)$, and perform an eccentricity expansion to get them as functions of $(x,e_t,l)$.

An important feature of this parametrization, and especially of $\phi(x,e_t,l)$, is the fact that the phase variable can be split into two parts: a linear-in-$l$ part given by $\lambda \equiv K l$, and an oscillatory part $W(l)$
\begin{equation}
\phi= \lambda+W(l).
\end{equation}
After expressing $\phi$ as a function of $l$ using Eqs.~\eqref{eq:QKphi} and \eqref{eq:uofl}, and performing an eccentricity expansion, we identify $W(l)=\phi(l) - \lambda$. This split is useful in evaluating the hereditary integrals.

With this parametrization at hand, we compute the integrands in Eqs.~\eqref{eq:tailsgen} for generic orbits, then express them in terms of $(x,e_t,l)$. After simplifying the expressions, having performed an eccentricity expansion, we evaluate the following elementary integrals:
\begin{equation}
\int_0^\infty \dd\tau \, \tau^k \ln\left(\frac{\tau}{2b}\right) \e^{\di\, \omega\tau} = \frac{\di^k k!}{\omega^{k+1}}\left[ -\frac{\pi}{2}\text{sign}(\omega) -\di\bigl[\ln(2b|\omega|) + \gamma_E - H_k \bigr]   \right]\,,    
\end{equation}
where $\omega\in\mathbb{R}^*$, $k\in\mathbb{N}$, $\gamma_E$ is the Euler constant and $H_k$ is the $k^\text{th}$ harmonic number. Note that the case $\omega = 0$ does not appear in this computation.

\subsubsection{Memory integrals}

Regarding the memory contributions, we start from Eq.~\eqref{eq:hlmmemdotgen} and we need to compute $\dot{h}_{\ell m}$. We computed them in two different ways: the first is to take the derivative of the radiative moments and project them on the spherical harmonic basis defined above; the second is to start from the modes themselves and differentiate them using~\cite{Ebersold:2019kdc}
\begin{equation}
\frac{\dd}{\dd t} = n\left[\frac{\dd}{\dd l}+K\frac{\dd}{\dd \lambda}\right] +\frac{\dd x}{\dd t} \frac{\dd}{\dd x}+\frac{\dd e}{\dd t} \frac{\dd}{\dd e}\,.
\end{equation}
Both approaches yield the same results. Once the source of the time integrals is known, we get two types of elementary integrals: oscillatory and non-oscillatory.

The oscillatory memory integrals are of the form
\begin{equation}
I_{r,s} = \int_{-\infty}^{T_R} \dd t \, x^p(t) e^q(t) \e^{\di [r\, l(t) + s \lambda(t)]}\,,
\end{equation}
which was derived in Appendix C of Ref.~\cite{Ebersold:2019kdc} in the case of nonspinning binaries. After generalizing this method to spins, we found that at the considered order for the spin terms, the form of the integral remains the same, namely
\begin{equation}\label{eq:OscMemIntegral}
I_{r,s} = -\frac{\di}{n(r +s K)}x^p(T_R) e^q(T_R) \e^{\di [r\, l(T_R) + s \lambda(T_R)]}\,.
\end{equation}
This is due to the fact that we do not need to include post-adiabatic corrections for the computation of the spin terms to 3PN order in the waveform modes. However, if one were to compute the memory integrals for the 3.5PN waveform, one would need to include the radiation reaction and thus $I_{r,s}$ would need to be generalized. 

A crucial point to note is that when $r=-s$ in Eq.~\eqref{eq:OscMemIntegral}, the PN order of $I_{r,s}$ is lowered compared to the source of this integral, since $K = 1 + \calO(2)$, which complicates the power counting when trying to be consistent to a given PN order. The oscillatory memory effects only enter the even $\ell+m$ modes, and we derived them up to the $\ell = 7$ modes. The results are displayed in Subsection~\ref{subsec:memoryres} and in the Supplemental Material.

The second type of memory integrals are the non-oscillating (or DC) integrals, which take the form
\begin{equation}
J_{p,q} = \int_{-\infty}^{T_R} \dd t \, x^p(t) e^q(t)\,.
\end{equation}
These effects only enter the $(\ell,0)$ modes. For the purpose of this computation, we need to introduce an initial eccentricity $e_0$, and substitute the time variable in terms of the eccentricity
\begin{equation}\label{eq:DCmemoryIntegrals}
J_{p,q} = \int_{e_0}^{e(T_R)} \dd e \, \frac{x^p(e) e^q}{\dot{e}} \,.
\end{equation}
Note that in the practical computations, we used for $\dot{e}$ the value of its orbit average, since we are interested in the secular contribution to these integrals. 
In Subsections~\ref{subsec:SecOrb} and \ref{subsec:horiz}, we derive the secular evolution of the orbital elements and notably $\langle \dot{e} \rangle$. Then, by knowing the expression of $\langle \dot{e} \rangle$ as a function of $(x,e)$ and the explicit value of $x(e)$, we can compute $J_{p,q}$.

\section{3PN energy and angular momentum fluxes for nonprecessing spins}
\label{sec:fluxes}
In this section, we present the 3PN energy and angular momentum fluxes, which are related to the radiative moments via Eqs.~\eqref{eq:fluxes_moments}.
We split the fluxes into instantaneous and tail contributions, such that
\begin{align}
\mathcal{F} &\equiv \mathcal{F}^\text{inst} + \mathcal{F}^\text{tail}, \\
\mathcal{G}_i &\equiv \mathcal{G}_i^\text{inst} + \mathcal{G}_i^\text{tail},
\end{align}
where the instantaneous contributions are valid for generic motion, while the tail contributions are computed for bound orbits in a small-eccentricity expansion to $\calO(e^8)$.

\subsection{Instantaneous contributions to the fluxes}
Using the instantaneous part of the radiative moments from Eqs.~\eqref{eq:instMoments} and plugging them in Eqs.~\eqref{eq:fluxes_moments}, we obtain the instantaneous contributions to the fluxes to 3PN order for the spin terms.
The 3PN nonspinning part of the energy flux is given by Eqs.~(5.2) of Ref.~\cite{Arun:2007sg}, while the angular momentum flux is given by Eqs.~(3.4) of Ref.~\cite{Arun:2009mc}. We only computed the nonspinning part to 2PN, and found agreement with the results of those references.

For the energy flux, we obtain the following SO and SS parts:
\begin{subequations}
\begin{align}
\mathcal{F}^\text{inst} &= \mathcal{F}^\text{inst}_\text{nS} + \mathcal{F}^\text{inst}_\text{SO} + \mathcal{F}^\text{inst}_\text{SS}, \\
\mathcal{F}^\text{inst}_\text{SO} &= \frac{32 \nu^2 G^4 M^5 \dot{\phi }}{5 c^5 r^4} \Bigg[ 
\frac{1}{c^3}\bigg\lbrace
\chi_S \left[\frac{G M}{r} \left(\frac{8 \nu }{3}-1\right)-\left(\nu +\frac{37}{12}\right) r^2 \dot{\phi }^2+\left(3 \nu -\frac{5}{6}\right) \dot{r}^2\right] 
-\delta  \chi _A \left(\frac{G M}{r}+\frac{37}{12} r^2 \dot{\phi }^2+\frac{5}{6} \dot{r}^2\right)  \bigg\rbrace\nonumber\\
&\quad
+ \frac{1}{c^5} \bigg\lbrace
\chi_S \bigg[
\frac{G^2 M^2}{r^2} \left(\frac{43 \nu ^2}{21}-\frac{571 \nu }{21}+\frac{1081}{84}\right)
+\frac{G M \dot{r}^2}{r} \left(\frac{187 \nu ^2}{7}-\frac{1366 \nu }{21}+\frac{388}{21}\right)
+ G M r \dot{\phi }^2 \left(\frac{741}{56}-\frac{55 \nu ^2}{3}+\frac{491 \nu }{12}\right) \nonumber\\
&\qquad\quad
+\left(\frac{40 \nu ^2}{7}-\frac{1369 \nu }{84}+\frac{85}{21}\right) \dot{r}^4
+\left(\frac{45 \nu ^2}{7}+\frac{83 \nu }{12}-\frac{2225}{336}\right) r^4 \dot{\phi }^4
+\left(-\frac{283 \nu ^2}{7}+\frac{2197 \nu }{84}+\frac{85}{168}\right) r^2 \dot{r}^2 \dot{\phi }^2
\bigg] \nonumber\\
&\qquad
+ \delta\chi_A \bigg[
\frac{G^2 M^2}{r^2} \left(\frac{38 \nu }{21}+\frac{1081}{84}\right)
+G M r \dot{\phi }^2\left(\frac{13 \nu }{4}+\frac{741}{56}\right) 
+\frac{G M \dot{r}^2}{r} \left(\frac{388}{21}-5 \nu \right)
+\left(\frac{85}{21}-\frac{145 \nu }{84}\right) \dot{r}^4 \nonumber\\
&\qquad\quad
+\left(\frac{55 \nu }{6}-\frac{2225}{336}\right) r^4 \dot{\phi }^4
+\left(\frac{157 \nu }{84}+\frac{85}{168}\right) r^2 \dot{r}^2 \dot{\phi }^2
\bigg] 
\bigg\rbrace \nonumber\\
&\quad
+ \frac{1}{c^6}\bigg\lbrace
\chi_S \left[\frac{G^2 M^2 \dot{r}}{r^2} (2 \nu -1)+G M r \dot{r} \dot{\phi }^2 \left(19 \nu -\frac{19}{2}\right)+\frac{G M \dot{r}^3}{r} \left(\frac{1}{2}-\nu \right)\right] \nonumber\\
&\qquad
+\delta  \chi _A\left[\frac{G M \dot{r}^3}{2 r}-\frac{G^2 M^2 \dot{r}}{r^2}-\frac{19}{2} G M r \dot{r} \dot{\phi }^2\right]
\bigg\rbrace
+ \calO(7)
\Bigg], \\
\mathcal{F}^\text{inst}_\text{SS} &= \frac{32 \nu ^2 G^5 M^6}{5 c^5 r^6} \Bigg[ 
\frac{1}{c^4}\bigg\lbrace
\chi _A^2 \left[\left(\frac{49}{16}-12 \nu \right) r^2 \dot{\phi }^2+\nu  \dot{r}^2\right]
+\chi _S^2 \left[\left(\frac{49}{16}-\frac{\nu }{4}\right) r^2 \dot{\phi }^2-\nu  \dot{r}^2\right]
+\frac{49}{8} \delta  r^2 \dot{\phi }^2 \chi _A \chi _S \nonumber\\
&\qquad\quad
+ \delta  \kapA \left(3 r^2 \dot{\phi }^2-\frac{\dot{r}^2}{4}\right)
+\kapS \left[(3-6 \nu ) r^2 \dot{\phi }^2+\left(\frac{\nu }{2}-\frac{1}{4}\right) \dot{r}^2\right]\bigg\rbrace \nonumber\\
&\quad
+ \frac{1}{c^6} \bigg\lbrace
\chi_A^2 \bigg[
\frac{G^2 M^2}{r^2} \left(\frac{24 \nu ^2}{7}-\frac{59 \nu }{21}+\frac{41}{84}\right)
+\frac{G M \dot{r}^2}{r}  \left(\frac{25 \nu ^2}{7}-\frac{144 \nu }{7}+\frac{19}{84}\right)
+ G M r \dot{\phi }^2 \left(-\frac{30 \nu ^2}{7}+\frac{7577 \nu }{84}-\frac{989}{42}\right)\nonumber\\
&\qquad\quad
+\left(\frac{32 \nu ^2}{7}-\frac{104 \nu }{21}-\frac{4}{7}\right) \dot{r}^4
+\left(\frac{191 \nu ^2}{7}-\frac{7327 \nu }{168}+\frac{555}{56}\right) r^4 \dot{\phi }^4
+\left(-\frac{365 \nu ^2}{7}+\frac{5083 \nu }{168}-\frac{569}{112}\right) r^2 \dot{r}^2 \dot{\phi }^2
\bigg] \nonumber\\
&\qquad
+ \chi_S^2 \bigg[
\frac{G^2 M^2}{r^2} \left(\frac{38 \nu ^2}{21}-\frac{16 \nu }{7}+\frac{41}{84}\right)
+G M r \dot{\phi }^2 \left(-\frac{47 \nu ^2}{21}-\frac{2437 \nu }{84}-\frac{989}{42}\right)
+\frac{G M \dot{r}^2}{r} \left(-\frac{5 \nu ^2}{7}+\frac{362 \nu }{21}+\frac{19}{84}\right) \nonumber\\
&\qquad\quad
+\left(-\frac{2 \nu ^2}{7}+5 \nu -\frac{4}{7}\right) \dot{r}^4
+ \left(-\frac{\nu ^2}{14}-\frac{239 \nu }{56}+\frac{555}{56}\right) r^4 \dot{\phi }^4
+\left(\frac{85 \nu ^2}{14}-\frac{107 \nu }{8}-\frac{569}{112}\right) r^2 \dot{r}^2 \dot{\phi }^2
\bigg] \nonumber\\
&\qquad
+ \delta \chi_S\chi_A \bigg[
\frac{G^2 M^2}{r^2} \left(\frac{41}{42}-\frac{22 \nu }{7}\right)
+G M r \dot{\phi }^2 \left(-33 \nu -\frac{989}{21}\right)
+\frac{G M \dot{r}^2}{r} \left(\frac{19}{42}-\frac{17 \nu }{7}\right)
+\left(\frac{555}{28}-\frac{173 \nu }{21}\right) r^4 \dot{\phi }^4 \nonumber\\
&\qquad\quad
+\left(-\frac{289 \nu }{84}-\frac{569}{56}\right) r^2 \dot{r}^2 \dot{\phi }^2+\left(-\frac{47 \nu }{21}-\frac{8}{7}\right) \dot{r}^4
\bigg] 
+ \delta \kapA \bigg[
\frac{G^2 M^2}{r^2} \left(\frac{5}{21}-\frac{19 \nu }{21}\right)
+G M r \dot{\phi }^2 \left(-\frac{79 \nu }{12}-\frac{769}{28}\right)\nonumber\\
&\qquad\quad
+\frac{G M \dot{r}^2}{r} \left(\frac{109}{28}-\frac{25 \nu }{84}\right)
+\left(\frac{225}{28}-\frac{443 \nu }{56}\right) r^4 \dot{\phi }^4
+\left(\frac{130 \nu }{7}-\frac{389}{28}\right) r^2 \dot{r}^2 \dot{\phi }^2
+\left(\frac{19}{28}-\frac{9 \nu }{7}\right) \dot{r}^4
\bigg] \nonumber\\
&\qquad
+ \kapS \bigg[
\frac{G^2 M^2}{r^2} \left(\frac{12 \nu ^2}{7}-\frac{29 \nu }{21}+\frac{5}{21}\right)
+\frac{G M \dot{r}^2}{r} \left(\frac{25 \nu ^2}{14}-\frac{97 \nu }{12}+\frac{109}{28}\right)
+ G M r \dot{\phi }^2 \left(-\frac{15 \nu ^2}{7}+\frac{4061 \nu }{84}-\frac{769}{28}\right)
 \nonumber\\
&\quad\qquad
+\left(\frac{16 \nu ^2}{7}-\frac{37 \nu }{14}+\frac{19}{28}\right) \dot{r}^4
+\left(\frac{191 \nu ^2}{14}-\frac{1343 \nu }{56}+\frac{225}{28}\right) r^4 \dot{\phi }^4
+\left(-\frac{365 \nu ^2}{14}+\frac{649 \nu }{14}-\frac{389}{28}\right) r^2 \dot{r}^2 \dot{\phi }^2
\bigg]
\bigg\rbrace \nonumber\\
&\quad
+\calO(7)
\Bigg].
\end{align}
\end{subequations}
This result agrees with Refs.~\cite{Bohe:2015ana,Cho:2021mqw,Cho:2022syn}. However, the flux is defined in Ref.~\cite{Cho:2022syn} in terms of the source moments instead of the radiative moments, which means the 3PN SO contribution was not taken into account, while we find agreement on the overlapping terms. Nonetheless, these terms vanish when taking an orbit average.

For the angular momentum flux, we obtain the following SO and SS parts:
\begin{subequations}
\begin{align}
\mathcal{G}^\text{inst}_i &= \hat{l}_i \left(\mathcal{G}^\text{inst}_\text{nS} + \mathcal{G}^\text{inst}_\text{SO} + \mathcal{G}^\text{inst}_\text{SS}\right), \\
\mathcal{G}^\text{inst}_\text{SO} &= \frac{16 \nu ^2 G^3 M^4}{5 c^5 r^3} \Bigg[ 
\frac{1}{c^3} \bigg\lbrace
\chi_S \bigg[
\frac{G^2 M^2}{r^2} \left(\frac{4 \nu }{3}-\frac{1}{2}\right)
+G M r \dot{\phi }^2 \left(\frac{14 \nu }{3}-\frac{25}{6}\right)
+\frac{2 \nu G M \dot{r}^2}{3 r}
-\left(\frac{2 \nu }{3}+\frac{3}{2}\right) r^4 \dot{\phi }^4 \nonumber\\
&\qquad\quad
+\left(\frac{11 \nu }{3}-\frac{3}{2}\right) r^2 \dot{r}^2 \dot{\phi }^2
-\frac{2}{3} \nu  \dot{r}^4
\bigg]
+ \delta\chi_A \left[-\frac{G^2 M^2}{2 r^2}-\frac{25}{6} G M r \dot{\phi }^2-\frac{3}{2} r^4 \dot{\phi }^4-\frac{3}{2} r^2 \dot{r}^2 \dot{\phi }^2\right]
\bigg\rbrace \nonumber\\
&\quad
+ \frac{1}{c^5} \bigg\lbrace
\chi_S \bigg[
\frac{G^3 M^3}{r^3} \left(\frac{4 \nu ^2}{21}-\frac{1273 \nu }{84}+\frac{295}{42}\right)
+\frac{G^2 M^2 \dot{r}^2}{r^2} \left(\frac{131 \nu ^2}{21}-\frac{473 \nu }{84}+\frac{17}{6}\right)
+\frac{G M \dot{r}^4}{r} \left(\frac{251 \nu }{42}-\frac{31 \nu ^2}{7}-\frac{115}{21}\right) \nonumber\\
&\qquad\quad
+ G^2 M^2 \dot{\phi }^2 \left(-\frac{415 \nu ^2}{42}+\frac{605 \nu }{168}+\frac{4147}{168}\right)
+G M r \dot{r}^2 \dot{\phi }^2 \left(\frac{45 \nu ^2}{2}-\frac{419 \nu }{14}+\frac{2083}{168}\right)
+\left(\frac{23 \nu ^2}{21}+\frac{13 \nu }{14}\right) \dot{r}^6 \nonumber\\
&\qquad\quad
+\left(\frac{107 \nu ^2}{14}+\frac{90 \nu }{7}-\frac{51}{14}\right) r^2 \dot{r}^4 \dot{\phi }^2
+ G M r^3 \dot{\phi }^4 \left(-\frac{827 \nu ^2}{28}+\frac{15641 \nu }{336}-\frac{37}{4}\right)
+\left(\frac{503 \nu ^2}{84}+\frac{727 \nu }{112}-\frac{87}{28}\right) r^6 \dot{\phi }^6 \nonumber\\
&\qquad\quad
+\left(-\frac{306 \nu ^2}{7}-\frac{123 \nu }{28}+\frac{3}{4}\right) r^4 \dot{r}^2 \dot{\phi }^4
\bigg] 
+ \delta\chi_A \bigg[
\frac{G^3 M^3}{r^3} \left(\frac{9 \nu }{28}+\frac{295}{42}\right)
+G^2 M^2 \dot{\phi }^2 \left(\frac{947 \nu }{168}+\frac{4147}{168}\right)
\nonumber\\
&\qquad\quad
+\frac{G^2  M^2 \dot{r}^2}{r^2}\left(\frac{67 \nu }{84}+\frac{17}{6}\right)
+\frac{G M \dot{r}^4}{r} \left(-\frac{23 \nu }{42}-\frac{115}{21}\right)
-\frac{5}{42} \nu  \dot{r}^6
+G  M r^3 \dot{\phi }^4\left(\frac{5995 \nu }{336}-\frac{37}{4}\right) \nonumber\\
&\qquad\quad
+G M r \dot{r}^2 \dot{\phi }^2 \left(\frac{2083}{168}-\frac{54 \nu }{7}\right)
+\left(\frac{375 \nu }{28}+\frac{3}{4}\right) r^4 \dot{r}^2 \dot{\phi }^4
-\frac{51}{14} r^2 \dot{r}^4 \dot{\phi }^2
+\left(\frac{1835 \nu }{336}-\frac{87}{28}\right) r^6 \dot{\phi }^6
\bigg]\bigg\rbrace \nonumber\\
&\quad
+ \frac{1}{c^6} \bigg\lbrace
\chi_S \bigg[
\frac{G^3 M^3 \dot{r}}{r^3} \left(\frac{3 \nu }{2}-\frac{3}{4}\right)
+G^2 M^2 \dot{r} \dot{\phi }^2 \left(15 \nu -\frac{15}{2}\right)
+G  M r \dot{r}^3 \dot{\phi }^2\left(\frac{39}{4}-\frac{39 \nu }{2}\right)
+G M r^3 \dot{r} \dot{\phi }^4 \left(\frac{33 \nu }{2}-\frac{33}{4}\right) \nonumber\\
&\qquad\quad
+\frac{G M \dot{r}^5}{r} \left(\frac{1}{2}-\nu \right)
\!\bigg]
+ \delta\chi_A \left[
\frac{G M \dot{r}^5}{2 r}-\frac{3 G^3 M^3 \dot{r}}{4 r^3}-\frac{15}{2} G^2 M^2 \dot{r} \dot{\phi }^2-\frac{33}{4} G M r^3 \dot{r} \dot{\phi }^4+\frac{39}{4} G M r \dot{r}^3 \dot{\phi }^2
\right]\!
\bigg\rbrace 
+ \calO(7)
\Bigg], \\
\mathcal{G}^\text{inst}_\text{SS} &= \frac{16 \nu ^2 G^4 M^5 \dot{\phi }}{5 c^5 r^3} \Bigg[ 
\frac{1}{c^4} \bigg\lbrace
\chi_S^2 \left[\frac{G M}{r} \left(\frac{25}{8}-\frac{\nu }{2}\right)+\frac{3}{2} r^2 \dot{\phi }^2-\frac{9}{4} \dot{r}^2\right]
+ \delta\chi_S\chi_A \left[\frac{25 G M}{4 r}+3 r^2 \dot{\phi }^2-\frac{9}{2} \dot{r}^2\right]\nonumber\\
&\qquad\quad
+\chi_A^2 \left[\frac{G  M}{r}\left(\frac{25}{8}-12 \nu \right)+\left(\frac{3}{2}-6 \nu \right) r^2 \dot{\phi }^2+\left(9 \nu -\frac{9}{4}\right) \dot{r}^2\right] 
+ \delta \kapA \left[\frac{3 G M}{r}+\frac{3}{2} r^2 \dot{\phi }^2-\frac{9}{4} \dot{r}^2\right] \nonumber\\
&\qquad\quad
+ \kapS \left[\frac{G (3-6 \nu ) M}{r}+\left(\frac{3}{2}-3 \nu \right) r^2 \dot{\phi }^2+\left(\frac{9 \nu }{2}-\frac{9}{4}\right) \dot{r}^2\right]
\bigg\rbrace \nonumber\\
&\quad
+ \frac{1}{c^6} \bigg\lbrace
\chi_A^2 \bigg[
\frac{G^2 M^2}{r^2} \left(\frac{18439 \nu }{168}-\frac{\nu ^2}{7}-\frac{1623}{56}\right)
+\frac{G M \dot{r}^2}{r} \left(-\frac{31 \nu ^2}{7}-\frac{4003 \nu }{56}+\frac{377}{21}\right)
+\left(\frac{173 \nu ^2}{14}-\frac{107 \nu }{8}+\frac{59}{14}\right) \dot{r}^4 \nonumber\\
&\qquad\quad
+ G M r \dot{\phi }^2 \left(\frac{234 \nu ^2}{7}-\frac{1923 \nu }{56}+\frac{1069}{168}\right)
+\left(17 \nu ^2-\frac{407 \nu }{28}+\frac{367}{112}\right) r^4 \dot{\phi }^4
+\left(\frac{747 \nu }{56}-\frac{639 \nu ^2}{14}-\frac{333}{56}\right) r^2 \dot{r}^2 \dot{\phi }^2\bigg] \nonumber\\
&\qquad
+ \chi_S^2 \bigg[
\frac{G^2 M^2}{r^2} \left(\frac{121 \nu ^2}{14}-\frac{4141 \nu }{168}-\frac{1623}{56}\right)
+\frac{G M \dot{r}^2}{r} \left(\frac{247 \nu ^2}{42}-\frac{29 \nu }{168}+\frac{377}{21}\right)
+\left(4 \nu ^2-\frac{685 \nu }{56}+\frac{59}{14}\right) \dot{r}^4 \nonumber\\
&\qquad\quad
+ G M r \dot{\phi }^2 \left(-\frac{83 \nu ^2}{42}-\frac{5993 \nu }{168}+\frac{1069}{168}\right)
+\left(-\nu ^2-5 \nu +\frac{367}{112}\right) r^4 \dot{\phi }^4
+\left(3 \nu ^2+\frac{1275 \nu }{56}-\frac{333}{56}\right) r^2 \dot{r}^2 \dot{\phi }^2
\bigg] \nonumber\\
&\qquad
+ \delta\chi_S\chi_A \bigg[
\frac{G^2 M^2}{r^2} \left(-\frac{863 \nu }{28}-\frac{1623}{28}\right)
+G M r \dot{\phi }^2 \left(\frac{1069}{84}-\frac{3743 \nu }{84}\right)
+\frac{G M \dot{r}^2}{r} \left(\frac{13 \nu }{84}+\frac{754}{21}\right) \nonumber\\
&\qquad\quad
+\left(\frac{59}{7}-\frac{35 \nu }{4}\right) \dot{r}^4
+\left(\frac{367}{56}-\frac{45 \nu }{7}\right) r^4 \dot{\phi }^4
+\left(\frac{345 \nu }{28}-\frac{333}{28}\right) r^2 \dot{r}^2 \dot{\phi }^2\bigg] \nonumber\\
&\qquad
+ \kapS \bigg[
\frac{G^2 M^2}{r^2} \left(-\frac{\nu ^2}{14}+\frac{1517 \nu }{28}-\frac{2599}{84}\right)
+\frac{G M \dot{r}^2}{r} \left(-\frac{31 \nu ^2}{14}-\frac{121 \nu }{7}+\frac{1151}{84}\right)
+\left(\frac{173 \nu ^2}{28}-\frac{885 \nu }{56}+\frac{59}{14}\right) \dot{r}^4 \nonumber\\
&\qquad\quad
+ G M r \dot{\phi }^2 \left(\frac{117 \nu ^2}{7}-\frac{461 \nu }{28}+\frac{47}{21}\right)
+\left(\frac{17 \nu ^2}{2}-\frac{645 \nu }{56}+\frac{367}{112}\right) r^4 \dot{\phi }^4
+\left(-\frac{639 \nu ^2}{28}+\frac{1725 \nu }{56}-\frac{333}{56}\right) r^2 \dot{r}^2 \dot{\phi }^2
\bigg]\nonumber\\
&\qquad
+ \delta\kapA \bigg[
\frac{G^2 M^2}{r^2} \left(-\frac{647 \nu }{84}-\frac{2599}{84}\right)
+G M r \dot{\phi }^2\left(\frac{47}{21}-\frac{1007 \nu }{84}\right)
+\frac{G M \dot{r}^2}{r} \left(\frac{425 \nu }{42}+\frac{1151}{84}\right)
+\left(\frac{59}{14}-\frac{59 \nu }{8}\right) \dot{r}^4 \nonumber\\
&\qquad\quad
+\left(\frac{367}{112}-\frac{139 \nu }{28}\right) r^4 \dot{\phi }^4
+\left(\frac{1059 \nu }{56}-\frac{333}{56}\right) r^2 \dot{r}^2 \dot{\phi }^2
\bigg]
\bigg\rbrace
+ \calO(7)
\Bigg].
\end{align}
\end{subequations}
where this result agrees with Ref.~\cite{Cho:2021mqw}, except for the 3PN SO contribution, which was not accounted for in that paper.

\subsection{Tail contributions to the fluxes}\label{subsec:tailflux}

To compute the tail contribution to the fluxes, we use the radiative moments from Eqs.~\eqref{eq:tailsgen} computed in a small-eccentricity expansion, as explained in Subsection~\ref{subsec:heredEval}, then plug the moments in Eqs.~\eqref{eq:fluxes_moments}.

To the 3PN order, the only spin terms in the tail part of the fluxes are SO contributions at 3PN order. The nonspinning tail contributions, starting at 1.5PN, were derived to 3PN in Refs.~\cite{Arun:2007rg,Arun:2009mc,Ebersold:2019kdc}.
We write here the SO part of the tail to $\calO(e^2)$ and provide the full expressions to $\calO(e^8)$ in the Supplemental Material.
Henceforth, we use $e$ instead of $e_t$ to ease the notation.

For the energy flux, we obtain
\begin{subequations}
\begin{align}
\mathcal{F}^\text{tail} &= \mathcal{F}^\text{tail}_\text{nS} + \mathcal{F}^\text{tail}_\text{SO}, \\
\mathcal{F}^\text{tail}_\text{SO} &= \frac{32 \nu^2 c^5}{15G} x^8
\Bigg[
\pi  \left(34 \nu -\frac{65}{2}\right) \chi _S-\frac{65}{2} \pi  \delta  \chi _A
+ e \bigg\lbrace
\chi_S \bigg[
\frac{\pi}{2} \left(\e^{-\di l}+\e^{\di l}\right) (321 \nu -465) 
+ \frac{\di}{2} \left(\e^{-\di l}-\e^{\di l}\right)
\bigg((64 \nu -46) \ln b\nonumber\\
&\qquad\quad
+\left(64 \gamma_E-\frac{404}{3}\right) \nu -46 \gamma_E+(654-312 \nu ) \ln 2+(405 \nu -648) \ln 3+(96 \nu -69) \ln x+\frac{1123}{6}\bigg)
\bigg] \nonumber\\
&\qquad
+ \delta\chi_A \bigg[
-\frac{465}{2} \pi  \left(\e^{-\di l}+\e^{\di l}\right)
+ \frac{\di}{2} \left(\e^{-\di l}-\e^{\di l}\right)
\bigg(-46 \ln b-46 \gamma_E-69 \ln x+\frac{1123}{6}+654 \ln 2-648 \ln 3\bigg)
\bigg]
\bigg\rbrace \nonumber\\
&\quad
+ e^2 \bigg\lbrace
\chi_S \bigg[\pi  \left(\frac{1}{2} \left(\e^{-2 \di l}+\e^{2 \di l}\right) \left(\frac{22975 \nu }{24}-\frac{41591}{24}\right)+\frac{15173 \nu }{24}-\frac{6481}{6}\right)
+ \frac{\di}{2}  \left(\e^{-2 \di l}-\e^{2 \di l}\right)
\bigg(\frac{41377}{36} -\frac{1631 \gamma_E}{3} \nonumber\\
&\qquad\quad
+\left(\frac{7135 \nu }{3}-\frac{13727}{3}\right) \ln 2
+\left(\frac{1087 \nu }{3}-\frac{1631}{3}\right) \ln b+\left(\frac{1087 \gamma_E}{3}-\frac{24305}{36}\right) \nu +\left(\frac{7533}{8}-\frac{2673 \nu }{8}\right) \ln 3\nonumber\\
&\qquad\quad
+\left(\frac{1087 \nu }{2}-\frac{1631}{2}\right) \ln x
\bigg)
\bigg] 
+ \delta\chi_A \bigg[
\pi  \left(-\frac{6481}{6}-\frac{1}{48} 41591 \left(\e^{-2 \di l}+\e^{2 \di l}\right)\right) \nonumber\\
&\qquad\quad
+ \frac{\di}{2}  \left(\e^{-2 \di l}-\e^{2 \di l}\right) \left(-\frac{1631 \ln b}{3}-\frac{1631 \gamma_E}{3}-\frac{1631 \ln x}{2}+\frac{41377}{36}-\frac{13727}{3}  \ln 2+\frac{7533 \ln 3}{8}\right)
\bigg]
\bigg\rbrace \nonumber\\
&\quad 
+ \calO(e^3)\Bigg]+ \calO(x^9),
\end{align}
\end{subequations}
and for the angular-momentum flux
\begin{subequations}
\begin{align}
\mathcal{G}^\text{tail}_i &= \hat{l}_i \left(\mathcal{G}^\text{tail}_\text{nS} + \mathcal{G}^\text{tail}_\text{SO}\right), \\
\mathcal{G}^\text{tail}_\text{SO} &= \frac{32 M \nu^2 c^2}{15} x^{13/2}
\Bigg[
\pi  \left(34 \nu -\frac{65}{2}\right) \chi _S-\frac{65}{2} \pi  \delta  \chi _A
+ e \bigg\lbrace
\chi_S \bigg[
\frac{\pi}{2}  \left(\e^{-\di l}+\e^{\di l}\right) \left(\frac{513 \nu }{2}-369\right) 
+ \frac{\di}{2} \left(\e^{-\di l}-\e^{\di l}\right) \bigg(
\frac{545}{4}\nonumber\\
&\qquad\qquad
+(48 \nu -30) \ln b-101\nu +(48\nu - 30) \gamma_E +(666-324 \nu ) \ln 2+\left(\frac{729 \nu }{2}-594\right) \ln 3 
+(72 \nu -45) \ln x\bigg)
\bigg] \nonumber\\
&\qquad
+ \delta\chi_A \bigg[
-\frac{369}{2} \pi \left(\e^{-\di l}+\e^{\di l}\right)
+ \frac{\di}{2} \left(e^{-\di l}-\e^{\di l}\right) \left(-30 \ln b-30 \gamma_E-45 \ln x+\frac{545}{4}+666 \ln 2-594 \ln 3\right)
\bigg]
\bigg\rbrace \nonumber\\
&\quad
+ e^2 \bigg\lbrace
\chi_S\bigg[
\pi  \left(\frac{1}{2} \left(\e^{-2 \di l}+\e^{2 \di l}\right) \left(648 \nu -\frac{2295}{2}\right)+385 \nu -\frac{2575}{4}\right)
+ \frac{\di}{2}  \left(\e^{-2 \di l}-\e^{2 \di l}\right)
\bigg(
\left(213 \nu -\frac{555}{2}\right) \ln b\nonumber\\
&\qquad\qquad
+\gamma_E \left(213 \nu -\frac{555}{2}\right)+\left(1869 \nu -\frac{7179}{2}\right) \ln 2+\left(\frac{4455}{4}-\frac{891 \nu }{2}\right) \ln 3+\left(\frac{639 \nu }{2}-\frac{1665}{4}\right) \ln x\nonumber\\
&\qquad\qquad
+\frac{2687}{4}-\frac{1663 \nu }{4}
\bigg)
\bigg] 
+ \delta\chi_A \bigg[
\pi  \left(-\frac{2575}{4}-\frac{2295}{4} \left(\e^{-2 \di l}+\e^{2 \di l}\right)\right)
+ \frac{\di}{2}  \left(\e^{-2 \di l}-\e^{2 \di l}\right) \bigg(-\frac{555 \ln b}{2}\nonumber\\
&\qquad\qquad
-\frac{555 \gamma_E}{2} -\frac{1665 \ln x}{4}+\frac{2687}{4}-\frac{7179}{2}  \ln 2+\frac{4455 \ln 3}{4}\bigg)
\bigg]
\bigg\rbrace
+ \calO(e^3)
\Bigg]+ \calO(x^{15/2}).
\end{align}
\end{subequations}
We checked that our results agree in the circular-orbit limit with Ref.~\cite{Cho:2022syn}.

\subsection{Orbit-averaged fluxes}
To compute the orbit average of the instantaneous fluxes, we express them using the QK parametrization in terms of $(x,e,u)$, then evaluate the integral
\begin{equation}
\langle \mathcal{F}^\text{inst} \rangle = \frac{1}{P} \int_{0}^{P} \mathcal{F}^\text{inst} \dd t = \frac{1}{2\pi} \int_{0}^{2\pi} \mathcal{F}^\text{inst} \frac{\dd l}{\dd u} \dd u,
\end{equation}
where $P$ is the radial period and $\langle ... \rangle$ stands for the average over one orbit.
To compute this integral, we made use of the following formulas, with $k$ and $n\in\mathbb{N}$: 
\begin{subequations}
\begin{align}
\int_0^{2\pi}\frac{\dd u}{2\pi} \frac{\cos{(ku)}}{(1-e\cos{u})^n} &= \frac{(-1)^k}{(1-e)^n} \, {}_3\widetilde{F}_2\left(n,1,\frac{1}{2};k+1,1-k;-\frac{2e}{1-e}\right)\label{eq:intcosku}\,,\\
\int_0^{2\pi}\frac{\dd u}{2\pi} \frac{\sin{(ku)}}{(1-e\cos{u})^n} &= 0\,,
\end{align}
\end{subequations}
where ${}_3\widetilde{F}_2(\alpha_1,\alpha_2,\alpha_3;\beta_1,\beta_2;z)\equiv{}_3F_2(\alpha_1,\alpha_2,\alpha_3;\beta_1,\beta_2;z)/(\Gamma(\beta_1)\Gamma(\beta_2))$ is the usual normalized hypergeometric function. Note that in the case of $k=0$, this integral reduces to the well known result
\begin{equation}
\int_0^{2\pi}\frac{\dd u}{2\pi} \frac{1}{(1-e\cos{u})^n} = \frac{1}{(1-e^2)^{n/2}}P_{n-1}\left(\frac{1}{\sqrt{1-e^2}}\right)\,,
\end{equation}
where $P_n$ is the Legendre polynomials.

The nonspinning part of the orbit-averaged fluxes (instantaneous and tail contributions) were computed in Refs.~\cite{Arun:2007sg,Arun:2007rg,Arun:2009mc,Ebersold:2019kdc}, and we find agreement with their results to 2PN order.
Our result for the 3PN spin contributions to the orbit-averaged instantaneous energy flux reads
\begin{subequations}
\label{FinstAvg}
\begin{align}
\langle \mathcal{F}^\text{inst} \rangle &= \langle \mathcal{F}^\text{inst}_\text{nS} \rangle + \langle \mathcal{F}^\text{inst}_\text{SO} \rangle + \langle \mathcal{F}^\text{inst}_\text{SS} \rangle\\
\langle \mathcal{F}^\text{inst}_\text{SO} \rangle &= 
-\frac{x^{13/2}c^5\nu^2}{G\left(1-e^2\right)^5} \bigg\lbrace
\chi _S \left[e^6 \left(\frac{4 \nu }{5}+\frac{13}{2}\right)+e^4 \left(\frac{9059}{45}-\frac{2392 \nu }{45}\right)+e^2 \left(\frac{11612}{45}-\frac{5584 \nu }{45}\right)-\frac{96 \nu }{5}+\frac{88}{5}\right] \nonumber\\
&\qquad
+\delta \chi _A \left[\frac{13 e^6}{2}+\frac{9059 e^4}{45}+\frac{11612 e^2}{45}+\frac{88}{5}\right]
\bigg\rbrace \nonumber\\
&\quad
+\frac{x^{15/2}c^5\nu^2}{G \left(1-e^2\right)^6} \bigg\lbrace
\frac{2\sqrt{1-e^2}}{45}  \left(74 e^6+3039 e^4+5140 e^2+672\right)
\left[2 \delta  (\nu -3) \chi _A-\left(\nu ^2-8 \nu +6\right) \chi _S\right] \nonumber\\
&\qquad
+ \chi_S \bigg[
e^8 \left(\frac{58 \nu ^2}{15}+\frac{10001 \nu }{336}-\frac{10687}{448}\right)
+e^6 \left(-\frac{6317 \nu ^2}{15}+\frac{16369 \nu }{7}-\frac{37390}{21}\right)\nonumber\\
&\qquad\quad
+e^4 \left(\frac{871582 \nu }{105}-\frac{18164 \nu ^2}{9}-\frac{200365}{42}\right)
+e^2 \left(\frac{221356 \nu }{63}-\frac{3656 \nu ^2}{3}-\frac{29304}{35}\right)
-\frac{736 \nu ^2}{9}-\frac{3488 \nu }{45}+\frac{778}{5}
\bigg] \nonumber\\
&\qquad
+ \delta\chi_A \bigg[
e^8 \left(\frac{377 \nu }{12}-\frac{10687}{448}\right)+e^6 \left(\frac{26051 \nu }{15}-\frac{37390}{21}\right)+e^4 \left(\frac{25908 \nu }{5}-\frac{200365}{42}\right)+e^2 \left(\frac{97268 \nu }{45}-\frac{29304}{35}\right)\nonumber\\
&\qquad\quad
+\frac{584 \nu }{9}+\frac{778}{5}
\bigg]
\bigg\rbrace + \calO\left(x^{17/2}\right),\\
\langle \mathcal{F}^\text{inst}_\text{SS} \rangle &= 
\frac{x^7c^5\nu^2}{G\left(1-e^2\right)^{11/2}} \bigg\lbrace
\chi_A^2 \left[e^6 \left(\frac{49}{8}-\frac{118 \nu }{5}\right)+ e^4 \left(\frac{8203}{60}-\frac{7996 \nu }{15}\right)+ e^2 \left(\frac{827}{5}-\frac{3232 \nu }{5}\right)-\frac{256 \nu }{5}+\frac{66}{5}\right]\nonumber\\
&\qquad
+ \chi_S^2 \left[e^6 \left(\frac{49}{8}-\frac{9 \nu }{10}\right)+e^4 \left(\frac{8203}{60}-\frac{69 \nu }{5}\right)+e^2 \left(\frac{827}{5}-\frac{76}{5} \nu\right)+ \frac{66}{5}-\frac{8}{5} \nu\right]\nonumber\\
&\qquad
+ \delta\chi_S\chi_A \left[\frac{49 e^6}{4}+\frac{8203 e^4}{30}+\frac{1654 e^2}{5}+\frac{132}{5}\right] 
+\left(\delta \kapA +\kapS(1-2 \nu)\right) \left(\frac{59 e^6}{10}+\frac{1999 e^4}{15}+\frac{808 e^2}{5}+\frac{64}{5}\right)
\bigg\rbrace \nonumber\\
&\quad
+ \frac{x^8c^5\nu^2}{\left(1-e^2\right)^{13/2}} \bigg\lbrace
\frac{\sqrt{1-e^2}}{90}\left(74 e^6+3039 e^4+5140 e^2+672\right)
\Big[
-2\left(6 \nu ^2-46 \nu +11\right) \chi _A^2
- 2\left(4 \nu ^2-16 \nu +11\right) \chi _S^2  \nonumber\\
&\qquad
+4 \delta  (9 \nu -11) \chi _A \chi _S
+\delta  \kapA (5 \nu -14)+\kapS \left(-6 \nu ^2+33 \nu -14\right)\Big]
+\chi_A^2 \bigg[
e^8 \left(\frac{2183 \nu ^2}{15}-\frac{64675 \nu }{336}+\frac{90693}{2240}\right) \nonumber\\
&\qquad\quad
+e^6 \left(\frac{243961 \nu ^2}{45}-\frac{12907597 \nu }{1260}+\frac{1899671}{840}\right)+e^4 \left(\frac{633316 \nu ^2}{45}-\frac{57077599 \nu }{1890}+\frac{25466111}{3780}\right) \nonumber\\
&\qquad\quad
+e^2 \left(\frac{245032 \nu ^2}{45}-\frac{8869958 \nu }{945}+\frac{1895633}{945}\right)
+\frac{928 \nu ^2}{5}+\frac{183004 \nu }{315}-\frac{9860}{63}
\bigg] \nonumber\\
&\qquad
+ \chi_S^2 \bigg[
e^8 \left(\frac{111 \nu ^2}{20}-\frac{10139 \nu }{240}+\frac{90693}{2240}\right)+e^6 \left(\frac{2323 \nu ^2}{15}-\frac{376537 \nu }{180}+\frac{1899671}{840}\right)
+\frac{16 \nu ^2}{9}+\frac{236 \nu }{9}-\frac{9860}{63}\nonumber\\
&\qquad\quad
+e^4 \left(\frac{28930 \nu ^2}{27}-\frac{2280811 \nu }{270}+\frac{25466111}{3780}\right)+e^2 \left(\frac{23368 \nu ^2}{27}-\frac{622742 \nu }{135}+\frac{1895633}{945}\right)\bigg] \nonumber\\
&\qquad
+ \delta\chi_A\chi_S \bigg[
e^8 \left(\frac{90693}{1120}-\frac{17467 \nu }{240}\right)
+e^6 \left(\frac{1899671}{420}-\frac{59219 \nu }{18}\right)+e^4 \left(\frac{25466111}{1890}-\frac{1579361 \nu }{135}\right)\nonumber\\
&\qquad\quad
+e^2 \left(\frac{3791266}{945}-\frac{161332 \nu }{27}\right)-\frac{848 \nu }{45}-\frac{19720}{63}
\bigg]
+ \delta\kapA \bigg[
e^8 \left(\frac{19203}{560}-\frac{3545 \nu }{96}\right)+e^6 \left(\frac{66181}{42}-\frac{265807 \nu }{180}\right)\nonumber\\
&\qquad\quad
+e^4 \left(\frac{134213}{35}-\frac{39362 \nu }{9}\right)+e^2 \left(\frac{6012}{7}-\frac{91582 \nu }{45}\right)-\frac{964 \nu }{15}-\frac{2176}{21}
\bigg]\nonumber\\
&\qquad
+ \kapS \bigg[
e^8 \left(-\frac{2183 \nu ^2}{30}+\frac{354511 \nu }{3360}-\frac{19203}{560}\right)+e^6 \left(-\frac{243961 \nu ^2}{90}+\frac{5831509 \nu }{1260}-\frac{66181}{42}\right)\nonumber\\
&\qquad\quad
+e^4 \left(-\frac{316658 \nu ^2}{45}+\frac{3793504 \nu }{315}-\frac{134213}{35}\right)+e^2 \left(-\frac{122516 \nu ^2}{45}+\frac{1182154 \nu }{315}-\frac{6012}{7}\right)\nonumber\\
&\qquad\quad
-\frac{464 \nu ^2}{5}-\frac{5004 \nu }{35}+\frac{2176}{21}
\bigg]
\bigg\rbrace + \calO\left(x^9\right),
\end{align}
\end{subequations}
and for the angular momentum flux
\begin{subequations}
\label{GinstAvg}
\begin{align}
\langle \mathcal{G}_i^\text{inst} \rangle &= \hat{l}_i \left(\langle \mathcal{G}_\text{nS}^\text{inst} \rangle + \langle \mathcal{G}_\text{SO}^\text{inst} \rangle + \langle \mathcal{G}_\text{SS}^\text{inst} \rangle\right), \\
\langle \mathcal{G}_\text{SO}^\text{inst} \rangle &= -\frac{x^5 c^2 \nu^2 M}{15 \left(1-e^2\right)^{7/2}}\left[\delta  \left(297 e^4+1664 e^2+264\right) \chi _A+\left(297 e^4+1664 e^2-8 \left(5 e^4+94 e^2+36\right) \nu +264\right) \chi _S\right] \nonumber\\
&\quad
+ \frac{16 x^6 c^2\nu^2M}{15 \left(e^2-1\right)^4} \left(7 e^4+67 e^2+16\right) \left[2 \delta  (\nu -3) \chi _A-((\nu -8) \nu +6) \chi _S\right]\nonumber\\
&\quad
+ \frac{x^6 c^2 \nu^2M}{\left(1-e^2\right)^{9/2}} \bigg\lbrace
\frac{16}{15}\sqrt{1-e^2} \left(7 e^4+67 e^2+16\right) \Big[2 \delta  (\nu -3) \chi _A-((\nu -8) \nu +6) \chi _S\Big] \nonumber\\
&\qquad
+\delta\chi_A \left[e^6 \left(\frac{2509 \nu }{30}-\frac{2603}{40}\right)+e^4 \left(\frac{6031 \nu }{5}-\frac{59077}{60}\right)+e^2 \left(\frac{55136 \nu }{45}-\frac{8027}{15}\right)+\frac{4072 \nu }{45}+\frac{394}{5}\right]\nonumber\\
&\qquad
+\chi_S \bigg[e^6 \left(-\frac{646 \nu ^2}{45}+\frac{10343 \nu }{105}-\frac{2603}{40}\right)+e^4 \left(-\frac{2498 \nu ^2}{5}+\frac{64304 \nu }{35}-\frac{59077}{60}\right)\nonumber\\
&\qquad\quad
+e^2 \left(-\frac{2192 \nu ^2}{3}+\frac{670892 \nu }{315}-\frac{8027}{15}\right)-\frac{4256 \nu ^2}{45}+\frac{224 \nu }{9}+\frac{394}{5}\bigg]
\bigg\rbrace + \calO\left(x^7\right),
\\
\langle \mathcal{G}_\text{SS}^\text{inst} \rangle &= \frac{x^{11/2} c^2\nu^2 M}{\left(1-e^2\right)^4} \bigg\lbrace\!\!
\left(\frac{63}{5} e^4+\frac{348}{5} e^2+\frac{64}{5}\right) [\delta \kapA + (1-2  \nu)\kapS]
+ \chi_S^2 \left[e^4 \!\left(\frac{51}{4}-\frac{3 \nu }{5}\right)+e^2 \!\left(\frac{354}{5}-\frac{24 \nu }{5}\right)-\frac{8 \nu }{5}+\frac{66}{5}\right] \nonumber\\
&\qquad
+ \chi_A^2 \left[e^4 \left(\frac{51}{4}-\frac{252 \nu }{5}\right)+e^2 \left(\frac{354}{5}-\frac{1392 \nu }{5}\right)-\frac{256 \nu }{5}+\frac{66}{5}\right]
+ \delta\chi_S\chi_A \left[\frac{51 e^4}{2}+\frac{708 e^2}{5}+\frac{132}{5}\right]
\bigg\rbrace \nonumber\\
&\quad
+ \frac{x^{13/2}c^2\nu^2M}{(1-e^2)^5}  \bigg\lbrace
\frac{4}{15}\sqrt{1-e^2} \left(7 e^4+67 e^2+16\right) \Big[2 \left(6 \nu ^2-46 \nu +11\right) \chi _A^2+4 \delta  (11-9 \nu ) \chi _A \chi _S+\delta  \kapA (14-5 \nu ) \nonumber\\
&\qquad
+\kapS \left(6 \nu ^2-33 \nu +14\right)+2 (4\nu^2 -16 \nu +11) \chi _S^2\Big]
+ \chi_A^2 \bigg[
e^6 \left(284 \nu ^2-\frac{31441 \nu }{70}+\frac{26723}{280}\right)\nonumber\\
&\qquad\quad
+e^4 \!\left(\! \frac{51502 \nu ^2}{15}-\frac{264053 \nu }{42}+\frac{144076}{105}\right)
+e^2 \!\left(\! \frac{46592 \nu ^2}{15}-\frac{1699484 \nu }{315}+\frac{73063}{63}\right)+224 \nu ^2+\frac{90268 \nu }{315}-\frac{27124}{315}
\bigg]\nonumber\\
&\qquad
+ \delta\chi_A\chi_S \bigg[
e^6 \left(\frac{26723}{140}-\frac{1577 \nu }{10}\right)+e^4 \left(\frac{288152}{105}-\frac{110798 \nu }{45}\right)
+e^2 \left(\frac{146126}{63}-\frac{30236 \nu }{9}\right)-\frac{6032 \nu }{45}-\frac{54248}{315}
\bigg]\nonumber\\
&\qquad
+ \chi_S^2 \bigg[
e^6 \left(\frac{12 \nu ^2}{5}-\frac{903 \nu }{10}+\frac{26723}{280}\right)+e^4 \left(\frac{6418 \nu ^2}{45}-\frac{149743 \nu }{90}+\frac{144076}{105}\right)
+\frac{1232 \nu ^2}{45}-\frac{3428 \nu }{45} -\frac{27124}{315}
\nonumber\\
&\qquad\quad
+e^2 \left(\frac{7136 \nu ^2}{15}-\frac{117148 \nu }{45}+\frac{73063}{63}\right) \bigg]
+ \kapS \bigg[
e^6 \left(142 \nu ^2-\frac{32449 \nu }{140}+\frac{11069}{140}\right)\nonumber\\
&\qquad\quad
+e^4 \left(\frac{25751 \nu ^2}{15}-\frac{98664 \nu }{35}+\frac{19256}{21}\right)+e^2 \left(\frac{23296 \nu ^2}{15}-\frac{78182 \nu }{35}+\frac{59362}{105}\right)
+112 \nu ^2+\frac{1308 \nu }{35}-\frac{6176}{105}
\bigg]\nonumber\\
&\qquad
+ \delta\kapA \bigg[
e^6 \left(\frac{11069}{140}-\frac{1473 \nu }{20}\right)
+e^4 \left(\frac{19256}{21}-\frac{14776 \nu }{15}\right)
+e^2 \left(\frac{59362}{105}-\frac{16546 \nu }{15}\right)-\frac{1204 \nu }{15}-\frac{6176}{105}
\bigg]
\bigg\rbrace \nonumber\\
&\quad + \calO\left(x^{15/2}\right).
\end{align}
\end{subequations}

The orbit average of the tail contribution to the fluxes is easier to compute since the only dependence on $l$ is through the exponential, which averages to zero, leading to
\begin{subequations}
\label{FtailAvg}
\begin{align}
\langle \mathcal{F}^\text{tail} \rangle &= \langle \mathcal{F}_\text{nS}^\text{tail} \rangle + \langle \mathcal{F}_\text{SO}^\text{tail} \rangle, \\
\langle \mathcal{F}_\text{nS}^\text{tail} \rangle &=
\frac{128 \pi \nu^2 c^5}{5G} x^{13/2} \left[1+\frac{2335 e^2}{192}+\frac{42955 e^4}{768}+\frac{6204647 e^6}{36864}+\frac{352891481 e^8}{884736}+\calO(e^{10})\right] + \calO(x^{15/2}), \\
\langle \mathcal{F}_\text{SO}^\text{tail} \rangle &= \frac{128 \pi \nu^2 c^5}{5G} x^8 \bigg\lbrace
\delta \chi_A \left[-\frac{65}{24}-\frac{6481 e^2}{72}-\frac{1582919 e^4}{2304}-\frac{27374225 e^6}{9216}-\frac{49776547565 e^8}{5308416} +\calO(e^{10})\right] \nonumber\\
&\qquad
+ \chi_S \bigg[
-\frac{65}{24}+\frac{17 \nu }{6}
+e^2 \left(\frac{15173 \nu }{288}-\frac{6481}{72}\right)
+e^4 \left(\frac{385193 \nu }{1152}-\frac{1582919}{2304}\right)
+e^6 \left(\frac{72544789 \nu }{55296}-\frac{27374225}{9216}\right)\nonumber\\
&\qquad\quad
+e^8 \left(\frac{1292814385 \nu }{331776}-\frac{49776547565}{5308416}\right)
+\calO(e^{10})
\bigg]
\bigg\rbrace + \calO(x^9),
\end{align}
\end{subequations}
and
\begin{subequations}
\label{GtailAvg}
\begin{align}
\langle\mathcal{G}^\text{tail}_i\rangle &= \hat{l}_i \left[\langle\mathcal{G}^\text{tail}_\text{nS}\rangle + \langle\mathcal{G}^\text{tail}_\text{SO}\rangle\right], \\
\langle\mathcal{G}^\text{tail}_\text{nS}\rangle &=
\frac{128 \pi \nu^2 c^2 M}{5} x^5 \left[
1+\frac{209 e^2}{32}+\frac{2415 e^4}{128}+\frac{730751 e^6}{18432}+\frac{10355719 e^8}{147456}+\calO(e^{10})\right] + \calO(x^6), \\
\langle\mathcal{G}^\text{tail}_\text{SO}\rangle &= 
\frac{128 \pi \nu^2 c^4 M}{5} x^{13/2} \bigg\lbrace
\delta\chi_A \left[-\frac{65}{24}-\frac{2575 e^2}{48}-\frac{209203 e^4}{768}-\frac{11800127 e^6}{13824}-\frac{3657249571 e^8}{1769472}+\calO(e^{10})\right]\nonumber\\
&\qquad
+\chi_S \bigg[
-\frac{65}{24}+\frac{17 \nu }{6}
+e^2 \left(\frac{385 \nu }{12}-\frac{2575}{48}\right)
+e^4 \left(\frac{26705 \nu }{192}-\frac{209203}{768}\right)
+e^6 \left(\frac{695917 \nu }{1728}-\frac{11800127}{13824}\right) \nonumber\\
&\qquad\quad
+e^8 \left(\frac{411146539 \nu }{442368}-\frac{3657249571}{1769472}\right) +\calO(e^{10})
\bigg]
\bigg\rbrace + \calO\left(x^{15/2}\right).
\end{align}
\end{subequations}

\subsection{Resummation of the tail contributions}

The eccentricity expansion in the tail part of the fluxes implies that it is valid for small eccentricities only. 
However, a simple resummation of the tail can make it accurate for high eccentricities; from Eqs.~\eqref{FinstAvg} and \eqref{GinstAvg}, we see that each PN order in the instantaneous part of the fluxes contains a factor $1/(1-e^2)^k$, for some power $k$ depending on the PN order, meaning that it diverges as $e \to 1$.
Such a divergence should also occur for the tail part of the fluxes, as argued in Ref.~\cite{Forseth:2015oua}, based on the known closed-form quadrupole formula and how it relates to the LO tail~\cite{Blanchet:1993ec}. 
Therefore, pulling out a factor of $1/(1-e^2)^k$ from the tail is expected to improve the series expansion for high eccentricities.\footnote{
The importance of factoring out the singular part in a series expansion was demonstrated in other contexts in, e.g., Ref.~\cite{Damour:2022ybd} for the scattering angle, Ref.~\cite{Taracchini:2013wfa} for the horizon flux, and Ref.~\cite{Antonelli:2019fmq} for the potential in the effective-one-body Hamiltonian.}
Reference~\cite{Forseth:2015oua} performed such a resummation of the tail and compared the analytical resummed expressions with numerical gravitational-self-force calculations and with analytical expansions to very high orders in eccentricities. 
Thereby, confirming the accuracy of the resummation that pulls a factor $(1-e^2)^{-5}$ at 1.5PN, $(1-e^2)^{-6}$ at 2.5PN, and $(1-e^2)^{-13/2}$ at 3PN in the energy flux.

Following this procedure, we advocate for using the following resummed form of the tail in the energy flux:
\begin{subequations}
\label{FtailResum}
\begin{align}
\langle \mathcal{F}^\text{tail} \rangle^\text{resum} &= \langle \mathcal{F}_\text{nS}^\text{tail} \rangle^\text{resum} + \langle \mathcal{F}_\text{SO}^\text{tail} \rangle^\text{resum}, \\
\langle \mathcal{F}_\text{nS}^\text{tail} \rangle^\text{resum} &=
\frac{128 \pi \nu^2 c^5  x^{13/2}}{5G \left(1-e^2\right)^5} \left[1+\frac{1375 e^2}{192} +\frac{3935 e^4}{768} +\frac{10007 e^6}{36864} + \frac{2321 e^8}{884736} +\calO(e^{10})\right] + \calO(x^{15/2}), \\
\langle \mathcal{F}_\text{SO}^\text{tail} \rangle^\text{resum} &= \frac{128 \pi \nu^2 c^5 x^8}{5G \left(1-e^2\right)^{13/2}} \bigg\lbrace
\delta \chi_A \left[-\frac{208}{3} -\frac{83416 e^2}{45}-\frac{346411 e^4}{90}-\frac{125873 e^6}{120} -\frac{3834029 e^8}{207360}  +\calO(e^{10})\right] \nonumber\\
&\qquad
+ \chi_S \bigg[\frac{1088 \nu }{15}-\frac{208}{3}+e^2 \left(\frac{39476 \nu }{45}-\frac{83416}{45}\right)+e^4 \left(\frac{49039 \nu }{45}-\frac{346411}{90}\right)+e^6 \left(\frac{237541 \nu }{2160}-\frac{125873}{120}\right) \nonumber\\
&\qquad\quad
+e^8 \left(-\frac{34207 \nu }{6480}-\frac{3834029}{207360}\right)
+\calO(e^{10})
\bigg]
\bigg\rbrace + \calO(x^9).
\end{align}
\end{subequations}
Similarly, for angular-momentum flux, the resummation is given by
\begin{subequations}
\label{GtailResum}
\begin{align}
\langle\mathcal{G}^\text{tail}_i\rangle^\text{resum} &= \hat{l}_i \left[\langle\mathcal{G}^\text{tail}_\text{nS}\rangle^\text{resum} + \langle\mathcal{G}^\text{tail}_\text{SO}\rangle^\text{resum}\right], \\
\langle\mathcal{G}^\text{tail}_\text{nS}\rangle^\text{resum} &=
\frac{128 \pi \nu^2 c^2 M x^5}{5 \left(1-e^2\right)^{7/2}} \left[
1+\frac{97 e^2}{32}+\frac{49 e^4}{128}-\frac{49 e^6}{18432} -\frac{109 e^8}{147456} +\calO(e^{10})\right] + \calO(x^6), \\
\langle\mathcal{G}^\text{tail}_\text{SO}\rangle^\text{resum} &= 
\frac{128 \pi \nu^2 c^2 M x^{13/2}}{5 \left(1-e^2\right)^5} \bigg\lbrace
\delta\chi_A \left[-\frac{65}{24}-\frac{1925 e^2}{48}-\frac{8001 e^4}{256}-\frac{13457 e^6}{13824} + \frac{80989 e^8}{1769472} +\calO(e^{10})\right]\nonumber\\
&\qquad
+\chi_S \bigg[
\frac{17 \nu }{6}-\frac{65}{24} +e^2 \left(\frac{215 \nu }{12}-\frac{1925}{48}\right) +e^4 \left(\frac{1345 \nu }{192}-\frac{8001}{256}\right) +e^6 \left(-\frac{23 \nu }{108}-\frac{13457}{13824}\right) \nonumber\\
&\qquad\quad
+e^8 \left(\frac{80989}{1769472}-\frac{3541 \nu }{442368}\right) +\calO(e^{10})
\bigg]
\bigg\rbrace + \calO\left(x^{15/2}\right).
\end{align}
\end{subequations}
The power of each $1/(1-e^2)$ factor in the resummed expressions is such that it is consistent with the powers of those factors in the instantaneous part in Eqs.~\eqref{FinstAvg} and \eqref{GinstAvg}, and the rest of the expression is such that the $\calO(e^8)$ expansion agrees with Eqs.~\eqref{FtailAvg} and \eqref{GtailAvg}.

\begin{figure*}
\centering
\includegraphics[width=\linewidth]{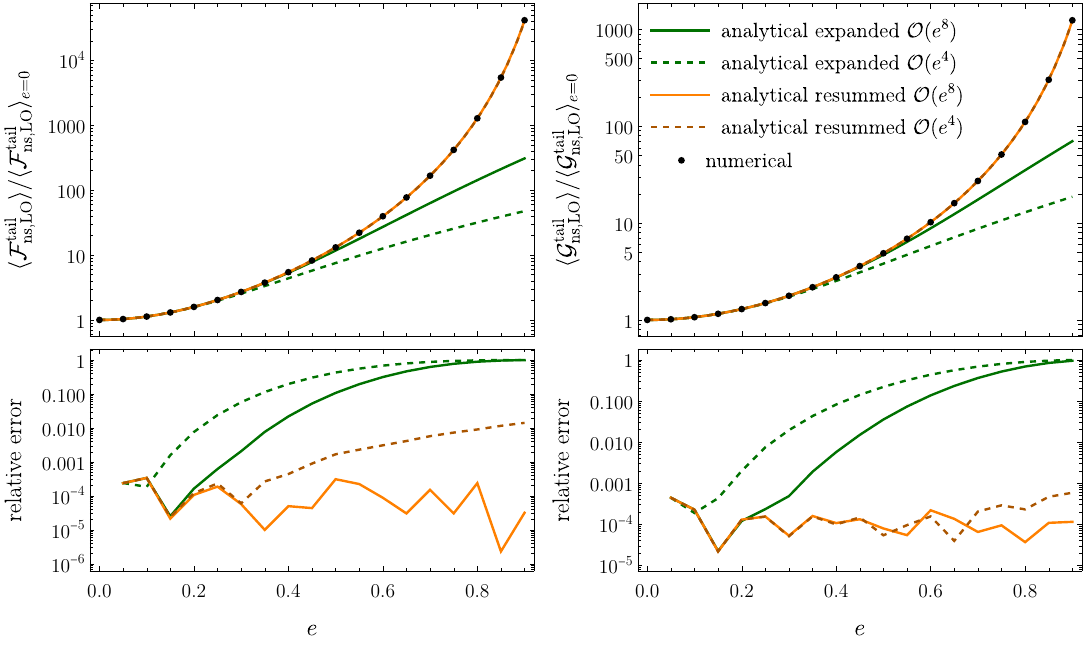}
\caption{The leading-order nonspinning tail part of the energy flux (left panel) and angular-momentum flux (right panel), for eccentricity expansions to $\calO(e^4)$ and $\calO(e^8)$, plotted on a log scale and normalized by the circular-orbit limit. 
The lower panels show the relative errors between the numerical results of Ref.~\cite{Arun:2009mc} and the analytical results.
We see that the resummed analytical expressions are much closer to the numerical results than the eccentricity-expanded expressions, even for eccentricities near unity.}
\label{fig:tail} 
\end{figure*}

As a check of this resummation, we compare the LO nonspinning part of Eqs.~\eqref{FtailResum} and~\eqref{GtailResum} with the numerical results for the tail integrals~\eqref{eq:tailsgen}, which were calculated numerically in Refs.~\cite{Arun:2007rg,Arun:2009mc} to 3PN in the nonspinning case.
Using the numerical values from Appendix~C of Ref.~\cite{Arun:2009mc}, which are only given to four significant digits, we plot in Fig.~\ref{fig:tail} the LO fluxes, and compare them to the eccentricity-expanded and resummed analytical expressions above. 
We see that the eccentricity-expanded fluxes provide a decent agreement with the numerical results for small eccentricities, but are several orders of magnitude different for large eccentricities.
However, the resummed fluxes show excellent agreement between the analytical and numerical results for all eccentricities, comparable to the numerical errors of Ref.~\cite{Arun:2009mc} in the case of the resummed $\calO(e^8)$ expansion, and less than $1\%$ for the $\calO(e^4)$ expansion.

We also checked that the higher-order nonspinning contributions are similarly close to their numerical solutions after factoring $(1-e^2)^{-6}$ at 2.5PN and $(1-e^2)^{-13/2}$ at 3PN in the energy flux, and factoring $(1-e^2)^{-9/2}$ at 2.5PN and $(1-e^2)^{-5}$ at 3PN in the angular momentum flux. 
While it would be good to compare the resummed spin part of the tail to numerical results, we consider such a calculation to be outside the scope of this paper, since it requires using a different method\footnote{
Reference~\cite{Arun:2007rg} computed the tail fluxes numerically by decomposing the source moments into a doubly-periodic Fourier expansion, expressing the Fourier coefficients numerically in terms of $e$ and the mean anomaly $l$, then numerically performing the integration of the multipole moments to obtain the flux.
} from what we followed here to compute the fluxes.
Despite not having numerical results for the spin contribution, we expect that, similarly to the nonspinning case, the resummation given in Eqs.~\eqref{FtailResum} and~\eqref{GtailResum} is also significantly more accurate than a pure eccentricity expansion.

As a different approach for resuming the hereditary contributions to the fluxes, Ref.~\cite{Loutrel:2016cdw} obtained superasymptotic and hyperasymptotic series that are accurate to better than $10^{-8}$, compared to the numerical PN results, for arbitrary eccentricities.
Therefore, if an extremely high accuracy is required, one could use the resummation of Ref.~\cite{Loutrel:2016cdw}, but we expect that the simple and physically-motivated resummation considered above is sufficient for most eccentric-orbit waveform models.

\subsection{Evolution of the secular orbital elements}\label{subsec:SecOrb}
The balance equations relate the orbit-averaged fluxes to the energy and angular momentum losses, such that
\begin{subequations}
\begin{align}
\left\langle \frac{\dd E}{\dd t}\right\rangle &= - \langle \mathcal{F} \rangle,\\
\left\langle \frac{\dd J_i}{\dd t} \right\rangle &= - \langle \mathcal{G}_i \rangle,
\end{align}
\end{subequations}
where the vectors $J_i$ and $\mathcal{G}_i$ can be replaced by their magnitudes in the case of planar orbits.

Using the balance equations and the orbit-averaged energy and angular momentum fluxes from the previous subsection, we compute the secular evolution of the orbital elements $x, a_r$, and $e$, due to radiation reaction. For example,
\begin{equation}
\frac{\dd x}{\dd t} = \frac{\partial x}{\partial E} \frac{\dd E}{\dd t}+\frac{\partial x}{\partial J_i} \frac{\dd J_i}{\dd t}\,,
\end{equation}
and taking the orbit average leads to
\begin{equation}
\label{xdotFluxes}
\langle \dot{x} \rangle = - \frac{\partial x}{\partial E} \langle \mathcal{F} \rangle -\frac{\partial x}{\partial J_i} \langle \mathcal{G}_i \rangle.
\end{equation}
The evolution of other secularly varying quantities, such as $e$, $a_r$ and $K$, can be similarly related to the fluxes.

The 3PN nonspinning contributions were obtained in Refs.~\cite{Arun:2009mc,Ebersold:2019kdc} in ADM and modified-harmonic coordinates, and we find agreement with their results to 2PN order.
For the 3PN spin contributions in $\langle \dot{x} \rangle$, we obtain
\begin{subequations}
\label{xdotInfty}
\begin{align}
\langle \dot{x} \rangle &= \langle \dot{x} \rangle^\text{nS} + \langle \dot{x} \rangle^\text{SO} +  \langle \dot{x} \rangle^\text{SS}, \\
\langle \dot{x} \rangle^\text{SO} &= \frac{\nu c^3 x^{13/2}}{G M \left(1-e^2\right)^5} \Bigg\lbrace
\chi_S \left[e^6 \left(-\frac{8 \nu }{5}-13\right)+e^4 \left(\frac{4072 \nu }{45}-\frac{16694}{45}\right)+e^2 \left(\frac{2672 \nu }{9}-\frac{27608}{45}\right)+\frac{1216 \nu }{15}-\frac{1808}{15}\right] \nonumber\\
&\qquad
- \delta\chi_A \left[13 e^6+\frac{16694 e^4}{45}+\frac{27608 e^2}{45}+\frac{1808}{15}\right]\Bigg\rbrace \nonumber\\
&\quad
+ \frac{\nu c^3 x^{15/2}}{G M \left(1-e^2\right)^6} \Bigg\lbrace
\delta\chi_A \bigg[
e^8 \left(65 \nu -\frac{17967}{224}\right)
+e^6 \left(\frac{99217 \nu }{30}-\frac{2581907}{630}\right)+e^4 \left(\frac{164579 \nu }{15}-\frac{505492}{45}\right)\nonumber\\
&\quad\qquad
+e^2 \left(\frac{56696 \nu }{9}-\frac{445612}{105}\right)
+\frac{9272 \nu }{15}-\frac{125276}{315}
\bigg] 
+ \chi_S \bigg[
e^8 \left(8 \nu ^2+\frac{3231 \nu }{56}-\frac{17967}{224}\right)\nonumber\\
&\quad\qquad
+e^6 \left(-\frac{2234 \nu ^2}{3}+\frac{2894533 \nu }{630}-\frac{2581907}{630}\right)+e^4 \left(-\frac{21948 \nu ^2}{5}+\frac{806942 \nu }{45} -\frac{505492}{45}\right)\nonumber\\
&\quad\qquad
+e^2 \left(-\frac{152024 \nu ^2}{45}+\frac{3314056 \nu }{315}-\frac{445612}{105}\right)-\frac{5056 \nu ^2}{15}+\frac{73520 \nu }{63}-\frac{125276}{315}
\bigg]
\nonumber\\
&\qquad
+\sqrt{1-e^2} \bigg\lbrace
\delta\chi_A \left[e^6 \left(\frac{728 \nu }{45}-\frac{728}{15}\right)+e^4 \left(\frac{11888 \nu }{15}-\frac{11888}{5}\right)+e^2 \left(\frac{35008 \nu }{45}-\frac{35008}{15}\right)\right] \nonumber\\
&\quad\qquad
+ \chi_S \bigg[e^6 \left(-\frac{364 \nu ^2}{45}+\frac{2912 \nu }{45}-\frac{728}{15}\right)+e^4 \left(-\frac{5944 \nu ^2}{15}+\frac{47552 \nu }{15}-\frac{11888}{5}\right) \nonumber\\
&\qquad\qquad
+e^2 \left(-\frac{17504 \nu ^2}{45}+\frac{140032 \nu }{45}-\frac{35008}{15}\right)\bigg]
\bigg\rbrace
\Bigg\rbrace + 
\frac{\pi \nu c^3 x^8}{G M \left(1-e^2\right)^{13/2}} \Bigg\lbrace
\chi_S \bigg[
e^8 \left(-\frac{3299 \nu }{324}-\frac{3911917}{103680}\right)\nonumber\\
&\quad\qquad
+e^6 \left(\frac{42985 \nu }{216}-\frac{1110241}{540}\right)+e^4 \left(\frac{101614 \nu }{45}-\frac{353483}{45}\right)+e^2 \left(\frac{110312 \nu }{45}-\frac{229552}{45}\right)+\frac{4736 \nu }{15}-480
\bigg] \nonumber\\
&\qquad
+\delta\chi_A \left[-\frac{3911917 e^8}{103680}-\frac{1110241 e^6}{540}-\frac{353483 e^4}{45}-\frac{229552 e^2}{45}-480\right] + \calO(e^{10})
\Bigg\rbrace + \calO(x^{17/2}), \\
\langle \dot{x} \rangle^\text{SS} &= 
\frac{\nu c^3 x^7}{G M \left(1-e^2\right)^{11/2}} \Bigg\lbrace
\chi_A^2 \left[e^6 \left(\frac{49}{4}-\frac{236 \nu }{5}\right)+e^4 \left(\frac{7679}{30}-\frac{14944 \nu }{15}\right)+e^2 \left(\frac{5498}{15}-\frac{21536 \nu }{15}\right)-256 \nu +\frac{324}{5}\right] \nonumber\\
&\qquad
+ \chi_S^2 \left[e^6 \left(\frac{49}{4}-\frac{9 \nu }{5}\right)+e^4 \left(\frac{7679}{30}-\frac{138 \nu }{5}\right)+e^2 \left(\frac{5498}{15}-\frac{152 \nu }{5}\right)-\frac{16 \nu }{5}+\frac{324}{5}\right]\nonumber\\
&\qquad
+ \delta\chi_S\chi_A \left[\frac{49 e^6}{2}+\frac{7679 e^4}{15}+\frac{10996 e^2}{15}+\frac{648}{5}\right]
+\delta\kapA \left[\frac{59 e^6}{5}+\frac{3736 e^4}{15}+\frac{5384 e^2}{15}+64\right]\nonumber\\
&\qquad
+ \kapS \left[e^6 \left(\frac{59}{5}-\frac{118 \nu }{5}\right)+e^4 \left(\frac{3736}{15}-\frac{7472 \nu }{15}\right)+e^2 \left(\frac{5384}{15}-\frac{10768 \nu }{15}\right)-128 \nu +64\right]
\Bigg\rbrace \nonumber\\
&\quad
+\frac{\nu c^3 x^8}{G M \left(1-e^2\right)^{13/2}} \Bigg\lbrace
\chi_A^2\bigg[
e^8 \left(\frac{4484 \nu ^2}{15}-\frac{42421 \nu }{84}+\frac{124993}{1120}\right)+e^6 \left(\frac{468544 \nu ^2}{45}-\frac{7544819 \nu }{360}+\frac{562207}{120}\right)\nonumber\\
&\quad\qquad
+e^4 \left(\frac{261152 \nu ^2}{9}-\frac{244995241 \nu }{3780}+\frac{54967327}{3780}\right)
+e^2 \left(\frac{641296 \nu ^2}{45}-\frac{941927 \nu }{27}+\frac{1054837}{135}\right)\nonumber\\
&\quad\qquad
+\frac{5632 \nu ^2}{5}-\frac{486058 \nu }{105}+\frac{111634}{105}
\bigg]
+\chi_S^2 \bigg[
e^8 \left(\frac{57 \nu ^2}{5}-\frac{2731 \nu }{30}+\frac{124993}{1120}\right)\nonumber\\
&\quad\qquad
+e^6 \left(\frac{9629 \nu ^2}{30}-\frac{1444823 \nu }{360}+\frac{562207}{120}\right)+e^4 \left(\frac{254731 \nu ^2}{135}-\frac{9055939 \nu }{540}+\frac{54967327}{3780}\right)\nonumber\\
&\quad\qquad
+e^2 \left(\frac{319564 \nu ^2}{135}-\frac{1737703 \nu }{135}+\frac{1054837}{135}\right)+\frac{1912 \nu ^2}{3}-\frac{11162 \nu }{5}+\frac{111634}{105}
\bigg] \nonumber\\
&\qquad
+ \delta\chi_A\chi_S \bigg[
e^8 \left(\frac{124993}{560}-\frac{17957 \nu }{120}\right)+e^6 \left(\frac{562207}{60}-\frac{1121579 \nu }{180}\right)+e^4 \left(\frac{54967327}{1890}-\frac{6322679 \nu }{270}\right)\nonumber\\
&\quad\qquad
+e^2 \left(\frac{2109674}{135}-\frac{445598 \nu }{27}\right)-\frac{13044 \nu }{5}+\frac{223268}{105}
\bigg]
+ \delta\kapA \bigg[
e^8 \left(\frac{27463}{280}-\frac{18197 \nu }{240}\right)\nonumber\\
&\quad\qquad
+e^6 \left(\frac{69419}{20}-\frac{25499 \nu }{9}\right)+e^4 \left(\frac{1793929}{210}-\frac{400876 \nu }{45}\right)+e^2 \left(\frac{46802}{15}-\frac{227596 \nu }{45}\right)-\frac{7832 \nu }{15}+\frac{35852}{105}
\bigg]\nonumber\\
&\qquad
+ \kapS \bigg[
e^8 \left(\frac{2242 \nu ^2}{15}-\frac{91387 \nu }{336}+\frac{27463}{280}\right)+e^6 \left(\frac{234272 \nu ^2}{45}-\frac{879761 \nu }{90}+\frac{69419}{20}\right)+\frac{2816 \nu ^2}{5}-\frac{42176 \nu }{35}\nonumber\\
&\quad\qquad
+e^4 \left(\frac{130576 \nu ^2}{9}-\frac{8187919 \nu }{315}+\frac{1793929}{210}\right)+e^2 \left(\frac{320648 \nu ^2}{45}-\frac{508408 \nu }{45}+\frac{46802}{15}\right)+\frac{35852}{105}\bigg]\nonumber\\
&\qquad
+\sqrt{1-e^2} \bigg\lbrace
\chi_A^2 \bigg[e^6 \left(\frac{28 \nu ^2}{15}-\frac{644 \nu }{45}+\frac{154}{45}\right)+e^4 \left(1208 \nu ^2-\frac{27784 \nu }{3}+\frac{6644}{3}\right)-\frac{128 \nu ^2}{5}+\frac{2944 \nu }{15}-\frac{704}{15} \nonumber\\
&\qquad\qquad 
+e^2 \left(\frac{17936 \nu ^2}{15}-\frac{412528 \nu }{45}+\frac{98648}{45}\right)
\bigg]
+\chi_S^2\bigg[
e^6 \left(\frac{56 \nu ^2}{45}-\frac{224 \nu }{45}+\frac{154}{45}\right)-\frac{256 \nu ^2}{15}+\frac{1024 \nu }{15}-\frac{704}{15}\nonumber\\
&\qquad\qquad 
+e^4 \left(\frac{2416 \nu ^2}{3}-\frac{9664 \nu }{3}+\frac{6644}{3}\right)
+e^2 \left(\frac{35872 \nu ^2}{45}-\frac{143488 \nu }{45}+\frac{98648}{45}\right)
\bigg] \nonumber\\
&\quad\qquad
+ \delta\chi_S\chi_A \bigg[
e^6 \left(\frac{308}{45}-\frac{28 \nu }{5}\right)+e^4 \left(\frac{13288}{3}-3624 \nu \right)+e^2 \left(\frac{197296}{45}-\frac{17936 \nu }{5}\right)+\frac{384 \nu }{5}-\frac{1408}{15}
\bigg]\nonumber\\
&\quad\qquad
+ \delta\kapA \left[e^6 \left(\frac{98}{45}-\frac{7 \nu }{9}\right)+e^4 \left(\frac{4228}{3}-\frac{1510 \nu }{3}\right)+e^2 \left(\frac{62776}{45}-\frac{4484 \nu }{9}\right)+\frac{32 \nu }{3}-\frac{448}{15}\right]\nonumber\\
&\quad\qquad
+ \kapS \bigg[
e^6 \left(\frac{14 \nu ^2}{15}-\frac{77 \nu }{15}+\frac{98}{45}\right)+e^4 \left(604 \nu ^2-3322 \nu +\frac{4228}{3}\right)+e^2 \left(\frac{8968 \nu ^2}{15}-\frac{49324 \nu }{15}+\frac{62776}{45}\right)\nonumber\\
&\qquad\qquad-\frac{64 \nu ^2}{5}+\frac{352 \nu }{5}-\frac{448}{15}
\bigg]
\bigg\rbrace
\Bigg\rbrace + \calO(x^{17/2}),
\end{align}
\end{subequations}
where the SO term of order $x^8$ comes from the tail contribution to the fluxes while the rest are instantaneous contributions.

For $\langle \dot{e} \rangle$, we obtain
\begin{subequations}
\label{edotInfty}
\begin{align}
\langle \dot{e} \rangle &= \langle \dot{e} \rangle^\text{nS} + \langle \dot{e} \rangle^\text{SO} +  \langle \dot{e} \rangle^\text{SS},\\
\langle \dot{e} \rangle^\text{SO} &= \frac{c^3\nu e x^{11/2}}{GM \left(1-e^2\right)^4} \Bigg\lbrace
\chi_S \left[e^4 \left(\frac{46 \nu }{15}+\frac{263}{10}\right)+e^2 \left(\frac{14128}{45}-\frac{3986 \nu }{45}\right)-\frac{6656 \nu }{45}+\frac{9844}{45}\right] \nonumber\\
&\qquad
+\delta \chi_A \left[\frac{263 e^4}{10}+\frac{14128 e^2}{45}+\frac{9844}{45}\right] 
\Bigg\rbrace 
+ \frac{c^3\nu e x^{13/2}}{GM \left(1-e^2\right)^5} \Bigg\lbrace
\delta\chi_A \bigg[e^6 \left(\frac{345811}{2240}-\frac{4267 \nu }{30}\right)\nonumber\\
&\quad\qquad
+e^4 \left(\frac{97535}{24}-\frac{1838173 \nu }{540}\right)+e^2 \left(\frac{6885449}{1260}-\frac{523579 \nu }{90}\right)-\frac{171062 \nu }{135}+\frac{36083}{105}\bigg] \nonumber\\
&\qquad
+ \chi_S \bigg[
e^6 \left(-\frac{752 \nu ^2}{45}-\frac{221909 \nu }{1680}+\frac{345811}{2240}\right)+e^4 \left(\frac{112736 \nu ^2}{135}-\frac{4593796 \nu }{945}+\frac{97535}{24}\right)\nonumber\\
&\quad\qquad
+e^2 \left(\frac{126833 \nu ^2}{45}-\frac{2152457 \nu }{210}+\frac{6885449}{1260}\right)+\frac{105832 \nu ^2}{135}-\frac{357772 \nu }{189}+\frac{36083}{105}
\bigg] \nonumber\\
&\qquad
+ \sqrt{1-e^2} \bigg\lbrace
\chi_S \left[e^4 \left(\frac{626 \nu ^2}{45}-\frac{5008 \nu }{45}+\frac{1252}{15}\right)+e^2 \left(\frac{4618 \nu ^2}{15}-\frac{36944 \nu }{15}+\frac{9236}{5}\right)+\frac{1184 \nu ^2}{9}-\frac{9472 \nu }{9}+\frac{2368}{3}\right] \nonumber\\
&\quad\qquad
+ \delta\chi_A \left[e^4 \left(\frac{1252}{15}-\frac{1252 \nu }{45}\right)+e^2 \left(\frac{9236}{5}-\frac{9236 \nu }{15}\right)-\frac{2368 \nu }{9}+\frac{2368}{3}\right]
\bigg\rbrace 
\Bigg\rbrace \nonumber\\
&\quad
+ \frac{c^3 \pi \nu e x^7}{GM \left(1-e^2\right)^{11/2}} \Bigg\lbrace
\chi_S\bigg[e^6 \left(\frac{631523 \nu }{25920}+\frac{2313613}{51840}\right)+e^4 \left(\frac{403697}{216}-\frac{95561 \nu }{1080}\right)+e^2 \left(\frac{207277}{45}-\frac{74954 \nu }{45}\right)\nonumber\\
&\quad\qquad
-\frac{39496 \nu }{45}+\frac{56096}{45}\bigg] 
+ \delta\chi_A \left[\frac{2313613 e^6}{51840}+\frac{403697 e^4}{216}+\frac{207277 e^2}{45}+\frac{56096}{45}\right] + \calO(e^{10})
\Bigg\rbrace
 + \calO(x^{15/2}), \\
\langle \dot{e} \rangle^\text{SS} &= 
\frac{c^3 \nu e x^6}{GM \left(1-e^2\right)^{9/2}} \Bigg\lbrace
\chi _A^2 \left[e^4 \left(74 \nu -\frac{151}{8}\right)+e^2 \left(\frac{2380 \nu }{3}-\frac{1217}{6}\right)+\frac{7504 \nu }{15}-\frac{1921}{15}\right] \nonumber\\
&\qquad
+ \chi _S^2 \left[e^4 \left(\frac{3 \nu }{2}-\frac{151}{8}\right)+e^2 \left(18 \nu -\frac{1217}{6}\right)+12 \nu -\frac{1921}{15}\right]
+ \delta \chi _A \chi _S \left[-\frac{151 e^4}{4}-\frac{1217 e^2}{3}-\frac{3842}{15}\right] \nonumber\\
&\qquad
+ \delta \kapA \left[-\frac{37 e^4}{2}-\frac{595 e^2}{3}-\frac{1876}{15}\right] 
+\kapS \left[e^4 \left(37 \nu -\frac{37}{2}\right)+e^2 \left(\frac{1190 \nu }{3}-\frac{595}{3}\right)+\frac{3752 \nu }{15}-\frac{1876}{15}\right]
\Bigg\rbrace \nonumber\\
&\quad
+\frac{c^3 \nu e x^7}{GM \left(1-e^2\right)^{11/2}} \Bigg\lbrace
\chi_A^2\bigg[
e^6 \left(-\frac{2574 \nu ^2}{5}+\frac{243291 \nu }{280}-\frac{422777}{2240}\right)+e^4 \left(-\frac{445066 \nu ^2}{45}+\frac{103061071 \nu }{5040}-\frac{2553951}{560}\right) \nonumber\\
&\quad\qquad
+e^2 \left(-\frac{43730 \nu ^2}{3}+\frac{132190589 \nu }{3780}-\frac{7460666}{945}\right)-\frac{115688 \nu ^2}{45}+\frac{12487753 \nu }{1890}-\frac{553121}{378}
\bigg] \nonumber\\
&\qquad
+ \chi_S^2 \bigg[
e^6 \left(-\frac{117 \nu ^2}{10}+\frac{589 \nu }{4}-\frac{422777}{2240}\right)+e^4 \left(-\frac{45373 \nu ^2}{180}+\frac{186715 \nu }{48}-\frac{2553951}{560}\right)\nonumber\\
&\quad\qquad
+e^2 \left(-\frac{185723 \nu ^2}{135}+\frac{5573423 \nu }{540}-\frac{7460666}{945}\right)-\frac{26254 \nu ^2}{27}+\frac{980419 \nu }{270}-\frac{553121}{378}
\bigg]\nonumber\\
&\qquad
+ \delta\chi_S\chi_A \bigg[
e^6 \left(\frac{4179 \nu }{16}-\frac{422777}{1120}\right)+e^4 \left(\frac{438913 \nu }{72}-\frac{2553951}{280}\right)+e^2 \left(\frac{3702421 \nu }{270}-\frac{14921332}{945}\right)\nonumber\\
&\quad\qquad
+\frac{592019 \nu }{135}-\frac{553121}{189}
\bigg] 
+ \kapS \bigg[
e^6 \left(-\frac{1287 \nu ^2}{5}+\frac{516661 \nu }{1120}-\frac{18447}{112}\right)
-\frac{57844 \nu ^2}{45}+\frac{527402 \nu }{315}-\frac{9977}{35}\nonumber\\
&\quad\qquad
+e^4 \left(-\frac{222533 \nu ^2}{45}+\frac{2935396 \nu }{315}-\frac{2772439}{840}\right)
+e^2 \left(-\frac{21865 \nu ^2}{3}+\frac{89722 \nu }{7}-\frac{847619}{210}\right)
\bigg] \nonumber\\
&\qquad
+ \delta\kapA \bigg[
e^6 \left(\frac{21103 \nu }{160}-\frac{18447}{112}\right)+e^4 \left(\frac{489181 \nu }{180}-\frac{2772439}{840}\right)
+e^2 \left(\frac{71173 \nu }{15}-\frac{847619}{210}\right)+\frac{49688 \nu }{45}-\frac{9977}{35}
\bigg]\nonumber\\
&\qquad
+ \sqrt{1-e^2} \bigg\lbrace
\delta\chi_A\chi_S \left[
e^4 \left(\frac{122 \nu }{5}-\frac{1342}{45}\right)+e^2 \left(\frac{13782 \nu }{5}-\frac{50534}{15}\right)+\frac{6496 \nu }{5}-\frac{71456}{45}\right] \nonumber\\
&\quad\qquad
+ \chi_S^2 \left[e^4 \left(\frac{976 \nu }{45}-\frac{244 \nu ^2}{45}-\frac{671}{45}\right)+e^2 \left(\frac{36752 \nu }{15}-\frac{9188 \nu ^2}{15}-\frac{25267}{15}\right)-\frac{12992 \nu ^2}{45}+\frac{51968 \nu }{45}-\frac{35728}{45}\right]\nonumber\\
&\quad\qquad
+ \chi_A^2 \left[e^4 \left(\frac{2806 \nu }{45}-\frac{122 \nu ^2}{15}-\frac{671}{45}\right)+e^2 \left(\frac{105662 \nu }{15}-\frac{4594 \nu ^2}{5}-\frac{25267}{15}\right)-\frac{6496 \nu ^2}{15}+\frac{149408 \nu }{45}-\frac{35728}{45}\right]\nonumber\\
&\quad\qquad
+ \kapS \left[e^4 \left(\frac{671 \nu }{30}-\frac{61 \nu ^2}{15}-\frac{427}{45}\right)+e^2 \left(\frac{25267 \nu }{10}-\frac{2297 \nu ^2}{5}-\frac{16079}{15}\right)-\frac{3248 \nu ^2}{15}+\frac{17864 \nu }{15}-\frac{22736}{45}\right] \nonumber\\
&\quad\qquad
+ \delta\kapA \left[e^4 \left(\frac{61 \nu }{18}-\frac{427}{45}\right)+e^2 \left(\frac{2297 \nu }{6}-\frac{16079}{15}\right)+\frac{1624 \nu }{9}-\frac{22736}{45}\right]
\bigg\rbrace
\Bigg\rbrace + \calO(x^{15/2}),
\end{align}
\end{subequations}
which is proportional to $e$.
We included the results for $\dot{a}_r$ and $\dot{K}$ in the Supplemental Material.

\subsection{Horizon flux contribution at leading order}
\label{subsec:horiz}

In addition to the radiated fluxes at infinity, which we derived in this section, the total 3PN fluxes receive a contribution through the horizon absorption starting at 2.5PN for the spin contributions and at 4PN for the nonspinning contributions, relative to the leading order of the flux at infinity.
For arbitrary mass ratios and quasi-circular inspirals, the horizon flux was derived in Ref.~\cite{Alvi:2001mx} at leading order, in Refs.~\cite{Poisson:2014gka,Taylor:2008xy} to 5PN for nonspinning and to 3.5PN for spinning binaries, and in Refs.~\cite{Chatziioannou:2012gq,Chatziioannou:2016kem,Saketh:2022xjb} to 4PN.
It was also derived to high PN orders in the test-mass limit for quasi-circular and eccentric orbits in Refs.~\cite{Poisson:1994yf,Tagoshi:1997jy,Fujita:2014eta,Shah:2014tka,Sago:2015rpa,Isoyama:2021jjd,Munna:2023vds}.
For eccentric orbits and arbitrary mass ratios, the horizon flux was computed at leading order in Ref.~\cite{Datta:2023wsn} and to next-to-next-to-leading order in Ref.~\cite{Chiaramello:2024unv}.

The rates of change to the black hole's mass and angular momentum due to the horizon absorption are given by Eq.~(17) of Ref.~\cite{Chiaramello:2024unv}.
Combining the rates of change for the binary, we obtain the following energy and angular momentum horizon fluxes at leading order:
\begin{subequations}
\begin{align}
\mathcal{F}^\text{horiz} &= \frac{8 \nu ^2 G^6 M^7 \dot{\phi }}{5 c^{10}  r^6} \left[\delta \chi _A  (\nu -1) \left(3 \chi _A^2+9 \chi _S^2+1\right)+\chi _S (3 \nu -1) \left(9 \chi _A^2+3 \chi _S^2+1\right)\right],\\
\mathcal{G}_i^\text{horiz} &= \frac{8 \nu ^2 G^6 M^7}{5 c^{10} r^6}  \left[\delta \chi _A  (\nu -1) \left(3 \chi _A^2+9 \chi _S^2+1\right)+\chi _S (3 \nu -1) \left(9 \chi _A^2+3 \chi _S^2+1\right)\right].
\end{align}
\end{subequations}
Expressing the result in terms of the quasi-Keplerian parametrization, taking the orbit average, and computing the contribution to $\dot{x}$ and $\dot{e}$ using Eq.~\eqref{xdotFluxes},  we obtain
\begin{subequations}
\label{xdotedotHoriz}
\begin{align}
\langle \dot{x}^\text{horiz}\rangle &= \frac{c^3 \nu x^{15/2} \left(5 e^6+90 e^4+120 e^2+16\right)}{5 GM\left(1-e^2\right)^6} 
\left[\delta \chi _A  (\nu -1) \left(3\chi _A^2+9 \chi _S^2+1\right)+  \chi _S(3 \nu -1) \left(9 \chi _A^2+3\chi _S^2+1\right)\right], \\
\langle \dot{e}^\text{horiz}\rangle &= -\frac{11 c^3 \nu x^{13/2}e \left(e^4+12 e^2+8\right) \left(1-e^2\right)^5}{10GM}
\left[\delta \chi _A  (\nu -1) \left(3\chi _A^2+9 \chi _S^2+1\right)+  \chi _S(3 \nu -1) \left(9 \chi _A^2+3\chi _S^2+1\right)\right].
\end{align}
\end{subequations}

The total 3PN evolution equations for $x$ and $e$ are the sums of the horizon flux contributions from Eqs.~\eqref{xdotedotHoriz} with the contributions due to the fluxes at infinity (Eqs.~\eqref{xdotInfty} and \eqref{xdotInfty}).
From these expressions, one can solve for $x(e)$ by first obtaining $\dd x/\dd e = \dot{x}/\dot{e}$, then solving this differential equation order by order, in a PN and an eccentricity expansion, starting from an initial frequency $x_0$ and eccentricity $e_0$.
For example, at leading order in eccentricity, and at leading PN order for the nonspinning and spin contributions, we get
\begin{align}
x(e) &= x_0 \left(\frac{e_0}{e}\right){}^{12/19} \bigg\lbrace
1 -\frac{3323 \left(e^2-e_0^2\right)}{2888}
- \frac{x_0^{3/2} (e^{18/19}-e_0^{18/19}) \left[157 \delta  \chi _A+(157-110 \nu ) \chi _S\right]}{171 e^{18/19}} + \dots
\bigg\rbrace.
\end{align}
The full result is provided in the Supplemental Material.

The expressions for $x(e)$ and $\dot{e}$ are needed in the derivation of the memory contributions to the modes, as discussed in Subsection~\ref{subsec:heredEval}. 
The horizon flux contributes to the DC memory, through the integral in Eq.~\eqref{eq:DCmemoryIntegrals}, and is needed to get agreement with the gravitational self-force results of Ref.~\cite{Cunningham:2024dog}. 
However, we do not account for possible horizon absorption effects on the other contributions to the waveform modes, since they are not straightforward to include in PN calculations within the MPM formalism.

\section{3PN waveform modes for nonprecessing spins}
\label{sec:modes}

In this section, we present our results for the 3PN spin contributions to the waveform modes, which are related to the radiative moments via Eqs.~\eqref{eq:hlm}.
We factor the modes such that
\begin{equation}
\label{eq:defHlm}
h_{\ell m} = \frac{8GM\nu}{c^4 R} \sqrt{\frac{\pi}{5}} H_{\ell m} \e^{-\di m \phi},
\end{equation}
and split $H_{\ell m}$ into instantaneous, tail, DC memory, and oscillatory memory contributions, i.e.,
\begin{equation}
H_{\ell m} \equiv H_{\ell m}^\text{inst} + H_{\ell m}^\text{tail} + H_{\ell m}^\text{DC} + H_{\ell m}^\text{osc}, 
\end{equation}
where the instantaneous contributions are valid for generic motion, while the hereditary contributions are computed for bound orbits in a small-eccentricity expansion to $\calO(e^6)$. 
To the order considered in this paper, there are no post-adiabatic contributions to the waveform, which start in the nonspinning part at 2.5PN, as computed for eccentric orbits in Ref.~\cite{Ebersold:2019kdc}, and start at 3.5PN for the spin contributions.

In this section, we write explicit expressions for the $\ell=2$ modes only and write eccentricity expansions to $\calO(e^2)$.
The full expressions for all modes up to the (6,6) mode, and to $\calO(e^6)$ for the hereditary contributions, are included in the Supplemental Material~\cite{ancMaterial}.

\subsection{Instantaneous contributions}\label{subsec:modesinst}

The instantaneous contributions to the modes are computed using the instantaneous part of the radiative moments from Eqs.~\eqref{eq:instMoments} and plugging them in Eq.~\eqref{eq:hlm}, which yields the spin part of the modes to 3PN.

The 3PN nonspinning instantaneous contributions were derived in Ref.~\cite{Mishra:2015bqa}, and we find agreement with the results of that reference to 2PN order. 
For the spin contributions, our 3PN results agree with, e.g., Ref.~\cite{Henry:2022dzx} in the circular-orbit limit, and agree at 2PN for general orbits with Refs.~\cite{Khalil:2021txt,Paul:2022xfy}.

For the (2,2) mode, we obtain the following instantaneous spin contributions
\begin{subequations}
\begin{align}
H_{22}^\text{inst} &\equiv H_{22}^\text{inst,nS} + H_{22}^\text{inst,SO} + H_{22}^\text{inst,SS}, \\
H_{22}^\text{inst,SO} &= \frac{G^2M^2}{c^3r^2} \bigg\lbrace
\delta  \chi _A \left(-r \dot{\phi }-\di \dot{r}\right)+\chi _S \left[\left(\frac{5 \nu }{3}-1\right) r \dot{\phi }+\di \left(\frac{8 \nu }{3}-1\right) \dot{r}\right]
\bigg\rbrace \nonumber\\
&\quad
+ \frac{G^2M^2}{c^5r^2} \bigg\lbrace
\delta\chi_A\bigg[
G M \dot{\phi } \left(\frac{17}{6}-\frac{47 \nu }{28}\right)
+\frac{\di G M \dot{r}}{r} \left(\frac{110}{21}-\frac{38 \nu }{21}\right)
+\left(\frac{275 \nu }{84}+\frac{3}{14}\right) r \dot{r}^2 \dot{\phi }
-\di \left(\frac{31 \nu }{21}+\frac{17}{14}\right) r^2 \dot{r} \dot{\phi }^2 \nonumber\\
&\quad\qquad
-\left(\frac{83 \nu }{84}+\frac{10}{7}\right) r^3 \dot{\phi }^3
+\di \left(\nu -\frac{9}{14}\right) \dot{r}^3\bigg]
+ \chi_S \bigg[
G M \dot{\phi } \left(\frac{79 \nu ^2}{14}-\frac{181 \nu }{84}+\frac{17}{6}\right)
+\left(\frac{23 \nu }{4}-\frac{39 \nu ^2}{14}-\frac{10}{7}\right) r^3 \dot{\phi }^3 \nonumber\\
&\quad\qquad
+\frac{\di G M \dot{r}}{r} \left(\frac{172 \nu ^2}{21}-\frac{200 \nu }{21}+\frac{110}{21}\right)
+\di r^2 \dot{r} \dot{\phi }^2\left(-\frac{220 \nu ^2}{21}+\frac{44 \nu }{7}-\frac{17}{14}\right) 
+\left(\frac{47 \nu ^2}{21}-\frac{15 \nu }{4}+\frac{3}{14}\right) r \dot{r}^2 \dot{\phi } \nonumber\\
&\quad\qquad
+\di \left(-\frac{2 \nu ^2}{21}+\frac{5 \nu }{7}-\frac{9}{14}\right) \dot{r}^3
\bigg]
\bigg\rbrace 
+ \frac{G^3M^3}{c^6r^3} \bigg\lbrace
\delta  \chi _A \left[-\frac{\di G M}{6 r}-\frac{7}{6} \di r^2 \dot{\phi }^2-\frac{5}{3} r \dot{r} \dot{\phi }-\frac{\di \dot{r}^2}{3}\right]\nonumber\\
&\qquad
+\chi _S \left[\frac{\di G M}{r} \left(\frac{\nu }{3}-\frac{1}{6}\right)+\di r^2 \dot{\phi }^2\left(\frac{7 \nu }{3}-\frac{7}{6}\right)+ r \dot{r} \dot{\phi } \left(\frac{10 \nu }{3}-\frac{5}{3}\right)+\di \dot{r}^2 \left(\frac{2 \nu }{3}-\frac{1}{3}\right)\right]
\bigg\rbrace + \calO(7),\\
H_{22}^\text{inst,SS} &= \frac{3G^3M^3}{4c^4r^3} \Big[(1-4 \nu ) \chi _A^2+2 \delta  \chi _A \chi _S+\delta  \kapA+\kapS (1-2 \nu )+\chi _S^2\Big]
+ \frac{3G^3M^3}{4c^6r^3} \bigg\lbrace
\chi _A^2 \bigg[
\left(\frac{17}{14}-\frac{268 \nu ^2}{21}-\frac{2 \nu }{9}\right) r^2 \dot{\phi }^2 \nonumber\\
&\quad\qquad
+\frac{G M}{r} \left(-\frac{24 \nu ^2}{7}+\frac{1726 \nu }{63}-\frac{146}{21}\right)
+\di \left(-\frac{260 \nu ^2}{21}-\frac{439 \nu }{63}+\frac{13}{7}\right) r \dot{r} \dot{\phi }+\left(-\frac{8 \nu ^2}{21}+\frac{470 \nu }{63}-\frac{97}{42}\right) \dot{r}^2\bigg] \nonumber\\
&\qquad
+ \delta\chi_S\chi_A\left[\frac{G M}{r} \left(-\frac{290 \nu }{63}-\frac{292}{21}\right)
+\left(\frac{38 \nu }{63}+\frac{17}{7}\right) r^2 \dot{\phi }^2+\di \left(\frac{26}{7}-\frac{446 \nu }{63}\right) r \dot{r} \dot{\phi }+\left(\frac{52 \nu }{9}-\frac{97}{21}\right) \dot{r}^2\right] \nonumber\\
&\qquad
+ \chi_S^2 \bigg[
\frac{G M}{r} \left(\frac{8 \nu ^2}{3}-\frac{88 \nu }{21}-\frac{146}{21}\right)
+\left(-\frac{8 \nu ^2}{9}-\frac{254 \nu }{63}+\frac{17}{14}\right) r^2 \dot{\phi }^2
+\di r \dot{r} \dot{\phi } \left(\frac{8 \nu ^2}{9}-\frac{475 \nu }{63}+\frac{13}{7}\right) \nonumber\\
&\qquad\quad
+\left(-\frac{16 \nu ^2}{9}+\frac{68 \nu }{9}-\frac{97}{42}\right) \dot{r}^2
\bigg] 
+\kapS \bigg[
\frac{G M}{r} \left(-\frac{12 \nu ^2}{7}+\frac{773 \nu }{63}-\frac{146}{21}\right)
+\left(-\frac{134 \nu ^2}{21}+\frac{62 \nu }{63}+\frac{17}{14}\right) r^2 \dot{\phi }^2 \nonumber\\
&\quad\qquad
+\di \left(-\frac{130 \nu ^2}{21}-\frac{23 \nu }{63}+\frac{13}{7}\right) r \dot{r} \dot{\phi }+\left(-\frac{4 \nu ^2}{21}+\frac{445 \nu }{63}-\frac{97}{42}\right) \dot{r}^2
\bigg]
+ \delta\kapS \bigg[
\frac{G M}{r} \left(-\frac{103 \nu }{63}-\frac{146}{21}\right)\nonumber\\
&\quad\qquad
+\left(\frac{215 \nu }{63}+\frac{17}{14}\right) r^2 \dot{\phi }^2+\di \left(\frac{211 \nu }{63}+\frac{13}{7}\right) r \dot{r} \dot{\phi }+\left(\frac{22 \nu }{9}-\frac{97}{42}\right) \dot{r}^2
\bigg]
\bigg\rbrace  + \calO(8).
\end{align}
\end{subequations}
For the (2,1) mode, we obtain
\begin{subequations}
\begin{align}
H_{21}^\text{inst} &\equiv H_{21}^\text{inst,nS} + H_{21}^\text{inst,SO} + H_{21}^\text{inst,SS} + H_{21}^\text{inst,S$^3$}, \\
H_{21}^\text{inst,SO} &= \frac{\di}{2}\frac{G^2 M^2}{r^2}\Bigg[
\frac{1}{c^2} \left(\chi _A+\delta  \chi _S\right) \nonumber\\
&\qquad
+ \frac{1}{c^4} \Bigg\lbrace 
\chi_A \left[\frac{G M}{r} \left(-\frac{59 \nu }{21}-\frac{11}{3}\right)+\left(2-\frac{44 \nu }{7}\right) r^2 \dot{\phi }^2+\di \left(7-\frac{83 \nu }{21}\right) r \dot{r} \dot{\phi }+\left(\frac{26 \nu }{21}-\frac{5}{2}\right) \dot{r}^2\right] \nonumber\\
&\quad\qquad
+ \delta\chi_S \left[\frac{G M}{r} \left(-\frac{11 \nu }{21}-\frac{11}{3}\right)+\left(2-\frac{8 \nu }{21}\right) r^2 \dot{\phi }^2+\di \left(\frac{13 \nu }{21}+7\right) r \dot{r} \dot{\phi }+\left(\frac{10 \nu }{7}-\frac{5}{2}\right) \dot{r}^2\right]
\Bigg\rbrace\nonumber\\
&\qquad
+ \frac{1}{c^6}\Bigg\lbrace
\chi_A \Bigg[
\frac{G^2 M^2}{r^2} \left(-\frac{11 \nu ^2}{36}+\frac{4289 \nu }{189}+\frac{629}{84}\right)
+\di G M \dot{r} \dot{\phi } \left(-\frac{3545 \nu ^2}{252}-\frac{37 \nu }{756}-\frac{227}{84}\right)\nonumber\\
&\quad\qquad
+ G M r\dot{\phi }^2 \left(-\frac{2591 \nu ^2}{252}-\frac{667 \nu }{756}-\frac{85}{84}\right) 
+\di r \dot{r}^3 \dot{\phi } \left(-\frac{10 \nu ^2}{21}-\frac{251 \nu }{63}+\frac{5}{2}\right) 
+\frac{G  M \dot{r}^2}{r}\left(\frac{23 \nu ^2}{9}-\frac{320 \nu }{27}-\frac{26}{21}\right) \nonumber\\
&\quad\qquad
+\left(\frac{190 \nu ^2}{21}-\frac{1315 \nu }{126}+2\right) r^4 \dot{\phi }^4
+\di \left(\frac{451 \nu ^2}{21}-\frac{4895 \nu }{126}+\frac{52}{7}\right) r^3 \dot{r} \dot{\phi }^3
+\left(-\frac{55 \nu ^2}{6}+\frac{521 \nu }{21}-\frac{9}{2}\right) r^2 \dot{r}^2 \dot{\phi }^2\nonumber\\
&\quad\qquad
+\left(-\frac{59 \nu ^2}{21}+\frac{22 \nu }{9}-\frac{9}{8}\right) \dot{r}^4
\Bigg]
+ \delta\chi_S \Bigg[
\frac{G^2 M^2}{r^2} \left(-\frac{775 \nu ^2}{756}+\frac{2293 \nu }{189}+\frac{629}{84}\right)
+\di r \dot{r}^3 \dot{\phi } \left(-\frac{2 \nu ^2}{63}-\frac{463 \nu }{63}+\frac{5}{2}\right) \nonumber\\
&\quad\qquad
+\di G M \dot{r} \dot{\phi } \left(-\frac{139 \nu ^2}{756}-\frac{371 \nu }{108}-\frac{227}{84}\right)
+G M r \dot{\phi }^2\left(\frac{245 \nu ^2}{108}-\frac{533 \nu }{108}-\frac{85}{84}\right)
+\left(2-\frac{86 \nu ^2}{63}-\frac{857 \nu }{126}\right) r^4 \dot{\phi }^4 \nonumber\\
&\quad\qquad
+\frac{G  M \dot{r}^2}{r}\left(\frac{1111 \nu ^2}{189}-\frac{3604 \nu }{189}-\frac{26}{21}\right)
+\di r^3 \dot{r} \dot{\phi }^3 \left(\frac{293 \nu ^2}{63}-\frac{1075 \nu }{126}+\frac{52}{7}\right)
+\left(\frac{293 \nu }{21}-\frac{23 \nu ^2}{6}-\frac{9}{2}\right) r^2 \dot{r}^2 \dot{\phi }^2\nonumber\\
&\quad\qquad
+\left(-\frac{173 \nu ^2}{63}+\frac{128 \nu }{63}-\frac{9}{8}\right) \dot{r}^4
\Bigg]
\Bigg\rbrace + \calO(7)
\Bigg], \\
H_{21}^\text{inst,SS} &= \frac{G^3 M^3}{6c^5r^3} \Big\lbrace
\delta  \chi _A^2 \left[-4 \dot{r}+\di (10-12 \nu ) r \dot{\phi }\right]
+\delta \chi _S^2 \left[-4 \dot{r}+10 \di r \dot{\phi }\right]
+\chi _A \chi _S \left[(16 \nu -8) \dot{r}+\di (20-52 \nu ) r \dot{\phi }\right] \nonumber\\
&\qquad
+\kapA \left[(8 \nu -4) \dot{r}+\di (1-8 \nu ) r \dot{\phi }\right]
+\delta  \kapS \left[-4 \dot{r}+\di (1-6 \nu ) r \dot{\phi }\right]
\Big\rbrace + \calO(7),\\
H_{21}^\text{inst,S$^3$} &= -\di \frac{3G^4M^4}{4c^6 r^4} \Bigr\lbrace
 \chi _A^3 (1-4 \nu ) + \chi _A \chi _S^2 (3-8 \nu ) +\delta  \chi _S^3
+ \delta \chi _A^2\chi _S (3-4 \nu )  +\kapA\chi _S (1-4 \nu )+\delta \chi _S \kapS (1-2 \nu ) \nonumber\\
&\qquad
+ \delta  \chi _A\kapA + \chi _A\kapS (1-2 \nu )
\Bigr\rbrace  + \calO(7).
\end{align}
\end{subequations}
Note that the (2,1) mode is the only mode containing cubic-in-spin terms for the 3PN waveform. Finally, we find for the (2,0) mode
\begin{subequations}
\begin{align}
H_{20}^\text{inst} &\equiv H_{20}^\text{inst,nS} + H_{20}^\text{inst,SO} + H_{20}^\text{inst,SS}, \\
H_{20}^\text{inst,SO} &= \sqrt{\frac{2}{3}} \frac{G^2 M^2}{r^2}r \dot{\phi} \bigg[
\frac{1}{c^3}\left\lbrace \delta  \chi _A+(\nu +1) \chi _S \right\rbrace
+ \frac{1}{c^5} \bigg\lbrace
\delta  \chi _A \left[\frac{G M}{r} \left(\frac{31 \nu }{28}-\frac{43}{14}\right)
-\left(\frac{15 \nu }{28}+\frac{9}{7}\right) r^2 \dot{\phi }^2
+\left(\frac{15 \nu }{28}+\frac{3}{2}\right) \dot{r}^2\right] \nonumber\\
&\qquad
+ \chi_S \left[
\left(\frac{3}{2}-\frac{45 \nu ^2}{7}-\frac{295 \nu }{28}\right) \dot{r}^2
-\frac{G M}{r} \left(\frac{3 \nu ^2}{14}+\frac{369 \nu }{28}+\frac{43}{14}\right)
+\left(\frac{263 \nu }{28}-\frac{45 \nu ^2}{14}-\frac{9}{7}\right) r^2 \dot{\phi }^2\right]
\bigg\rbrace
\bigg] + \calO(7),\\
H_{20}^\text{inst,SS} &= \sqrt{\frac{2}{3}} \frac{3 G^3 M^3}{4 r^3} \Bigg[
\frac{1}{c^4} \left\lbrace
(4 \nu -1) \chi _A^2-2 \delta  \chi _A \chi _S-\delta \kapA+\kapS (2 \nu -1)-\chi _S^2
\right\rbrace \nonumber\\
&\quad
+ \frac{1}{c^6} \bigg\lbrace
\chi _A^2 \left[\frac{G M}{r} \left(\frac{24 \nu ^2}{7}-\frac{526 \nu }{21}+\frac{104}{21}\right)
+\left(-\frac{52 \nu ^2}{21}+\frac{134 \nu }{21}-\frac{11}{42}\right) r^2 \dot{\phi }^2+\left(\frac{8 \nu ^2}{21}-\frac{58 \nu }{21}-\frac{71}{42}\right) \dot{r}^2\right]\nonumber\\
&\qquad
+\delta  \chi _A \chi _S \left[\frac{G M}{r} \left(\frac{34 \nu }{7}+\frac{208}{21}\right)+\left(-2 \nu -\frac{11}{21}\right) r^2 \dot{\phi }^2+\left(-\frac{12 \nu }{7}-\frac{71}{21}\right) \dot{r}^2\right]\nonumber\\
&\qquad
+\chi _S^2 \left[\frac{G M}{r} \left(\frac{8 \nu ^2}{3}+\frac{212 \nu }{21}+\frac{104}{21}\right)+\left(-\frac{8 \nu ^2}{3}-\frac{22 \nu }{3}-\frac{11}{42}\right) r^2 \dot{\phi }^2+\left(\frac{16 \nu ^2}{3}+\frac{164 \nu }{21}-\frac{71}{42}\right) \dot{r}^2\right]\nonumber\\
&\qquad
+\kapS \left[\frac{G M}{r} \left(\frac{12 \nu ^2}{7}-\frac{143 \nu }{21}+\frac{104}{21}\right)+\left(-\frac{26 \nu ^2}{21}+\frac{6 \nu }{7}-\frac{11}{42}\right) r^2 \dot{\phi }^2+\left(\frac{4 \nu ^2}{21}+\frac{27 \nu }{7}-\frac{71}{42}\right) \dot{r}^2\right]\nonumber\\
&\qquad
+ \delta \kapA \left[\frac{G M}{r} \left(\frac{65 \nu }{21}+\frac{104}{21}\right)+\left(\frac{\nu }{3}-\frac{11}{42}\right) r^2 \dot{\phi }^2+\left(\frac{10 \nu }{21}-\frac{71}{42}\right) \dot{r}^2\right]
\bigg\rbrace
\Bigg] + \calO(8).
\end{align}
\end{subequations}

Let us stress that the $(\ell,0)$ modes do not vanish for eccentric orbits, unlike the case for circular orbits. The (2,0) mode in particular could be important for highly-eccentric binaries, since it is proportional to the eccentricity and starts at the same PN order as the (2,2) mode.

\subsection{Tail contributions}\label{subsec:tailmodes}

The tail contributions to the modes are computed using the tail part of the radiative moments from Eqs.~\eqref{eq:tailsgen} and plugging them in Eq.~\eqref{eq:hlm}, which yields the spin part of the modes to 3PN.
The 3PN nonspinning tail contributions were derived in Ref.~\cite{Boetzel:2019nfw}, and we checked that our results for the 2.5PN tail are in agreement.
For the spin terms, we get agreement in the circular-orbit limit with Ref.~\cite{Henry:2022dzx}.

To 3PN order, there are SO tail contributions in the (2,0), (2,1), (2,2), (3,0), and (3,2) modes, starting at 2.5PN for the (2,1) mode and at 3PN for the other modes.
No SS tail contributions enter at this order, since they start at 3.5PN in the (2,2) mode.
We write here the SO part of the tail to $\calO(e^2)$ for the $\ell = 2$ modes, and provide the full expressions to $\calO(e^6)$ in the Supplemental Material.

For the (2,2) mode, we obtain
\begin{align}
H_{22}^\text{tail,SO} &= x^4 c^2 \Bigg[
\left(\delta  \chi _A-\nu  \chi _S+\chi _S\right) \left[-\frac{8}{3} \pi +\di \left(-\frac{16 \ln b}{3}-\frac{16 \gamma_E}{3}-8 \ln x+\frac{44}{9}-\frac{32}{3} \ln 2\right)\right] \nonumber\\
&\qquad
+ e \bigg\lbrace
\e^{-\di l} \bigg[\left(\delta  \chi _A+\chi _S\right) \left(-\frac{32}{3} \pi +\di \left(-\frac{64 \ln b}{3}-\frac{64 \gamma_E}{3}-32 \ln x+\frac{338}{9}+\frac{88 \ln 2}{3}-72 \ln 3\right)\right) \nonumber\\
&\quad\qquad
+\nu  \chi _S \left(\frac{14 \pi }{3}+\di \left(\frac{28 \ln b}{3}+\frac{28 \gamma_E}{3}+14 \ln x-\frac{158}{9}-\frac{52}{3} \ln 2+36 \ln 3\right)\right)\bigg] \nonumber\\
&\qquad
+\e^{\di l} \bigg[\left(\delta  \chi _A+\chi _S\right) \left(-\frac{37}{3} \pi+\di \left(-\frac{74 \ln b}{3}-\frac{74 \gamma_E}{3}-37 \ln x+\frac{515}{18}-\frac{226}{3}  \ln 2\right)\right)\nonumber\\
&\quad\qquad
+\nu  \chi _S \left(\frac{22 \pi }{3}+\di \left(\frac{44 \ln b}{3}+\frac{44 \gamma_E}{3}+22 \ln x-\frac{148}{9}+\frac{124 \ln 2}{3}\right)\right)\bigg]
\bigg\rbrace\nonumber\\
&\qquad
+ e^2 \bigg\lbrace
\left(\delta  \chi _A+\chi _S\right) \left[-\frac{47}{3} \pi+\di \left(-\frac{94 \ln b}{3}-\frac{94 \gamma_E}{3}-47 \ln x+\frac{949}{18}+\frac{850 \ln 2}{3}-279 \ln 3\right)\right]\nonumber\\
&\quad\qquad
+\nu  \chi _S \left[\frac{26 \pi }{3}+\di \left(\frac{52 \ln b}{3}+\frac{52 \gamma_E}{3}+26 \ln x-\frac{251}{9}-\frac{364}{3} \ln 2+126 \ln 3\right)\right]\nonumber\\
&\qquad
\e^{-2 \di l} \bigg[\left(\delta  \chi _A+\chi _S\right) \left(-\frac{131}{6} \pi +\di \left(-\frac{131 \ln b}{3}-\frac{131 \gamma_E}{3}-\frac{131 \ln x}{2}+\frac{2449}{36}-633 \ln 2+279 \ln 3\right)\right)\nonumber\\
&\quad\qquad
+\nu  \chi _S \left(6 \pi +\di \left(12 \ln b+12 \gamma_E+18 \ln x-25+262 \ln 2-126 \ln 3\right)\right)\bigg]\nonumber\\
&\qquad
+ \e^{2 \di l} \bigg[\left(\delta  \chi _A+\chi _S\right) \left(-\frac{175}{6} \pi+\di \left(-\frac{175 \ln b}{3}-\frac{175 \gamma_E}{3}-\frac{175 \ln x}{2}+\frac{2357}{36}-\frac{551}{3}  \ln 2\right)\right)\nonumber\\
&\quad\qquad
+\nu  \chi _S \left(\frac{44 \pi }{3}+\di \left(\frac{88 \ln b}{3}+\frac{88 \gamma_E}{3}+44 \ln x-\frac{296}{9}+\frac{266 \ln 2}{3}\right)\right)\bigg]
\bigg\rbrace
\Bigg] + \calO(e^3) + \calO(8),
\end{align}
for (2,1) mode, we obtain
\begin{align}
H_{21}^\text{tail,SO} &= x^{7/2} c^2 (\chi_A+\delta\chi_S) \bigg\lbrace
\ln (2 b x^{3/2})+\gamma_E-\frac{7}{6}-\frac{\di \pi }{2}
+ e \bigg[\e^{\di l} \left(\gamma_E+\ln (2 b x^{3/2})-\frac{7}{6}-\frac{\di \pi }{2}\right) \nonumber\\
&\qquad
+\e^{-\di l} \left(3 \gamma_E+\ln (128 b^3 x^{9/2})-\frac{7}{2}-\frac{3 \di \pi }{2}\right)\bigg] 
+ e^2 \bigg[
\frac{5 \ln b}{2}+\frac{5 \gamma_E}{2}+\frac{15 \ln x}{4}-\frac{35}{12}-\frac{5 \di \pi }{4}+\frac{13 \ln 2}{2} \nonumber\\
&\qquad
+ \e^{-2 \di l} \left(\ln (4 b^6 x^9)+6 \gamma_E-7-3 \di \pi +\frac{81 \ln 3}{8}\right)+\e^{2 \di l} \left(\gamma_E+\ln (2 b x^{3/2})-\frac{7}{6}-\frac{5 \di \pi }{8}\right)
\bigg]
\bigg\rbrace + \calO(e^3) + \calO(7),
\end{align}
and for (2,0) mode, we obtain
\begin{align}
H_{20}^\text{tail,SO} &= -\frac{4}{3} \sqrt{\frac{2}{3}} x^4 c^2 \bigg[
e \bigg\lbrace
\e^{-\di l} \bigg[\left(\delta  \chi _A+\chi _S\right) \left(\frac{13}{8} \pi+\di \left(-\frac{13}{8}\ln (4 b^2 x^3)-\frac{13 \gamma_E}{4}+\frac{71}{48}\right)\right)\nonumber\\
&\quad
+\nu  \chi _S \left(\pi +\di \left(\ln (4 b^2 x^3)+2 \gamma_E-\frac{13}{12}\right)\right)\bigg]
+\e^{\di l} \bigg[
\left(\delta  \chi _A+\chi _S\right) \left(-\frac{13}{8}\pi +\di \left(\frac{13}{8} \ln (4 b^2 x^3)+\frac{13 \gamma_E}{4}-\frac{71}{48}\right)\right)\nonumber\\
&\quad
+\nu  \chi _S \left(\pi +\di \left(-\ln (4 b^2 x^3)-2 \gamma_E+\frac{13}{12}\right)\right)\bigg]
\bigg\rbrace
+ e^2 \bigg\lbrace
\e^{-2 \di l} \bigg[
\nu  \chi _S \left(\frac{11 \pi }{4}+\di \left(\frac{11}{4} \ln (16 b^2 x^3)+\frac{11 \gamma_E}{2}-\frac{85}{24}\right)\right)\nonumber\\
&\quad
+\left(\delta  \chi _A+\chi _S\right) \left(-4 \pi +\di \left(\frac{13}{3}-4 \ln (16 b^2 x^3)-8 \gamma_E\right)\!\right)\!\!\bigg] 
+\e^{2 \di l} \bigg[
\nu  \chi _S \left(\frac{11 \pi }{4}+\di \left(\frac{85}{24}-\frac{11}{4}  \ln (16 b^2 x^3)-\frac{11 \gamma_E}{2}\right)\!\right)\nonumber\\
&\quad
+\left(\delta  \chi _A+\chi _S\right) \left(-4 \pi +\di \left(8 \gamma_E+4 \ln (16 b^2 x^3)-\frac{13}{3}\right)\right)\bigg]
\bigg\rbrace
\bigg] + \calO(e^3) + \calO(8).
\end{align}

\subsection{Memory contributions}\label{subsec:memoryres}

As explained in Subsection~\ref{subsec:heredEval}, the memory contributions to the modes can be computed from the energy flux, which can be written in terms of the modes.
The oscillatory memory contributions for eccentric orbits start at 1.5PN in the (2,2) mode for the nonspinning part, and at 2PN for spin.

For the (2,2) mode, we obtain
\begin{subequations}
\begin{align}
H_{22}^\text{osc} &\equiv H_{22}^\text{osc,nS} + H_{22}^\text{osc,SO} + H_{22}^\text{osc,SS} + H_{22}^\text{osc,S$^3$}, \\
H_{22}^\text{osc,nS} &= -\frac{13 \di c^2 \nu}{126} \bigg\lbrace
\frac{e^2 \e^{2 \di l} x^{5/2}}{2}
+x^{7/2}\left[
e \left(-\frac{8 \e^{-\di l}}{13}+\frac{24 \e^{\di l}}{13}\right)
+e^2 \left(\e^{2 \di l} \left(\frac{6055 \nu }{156}-\frac{2091}{104}\right)-\frac{5}{4} \e^{-2 \di l}-\frac{19}{13}\right)
\right] \nonumber\\
&\qquad
+\frac{29 \pi  e^2 \e^{2 \di l} x^4}{13}
\bigg\rbrace + \calO(e^3) + \calO(7), \\
H_{22}^\text{osc,SO} &= -\frac{13 \di c^2 \nu}{126} \bigg\lbrace
e^2 \e^{2 \di l} x^3 \left[\frac{2 \delta  \chi _A}{3}+\left(\frac{2}{3}-\frac{\nu }{3}\right) \chi _S\right]
+e^2 \e^{2 \di l} x^4 \bigg[
\left(-\frac{2053 \nu ^2}{78}+\frac{1139 \nu }{12}-\frac{20537}{624}\right) \chi _S \nonumber\\
&\qquad\quad
+\delta \chi _A \left(\frac{8095 \nu }{156}-\frac{20537}{624}\right)\bigg]
\bigg\rbrace + \calO(e^3) + \calO(7),\\
H_{22}^\text{osc,SS} &= 
-\frac{13 \di c^2 \nu}{126} e^2 \e^{2 \di l} x^{7/2} \bigg[
\left(\frac{23}{36}-\frac{23 \nu }{9}\right) \chi _A^2+\delta  \left(\frac{23}{18}-\frac{8 \nu }{9}\right) \chi _A \chi _S
+\left(\frac{2 \nu ^2}{9}-\frac{8 \nu }{9}+\frac{23}{36}\right) \chi _S^2
-\frac{\delta  \kapA}{4} \nonumber\\
&\qquad\quad
+\left(\frac{\nu }{2}-\frac{1}{4}\right)\kapS
\bigg] + \calO(e^3) + \calO(7),\\
H_{22}^\text{osc,S$^3$} &= -\frac{13 \di c^2 \nu}{126} e^2 \e^{2 \di l}  x^4 \bigg\lbrace
\delta \chi _A^3 \left(\frac{14}{27}-\frac{56 \nu }{27}\right)
+\chi _S^3\left(-\frac{4 \nu ^3}{27}+\frac{8 \nu ^2}{9}-\frac{13 \nu }{9}+\frac{14}{27}\right) 
+\chi _A^2 \chi _S\left(\frac{52 \nu ^2}{9}-\frac{23 \nu }{3}+\frac{14}{9}\right) \nonumber\\
&\qquad\quad 
+\chi _A \kapA \left(\frac{8 \nu }{3}-\frac{2}{3}\right)
+\delta \chi _A \kapS \left(\frac{4 \nu }{3}-\frac{2}{3}\right)
+\delta \chi _A \chi _S^2 \left(\frac{8 \nu ^2}{9}-\frac{26 \nu }{9}+\frac{14}{9}\right)
+\delta \chi _S \kapA \left(\frac{\nu }{3}-\frac{2}{3}\right)\nonumber\\
&\qquad\quad 
+\chi _S \kapS \left(-\frac{2 \nu ^2}{3}+\frac{5 \nu }{3}-\frac{2}{3}\right)
\bigg\rbrace + \calO(e^3) + \calO(7).
\end{align}
\end{subequations}
For the (2,0) mode, we obtain
\begin{align}
H_{20}^\text{osc} &= \frac{ \di c^2 \nu \sqrt{6}}{7}x^{7/2} \left[
e \left(16 \e^{\di l}-16 \e^{-\di l}\right)
+e^2 \left(-\frac{647}{36} \e^{-2 \di l}+\frac{647}{36} \e^{2 \di l}\right)
\right]  + \calO(e^3) + \calO(7),
\end{align}
which starts at 2.5PN, with no spin contributions at 3PN.

Note that while it might seem from the expressions above that the oscillatory memory vanishes when $e\to 0$ for quasi-circular orbits, some modes, namely the (3,3), (3,1), (4,4) and (5,~odd~$m$) modes, have a nonspinning oscillatory memory at 2.5 or 3PN for quasi-circular orbits.

In addition to the oscillatory memory, the $(\text{even}~\ell, m=0)$ modes contain a non-oscillatory (or DC) memory, which starts in the (2,0) mode at Newtonian order for the nonspinning part and at 1.5PN for the spin part.
At leading order in an eccentricity expansion, we get
\begin{subequations}
\begin{align}
H_{20}^\text{DC} &\equiv H_{20}^\text{DC,nS} + H_{20}^\text{DC,SO} + H_{20}^\text{DC,SS} + H_{20}^\text{DC,HF}, \\
H_{20}^\text{DC,SO} &= - \frac{95 c^2 x^{5/2}}{56448 \sqrt{6}} \Bigg\lbrace
\chi _S \left[\frac{54726}{95}+\frac{70336 e^{12/19}}{361 e_i^{12/19}}-\frac{1391474 e^{30/19}}{1805 e_i^{30/19}}+\nu  \left(-\frac{49280 e^{12/19}}{361 e_i^{12/19}}+\frac{319816 e^{30/19}}{1805 e_i^{30/19}}-\frac{3864}{95}\right) \right] \nonumber\\
&\qquad
+\delta  \chi _A \left[\frac{70336 e^{12/19}}{361 e_i^{12/19}}-\frac{1391474 e^{30/19}}{1805 e_i^{30/19}}+\frac{54726}{95}\right]
\Bigg\rbrace \nonumber\\
&\quad
- \frac{95 c^2 x^{7/2}}{56448 \sqrt{6}} \Bigg\lbrace
\chi_S \Bigg[
\frac{4888427}{3192}\frac{188274592 e^{12/19}}{102885 e_i^{12/19}}-\frac{45660938 e^{24/19}}{61731 e_i^{24/19}}-\frac{281574703 e^{30/19}}{164616 e_i^{30/19}}-\frac{7875278603 e^{42/19}}{8642340 e_i^{42/19}}\nonumber\\
&\quad\qquad
+\nu  \left(-\frac{314478088 e^{12/19}}{102885 e_i^{12/19}}+\frac{107136964 e^{24/19}}{61731 e_i^{24/19}}+\frac{76619726 e^{30/19}}{20577 e_i^{30/19}}-\frac{24969272 e^{42/19}}{2160585 e_i^{42/19}}-\frac{50210}{21}\right) \nonumber\\
&\quad\qquad
+\nu ^2 \left(\frac{14008736 e^{12/19}}{20577 e_i^{12/19}}-\frac{17549840 e^{24/19}}{20577 e_i^{24/19}}-\frac{15750938 e^{30/19}}{20577 e_i^{30/19}}+\frac{18813356 e^{42/19}}{20577 e_i^{42/19}}+\frac{442}{19}\right)
\Bigg] \nonumber\\
&\qquad
+ \delta\chi_A \Bigg[
\frac{4888427}{3192} +\frac{188274592 e^{12/19}}{102885 e_i^{12/19}}-\frac{45660938 e^{24/19}}{61731 e_i^{24/19}}-\frac{281574703 e^{30/19}}{164616 e_i^{30/19}}-\frac{7875278603 e^{42/19}}{8642340 e_i^{42/19}} \nonumber\\
&\quad\qquad
+ \nu  \left(\frac{257}{2}-\frac{97114304 e^{12/19}}{102885 e_i^{12/19}}+\frac{25048408 e^{24/19}}{20577 e_i^{24/19}}+\frac{137060189 e^{30/19}}{41154 e_i^{30/19}}-\frac{383998931 e^{42/19}}{102885 e_i^{42/19}}\right)
\Bigg]
\Bigg\rbrace \nonumber\\
&\quad
- \frac{95\pi  c^2 x^4}{56448 \sqrt{6}} \Bigg\lbrace
\chi_S \bigg[
-\frac{483}{38}-\frac{438718 e^{12/19}}{20577 e_i^{12/19}}
-\frac{328584529 e^{30/19}}{82308 e_i^{30/19}}+\frac{331385579 e^{48/19}}{82308 e_i^{48/19}} \nonumber\\
&\qquad\quad
+ \nu  \left(\frac{966}{19}-\frac{3294424 e^{12/19}}{20577 e_i^{12/19}}+\frac{26682929 e^{30/19}}{20577 e_i^{30/19}}-\frac{24434683 e^{48/19}}{20577 e_i^{48/19}}\right)
\bigg]\nonumber\\
&\qquad
+\delta\chi_A \left[-\frac{483}{38}-\frac{438718 e^{12/19}}{20577 e_i^{12/19}}-\frac{328584529 e^{30/19}}{82308 e_i^{30/19}}+\frac{331385579 e^{48/19}}{82308 e_i^{48/19}}\right]
\Bigg\rbrace + \calO(e^2) + \calO(7),\\
H_{20}^\text{DC,SS} &= \frac{55 c^2 x^3}{896 \sqrt{6}} \Bigg\lbrace
\chi _A^2 \left[1+\frac{382 e^{12/19}}{627 e_i^{12/19}}-\frac{1009 e^{36/19}}{627 e_i^{36/19}}+\nu  \left(-\frac{1424 e^{12/19}}{627 e_i^{12/19}}+\frac{3856 e^{36/19}}{627 e_i^{36/19}}-\frac{128}{33}\right)\right] \nonumber\\
&\qquad
+\chi _S^2 \left[1+\frac{382 e^{12/19}}{627 e_i^{12/19}}-\frac{1009 e^{36/19}}{627 e_i^{36/19}}+\nu  \left(-\frac{104 e^{12/19}}{627 e_i^{12/19}}+\frac{60 e^{36/19}}{209 e_i^{36/19}}-\frac{4}{33}\right)\right] \nonumber\\
&\qquad
+\delta  \chi _S\chi _A \left[2+\frac{764 e^{12/19}}{627 e_i^{12/19}}-\frac{2018 e^{36/19}}{627 e_i^{36/19}}\right]
\Bigg\rbrace \nonumber\\
&\quad 
+\frac{55 c^2 x^4}{896 \sqrt{6}} \Bigg\lbrace
\chi_A^2 \Bigg[
\nu ^2 \left(\frac{310960 e^{12/19}}{35739 e_i^{12/19}}-\frac{46102 e^{24/19}}{3249 e_i^{24/19}}-\frac{379816 e^{36/19}}{11913 e_i^{36/19}}+\frac{43186 e^{48/19}}{1083 e_i^{48/19}}-\frac{248}{99}\right) \nonumber\\
&\qquad\quad
+\nu  \left(-\frac{1421707 e^{12/19}}{50787 e_i^{12/19}}+\frac{37311565 e^{24/19}}{3002076 e_i^{24/19}}+\frac{31208774 e^{30/19}}{964953 e_i^{30/19}}+\frac{1374141 e^{36/19}}{55594 e_i^{36/19}}-\frac{2893925 e^{48/19}}{101574 e_i^{48/19}}-\frac{3277}{252}\right)\nonumber\\
&\qquad\quad
+\frac{176734909 e^{12/19}}{27018684 e_i^{12/19}}-\frac{27774647 e^{24/19}}{12008304 e_i^{24/19}}-\frac{15604387 e^{30/19}}{1929906 e_i^{30/19}}-\frac{2858497 e^{36/19}}{667128 e_i^{36/19}}+\frac{1097928245 e^{48/19}}{216149472 e_i^{48/19}}+\frac{29107}{9504}
\Bigg] \nonumber\\
&\qquad
+\chi_S^2 \Bigg[
\nu ^2 \left(\frac{2429620 e^{12/19}}{964953 e_i^{12/19}}-\frac{3367 e^{24/19}}{3249 e_i^{24/19}}-\frac{114220 e^{30/19}}{87723 e_i^{30/19}}-\frac{5910 e^{36/19}}{3971 e_i^{36/19}}-\frac{4279865 e^{48/19}}{964953 e_i^{48/19}}+\frac{1706}{297}\right)\nonumber\\
&\qquad\quad
+\nu  \left(-\frac{94071580 e^{12/19}}{6754671 e_i^{12/19}}+\frac{3329440 e^{24/19}}{750519 e_i^{24/19}}+\frac{7259759 e^{30/19}}{964953 e_i^{30/19}}+\frac{759448 e^{36/19}}{83391 e_i^{36/19}}+\frac{475406329 e^{48/19}}{54037368 e_i^{48/19}}-\frac{265073}{16632}\right)\nonumber\\
&\qquad\quad
+\frac{176734909 e^{12/19}}{27018684 e_i^{12/19}}-\frac{27774647 e^{24/19}}{12008304 e_i^{24/19}}-\frac{15604387 e^{30/19}}{1929906 e_i^{30/19}}-\frac{2858497 e^{36/19}}{667128 e_i^{36/19}}+\frac{1097928245 e^{48/19}}{216149472 e_i^{48/19}}+\frac{29107}{9504}
\Bigg]\nonumber\\
&\qquad
+\delta\chi_S\chi_A \Bigg[
\nu  \left(-\frac{1382126 e^{12/19}}{87723 e_i^{12/19}}+\frac{49469 e^{24/19}}{6498 e_i^{24/19}}+\frac{7259759 e^{30/19}}{964953 e_i^{30/19}}+\frac{198773 e^{36/19}}{11913 e_i^{36/19}}+\frac{2411891 e^{48/19}}{3859812 e_i^{48/19}}-\frac{19829}{1188}\right)\nonumber\\
&\qquad\quad
+\frac{176734909 e^{12/19}}{13509342 e_i^{12/19}}-\frac{27774647 e^{24/19}}{6004152 e_i^{24/19}}-\frac{15604387 e^{30/19}}{964953 e_i^{30/19}}-\frac{2858497 e^{36/19}}{333564 e_i^{36/19}}+\frac{1097928245 e^{48/19}}{108074736 e_i^{48/19}}+\frac{29107}{4752}
\Bigg] \nonumber\\
&\qquad
+ \kapS \Bigg[
\nu ^2 \left(\frac{155480 e^{12/19}}{35739 e_i^{12/19}}-\frac{23051 e^{24/19}}{3249 e_i^{24/19}}-\frac{189908 e^{36/19}}{11913 e_i^{36/19}}+\frac{21593 e^{48/19}}{1083 e_i^{48/19}}-\frac{124}{99}\right)\nonumber\\
&\qquad\quad
+\nu  \left(-\frac{500548 e^{12/19}}{35739 e_i^{12/19}}+\frac{23591675 e^{24/19}}{3002076 e_i^{24/19}}+\frac{1347431 e^{36/19}}{83391 e_i^{36/19}}+\frac{181063 e^{48/19}}{47652 e_i^{48/19}}-\frac{28712}{2079}\right)\nonumber\\
&\qquad\quad
+\frac{497792 e^{12/19}}{107217 e_i^{12/19}}-\frac{12942113 e^{24/19}}{6004152 e_i^{24/19}}-\frac{682753 e^{36/19}}{166782 e_i^{36/19}}-\frac{3243271 e^{48/19}}{857736 e_i^{48/19}}+\frac{4073}{756}
\Bigg] \nonumber\\
&\qquad
+ \delta\chi_A \Bigg[
\nu  \left(-\frac{506060 e^{12/19}}{107217 e_i^{12/19}}+\frac{23051 e^{24/19}}{6498 e_i^{24/19}}+\frac{94954 e^{36/19}}{11913 e_i^{36/19}}-\frac{403426 e^{48/19}}{107217 e_i^{48/19}}-\frac{601}{198}\right) +\frac{4073}{756}\nonumber\\
&\qquad\quad
+\frac{497792 e^{12/19}}{107217 e_i^{12/19}}-\frac{12942113 e^{24/19}}{6004152 e_i^{24/19}}-\frac{682753 e^{36/19}}{166782 e_i^{36/19}}-\frac{3243271 e^{48/19}}{857736 e_i^{48/19}}
\Bigg]
\Bigg\rbrace + \calO(e^2) + \calO(7), \\
H_{20}^\text{DC,HF} &= \frac{288 c^2 x^{7/2}}{19}
\left(1+\frac{98 e^{12/19}}{95 e_i^{12/19}}-\frac{193 e^{42/19}}{95 e_i^{42/19}}\right)
\left[\delta \chi _A  (1-\nu) \left(3\chi _A^2+9 \chi _S^2+1\right)+  \chi _S(1-3 \nu) \left(9 \chi _A^2+3\chi _S^2+1\right)\right] \nonumber\\
&\quad  + \calO(e^2) + \calO(7),
\end{align}
\end{subequations}
where the term $H_{20}^\text{DC,HF}$ is the horizon flux contribution.

Our results for the memory agree with Ref.~\cite{Ebersold:2019kdc} in the nonspinning case, which we computed to 3PN for the oscillatory memory as a check, and to 2PN for the DC memory. 
For the spin part in the circular-orbit limit, we find agreement with the 2PN results of Ref.~\cite{Mitman:2022kwt}, which were computed using a different method, by relating the memory to the Moreschi supermomentum, which is equivalent to the energy flux at null infinity.
For the 2.5PN and 3PN spin contributions, our results for the DC memory agree in the circular-orbit and test-mass limits with Ref.~\cite{Cunningham:2024dog}, in which the authors computed the memory contribution analytically and numerically using both PN and gravitational-self-force methods.

\begin{figure*}
\centering
\includegraphics[width=0.9\linewidth]{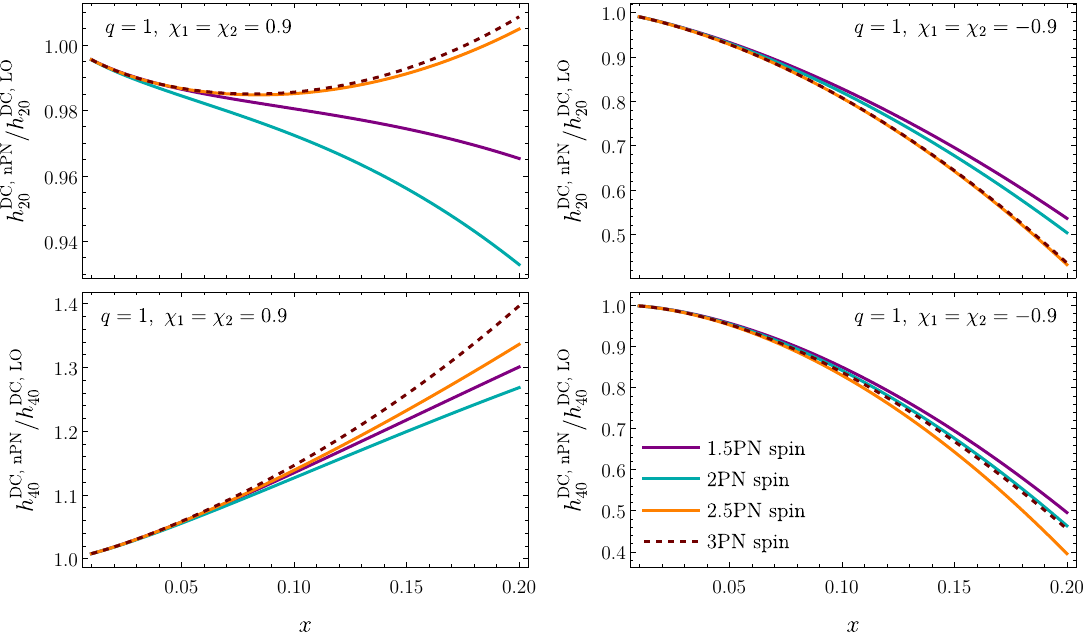}
\caption{Comparisons showing the effect of each spin PN order on the DC memory as a function of the frequency variable $x$.
The top panels are for the (2,0) mode, while the bottom ones are for the (4,0) mode. The left panels are for spins aligned with $\bm{L}$, while the right panels are for spins anti-aligned with $\bm{L}$, and all plots are for mass ratio $q = 1$. 
We see that the NLO SO term at 2.5PN can have a significant contribution to the memory, particularly for large aligned spins, while the 3PN spin terms have a smaller effect.}
\label{fig:memory} 
\end{figure*}

To estimate the effect of the 2.5PN and 3PN spin terms derived here on the memory, compared to the lower orders in the circular-orbit case, we plot in Fig.~\ref{fig:memory} the (2,0) and (4,0) modes, scaled by their LO, as functions of $x$ at each PN order for the spin part, while keeping the full 3PN nonspinning part.
We consider two equal-mass configurations: one with spin magnitudes $0.9$ aligned with the direction of orbital angular momentum, and the other with the same spin magnitudes but anti-aligned with the orbital angular momentum.
We see that for large aligned spins, the NLO SO term at 2.5PN has a significant effect on the DC memory, while the 3PN SO and SS terms have a smaller effect.
For anti-aligned spins, and also for small spin magnitudes, the effect of PN orders beyond the leading order is relatively small.

\subsection{Phase redefinition for the full waveform}
\label{sec:phaseRedef}

The waveform modes expressed as in Eq.~\eqref{eq:defHlm}, in terms of the orbital phase $\phi$, depend on the arbitrary gauge constant $b$. The logarithmic dependence on $b$ can be eliminated by a phase redefinition, or a shift in the coordinate time, as was done for circular orbits in Refs.~\cite{Arun:2004ff,Blanchet:1996pi,Kidder:2007rt}.
This procedure was generalized for eccentric orbits in Ref.~\cite{Boetzel:2019nfw} (see Sec.~V.C. there), and we follow the same steps here, except for not including the post-adiabatic corrections, which are not needed for the 3PN spin contributions.

First, one redefines the mean anomaly $l$ as
\begin{equation}
\label{eq:xil}
\xi \equiv l - \frac{3G\mathcal{M}}{c^3} n \ln \left(\frac{x}{x_0'}\right),
\end{equation}
where $\mathcal{M} = M (1 - \nu x/2) + \calO(x^2)$ is the ADM mass, and $x_0'$ is related to $b$ via
\begin{equation}
x_0' \equiv \left[(\e^{\frac{11}{12}-\gamma_E}) / (4 b)\right]^{2/3}.
\end{equation}
Then, one defines the new phase $\psi$ as
\begin{equation}
\psi = \phi - W(l) + W(\xi) - \frac{3 G \mathcal{M}}{c^3} K n \ln \left(\frac{x}{x_0'}\right)
\end{equation}
where $W(l)$ is the oscillatory part of the orbital phase $\phi = K l + W(l)$.

In terms of $\psi$, the waveform modes can be expressed as
\begin{equation}
h_{\ell m} = \frac{8GM\nu}{c^2 R} x \sqrt{\frac{\pi}{5}} H_{\ell m}^\psi \e^{-\di m \psi}.
\end{equation}
As an example, the (2,2) mode at leading order in eccentricity becomes
\begin{subequations}
\begin{align}
H_{22}^\psi &= H_{22}^{\psi,\text{nS}} + H_{22}^{\psi,\text{SO}} + H_{22}^{\psi,\text{SS}}, \\
H_{22}^{\psi,\text{SO}} &= x^{3/2} \bigg\lbrace
\left(\frac{4 \nu }{3}-\frac{4}{3}\right) \chi _S -\frac{4}{3}  \delta  \chi _A
+e \left[\e^{-\di \xi } \left(\left(\frac{5 \nu }{6}-\frac{7}{3}\right) \chi _S-\frac{7 \delta  \chi _A}{3}\right)+\e^{\di \xi } \left(\left(\frac{13 \nu }{6}-\frac{8}{3}\right) \chi _S-\frac{8 \delta  \chi _A}{3}\right)\right]
\bigg\rbrace \nonumber\\
&\quad
+ x^{5/2} \bigg\lbrace
\delta  \left(-\frac{16 \nu }{9}-\frac{160}{63}\right) \chi _A+\left(\frac{88 \nu ^2}{21}+\frac{202 \nu }{63}-\frac{160}{63}\right) \chi _S
+ e \bigg[
\e^{-\di \xi } \bigg(
\left(\frac{26 \nu ^2}{3}+\frac{9077 \nu }{504}-\frac{3253}{252}\right) \chi _S \nonumber\\
&\quad\qquad
+\delta  \left(-\frac{275 \nu }{72}-\frac{3253}{252}\right) \chi _A \bigg)
+\e^{\di \xi } \left(\delta  \left(-\frac{3013 \nu }{504}-\frac{1475}{252}\right) \chi _A+\left(\frac{61 \nu ^2}{7}+\frac{6949 \nu }{504}-\frac{1475}{252}\right) \chi _S\right)
\bigg]
\bigg\rbrace \nonumber\\
&\quad
+ x^6 \Bigg\lbrace
\left[\pi  \left(\frac{8 \nu }{3}-\frac{8}{3}\right)+\di \left(\frac{8 \nu }{3}-\frac{4}{3}\right)\right] \chi _S 
-\left(\frac{8 \pi }{3}+\frac{4}{3} \di\right) \delta  \chi _A
+ e \bigg\lbrace
\e^{-\di \xi } \bigg[
\delta \chi _A  \left(-\frac{32 \pi }{3}+\di \left(\frac{167}{12}-72 \ln (3/2)\right)\right)\nonumber\\
&\qquad\quad
+\left(\pi  \left(\frac{14 \nu }{3}-\frac{32}{3}\right)+\di \left(\frac{167}{12}-\frac{5 \nu }{6}+(36 \nu -72) \ln (3/2)\right)\right) \chi _S\bigg] 
+ \e^{\di \xi } \bigg[\delta  \left(-\frac{37 \pi }{3}+\di \left(\frac{43}{12}-26 \ln 2\right)\right) \chi _A\nonumber\\
&\qquad\quad
+\left(\pi  \left(\frac{22 \nu }{3}-\frac{37}{3}\right)+\di \left(\frac{11 \nu }{6}+2(6 \nu -13) \ln 2+\frac{43}{12}\right)\right) \chi _S\bigg]
\bigg\rbrace
\Bigg\rbrace + \calO(e^2) + \calO(7), \\
H_{22}^{\psi,\text{SS}} &= x^4 \Bigg\lbrace
(1-4 \nu ) \chi _A^2+2 \delta  \chi _A \chi _S+\chi _S^2+\delta  \kapA+\kapS (1-2 \nu )
+ e \bigg\lbrace
\e^{-\di \xi } \bigg[\left(\frac{15}{8}-\frac{15 \nu }{2}\right) \chi _A^2+\frac{15}{4} \delta  \chi _A \chi _S+\frac{15 \chi _S^2}{8} \nonumber\\
&\quad\qquad
+\frac{15 \delta  \kapA}{8}+\kapS \left(\frac{15}{8}-\frac{15 \nu }{4}\right)\bigg] 
+ \e^{\di \xi } \bigg[\left(\frac{15}{8}-\frac{15 \nu }{2}\right) \chi _A^2+\frac{15}{4} \delta  \chi _A \chi _S+\frac{15 \delta  \kapA}{8}+\kapS \left(\frac{15}{8}-\frac{15 \nu }{4}\right)\nonumber\\
&\quad\qquad
+\frac{15 \chi _S^2}{8}\bigg]
\bigg\rbrace
\Bigg\rbrace
+ x^3 \Bigg\lbrace
\left(-\frac{136 \nu ^2}{21}+\frac{73 \nu }{63}+\frac{8}{63}\right) \chi _A^2+\delta  \left(\frac{16}{63}-\frac{52 \nu }{9}\right) \chi _A \chi _S+\left(\frac{8 \nu ^2}{3}-\frac{67 \nu }{9}+\frac{8}{63}\right) \chi _S^2 \nonumber\\
&\quad\qquad
+\delta  \kapA \left(\frac{4}{7}-\nu \right)+\kapS \left(-\frac{68 \nu ^2}{21}-\frac{15 \nu }{7}+\frac{4}{7}\right) 
+ e \bigg\lbrace
\e^{\di \xi } \bigg[\left(-\frac{179 \nu ^2}{12}-\frac{45637 \nu }{1008}+\frac{1765}{144}\right) \chi _A^2\nonumber\\
&\quad\qquad
+\delta  \left(\frac{1765}{72}-\frac{14179 \nu }{504}\right) \chi _A \chi _S+\left(\frac{25 \nu ^2}{3}-\frac{32141 \nu }{1008}+\frac{1765}{144}\right) \chi _S^2
+\kapS \left(-\frac{179 \nu ^2}{24}-\frac{6481 \nu }{336}+\frac{359}{48}\right)\nonumber\\
&\quad\qquad
+\delta  \kapA \left(\frac{359}{48}-\frac{485 \nu }{112}\right) \bigg]
+\e^{-\di \xi } \bigg[\left(-\frac{1045 \nu ^2}{84}-\frac{8939 \nu }{144}+\frac{16763}{1008}\right) \chi _A^2
+\delta  \left(\frac{16763}{504}-\frac{15611 \nu }{504}\right) \chi _A \chi _S\nonumber\\
&\quad\qquad
+\left(\frac{25 \nu ^2}{3}-\frac{35701 \nu }{1008}+\frac{16763}{1008}\right) \chi _S^2
+\kapS \left(-\frac{1045 \nu ^2}{168}-\frac{7097 \nu }{336}+\frac{2713}{336}\right)\nonumber\\
&\quad\qquad
+\delta  \kapA \left(\frac{2713}{336}-\frac{557 \nu }{112}\right)\bigg]
\bigg\rbrace
\Bigg\rbrace + \calO(e^2) + \calO(7),
\end{align} 
\end{subequations}
where the nonspinning part $H_{22}^{\psi,\text{nS}}$ is given at 3PN by Eq.~(76) of Ref.~\cite{Boetzel:2019nfw}. The full expressions to $\calO(e^6)$ for all modes that have tail contributions are provided in the Supplemental Material.

\section{Conclusions}
\label{sec:conclusions}
In this paper, we completed the spin contributions in the radiative sector at 3PN order for eccentric orbits. 
We derived the energy and angular-momentum fluxes, as well as the waveform modes.
In the fluxes, our results for the tail contributions complement the instantaneous ones derived in Refs.~\cite{Bohe:2015ana,Cho:2021mqw,Cho:2022syn}, and the 3PN nonspinning results of Refs.~\cite{Arun:2007sg,Arun:2007rg,Arun:2009mc,Ebersold:2019kdc}. 
We also computed the orbit-averaged fluxes and from them, the time evolution of the secular orbital elements, namely the frequency, eccentricity, semi-major axis, and periastron advance.
For the modes, we obtained the instantaneous, tail, and memory (oscillatory and DC) contributions, which extends the 3PN nonspinning results of Refs.~\cite{Mishra:2015bqa,Boetzel:2019nfw,Ebersold:2019kdc} and the 2PN spin results of Refs.~\cite{Khalil:2021txt,Paul:2022xfy}.

To compute the hereditary contributions to the fluxes and modes, we first derived the quasi-Keplerian (QK) parametrization at 3PN for nonprecessing spins in harmonic coordinates using the covariant SSC, which complements the ADM-coordinates results of Refs.~\cite{Tessmer:2010hp,Tessmer:2012xr}.
Then, using the QK parametrization, we computed the hereditary contributions to the fluxes and modes in an eccentricity expansion for bound orbits, to $\calO(e^8)$ in the fluxes and to $\calO(e^6)$ in the modes. 
Thus, while the instantaneous contributions are valid for general planar orbits, the hereditary contributions are valid for small eccentricities.
However, resuming the eccentricity expansion can lead to very good agreement with numerical calculations, as shown in Fig.~\ref{fig:tail} for the orbit-averaged fluxes. A similar resummation can also be performed to the hereditary contributions in the waveform modes.

Recently, numerical-relativity (NR) simulations have been produced for eccentric orbits~\cite{Joshi:2022ocr,Ramos-Buades:2022lgf}, and several eccentric-orbit waveform models have been developed, using PN Taylor-expanded approximants~\cite{Cornish:2010cd,Huerta:2014eca,Tanay:2016zog,Moore:2018kvz,Moore:2019xkm}, 
hybrid PN-NR approaches~\cite{Huerta:2016rwp,Huerta:2017kez,Hinder:2017sxy,Ramos-Buades:2019uvh,Setyawati:2021gom,Chattaraj:2022tay,Chen:2020lzc}, the effective-one-body formalism~\cite{Hinderer:2017jcs,Ramos-Buades:2021adz,Chiaramello:2020ehz,Nagar:2021gss,Placidi:2021rkh,Albanesi:2022xge,Cao:2017ndf,Liu:2021pkr,Liu:2023dgl,Placidi:2023ofj}, and an NR surrogate~\cite{Islam:2021mha}.
However, these models were either restricted to nonspinning binaries, or included partial spin contributions at leading order in the radiative sector.
Hence, we expect the results of this paper will be important in improving eccentric-orbit waveform models for spinning binaries.

\section*{Acknowledgments}
We thank Luc Blanchet, Alessandra Buonanno, Guillaume Faye, Aldo Gamboa, François Larrouturou, and Chandra Mishra for interesting discussions.
We are also grateful to Kevin Cunningham, Chris Kavanagh, and Niels Warburton for comparing our results for the DC memory with theirs and pointing out the horizon flux contribution.

M.K.'s work is supported by Perimeter Institute for Theoretical Physics.
Research at Perimeter Institute is supported in part by the Government of Canada through the Department of Innovation, Science and Economic Development and by the Province of Ontario through the Ministry of Colleges and Universities.

\appendix

\section{Explicit expressions for the QK parametrization}\label{app:QK}

In this appendix, we include explicit expressions for the quantities that enter the QK parametrization discussed in Subsection~\ref{sec:QK}. We include the nonspinning contributions to 2PN and spin contributions to 3PN, in harmonic coordinates using the covariant SSC.

The semi-major axis $a_r$ and radial eccentricity $e_r$, which enter Eq.~\eqref{eq:QKr}, are given in terms of the energy and angular momentum by
\begin{subequations}
\begin{align}
\frac{a_r}{G M} &= \frac{1}{2 \tilde{E}} + \frac{\nu -7}{4 c^2} + \frac{1}{c^4} \left[\frac{7 \nu -4}{h^2}+\frac{1}{8} \left(\nu ^2+1\right) \tilde{E}\right]
+\frac{2}{c^3 h} \left(\delta  \chi _A-\nu  \chi _S+\chi _S\right) 
+ \frac{1}{c^5} \bigg[
\frac{2 (\nu -1) \tilde{E} \left(\delta  \chi _A-\nu  \chi _S+\chi _S\right)}{h} \nn\\
&\qquad
+\frac{\delta  (16-5 \nu ) \chi _A+\left(2 \nu ^2-21 \nu +16\right) \chi _S}{h^3}\bigg] 
+ \frac{1}{c^4 h^2} \left[\left(2 \nu -\frac{1}{2}\right) \chi _A^2-\delta  \chi _A \chi _S-\frac{\chi _S^2}{2}-\frac{\delta  \kapA}{2}+\kapS \left(\nu -\frac{1}{2}\right)\right] \nn\\
&\quad
+\frac{1}{c^6}\bigg\lbrace
\frac{\tilde{E}}{h^2} \Big[\left(4 \nu ^2-10 \nu +2\right) \chi _A^2+\delta  (4-8 \nu ) \chi _A \chi _S+2 \delta \kapA+2 \kapS (\nu -1)^2+\left(4 \nu ^2-6 \nu +2\right) \chi _S^2\Big] \nonumber\\
&\qquad
+\frac{1}{h^4} \bigg[\left(-6 \nu ^2+\frac{175 \nu }{2}-21\right) \chi _A^2+7 \delta  (5 \nu -6) \chi _A \chi _S+\left(-12 \nu ^2+\frac{63 \nu }{2}-21\right) \chi _S^2 +\delta  \kapA \left(\frac{5 \nu }{2}-9\right) \nn\\
&\qquad\quad
+\kapS \left(-3 \nu ^2+\frac{41 \nu }{2}-9\right)\bigg]
\bigg\rbrace, \\
e_r &= \sqrt{1-2 h^2 \tilde{E}} 
+ \frac{\tilde{E}[5 h^2 (\nu -3) \tilde{E}-2 (\nu -6)]}{2 c^2 \sqrt{1-2 h^2 \tilde{E}}}
+\frac{\tilde{E}}{8 c^4 h^2 (1-2 h^2 \tilde{E})^{3/2}} \Bigr[h^6 \left(7 \nu ^2-210 \nu +415\right) \tilde{E}^3 \nonumber\\
&\qquad
-4 h^4 \left(\nu ^2+148 \nu +50\right) \tilde{E}^2+8 h^2 (99 \nu -35) \tilde{E}+32 (4-7 \nu )\Bigr]\nonumber\\
&\quad
+\frac{4 \tilde{E}}{c^3 h \sqrt{1-2 h^2 \tilde{E}}} \Bigr[2 \delta  \chi _A \left(h^2 \tilde{E}-1\right)
+h^2 \tilde{E} (2-3 \nu ) \chi _S+2 (\nu -1) \chi _S\Bigr]\nonumber\\
&\quad
-\frac{\tilde{E}}{c^5 h^3 (1-2 h^2 \tilde{E})^{3/2}} \Bigr\lbrace\!
4 \left(2 \nu ^2-21 \nu +16\right) \chi _S +4 \delta  (16-5 \nu ) \chi _A
+2 \tilde{E} h^2 \left[5 \delta  (7 \nu -18) \chi _A+\left(125 \nu-14 \nu ^2 -90\right) \chi _S\right] \nonumber\\
&\qquad
-2 h^6 \tilde{E}^3 \left[5 \delta  (\nu -10) \chi _A+\left(-7 \nu ^2+70 \nu -50\right) \chi _S\right]
+3 h^4 \tilde{E}^2 \left[\delta  (12-19 \nu ) \chi _A+\left(6 \nu ^2-29 \nu +12\right) \chi _S\right] \Bigr\rbrace \nonumber\\
&\quad
+ \frac{2 \tilde{E} (1-h^2 \tilde{E})}{c^4 h^2 \sqrt{1-2 h^2 \tilde{E}}}  \left[(1-4 \nu ) \chi _A^2+2 \delta  \chi _A \chi _S+\delta \kapA+\kapS (1-2 \nu )+\chi _S^2\right]\nonumber\\
&\quad
+ \frac{\tilde{E}}{c^6 h^4 (1-2 h^2 \tilde{E})^{3/2}} \bigg\lbrace
\left(24 \nu ^2-350 \nu +84\right) \chi _A^2+28 \delta  (6-5 \nu ) \chi _A \chi _S+6 \left(8 \nu ^2-21 \nu +14\right) \chi _S^2
+2 \delta  \kapA (18-5 \nu )\nonumber\\
&\qquad
+\kapS \left(12 \nu ^2-82 \nu +36\right) 
+ h^2 \tilde{E} \Bigr[\left(-92 \nu ^2+1083 \nu -257\right) \chi _A^2+2 \delta  (227 \nu -257) \chi _A \chi _S
+11 \delta  \kapA (3 \nu -11)\nonumber\\
&\quad\qquad
+\kapS \left(-46 \nu ^2+275 \nu -121\right)
+\left(-152 \nu ^2+399 \nu -257\right) \chi _S^2\Bigr] 
+ h^4 \tilde{E}^2 \Bigr[\left(92 \nu ^2-535 \nu +119\right) \chi _A^2\nonumber\\
&\quad\qquad
+\delta  (238-278 \nu ) \chi _A \chi _S+\delta \kapA (87-23 \nu )+\kapS \left(46 \nu ^2-197 \nu +87\right)+\left(76 \nu ^2-219 \nu +119\right) \chi _S^2\Bigr] \nonumber\\
&\qquad
+ h^6 \tilde{E}^3 \Bigr[\left(-4 \nu ^2-299 \nu +77\right) \chi _A^2-22 \delta  (5 \nu -7) \chi _A \chi _S+\left(64 \nu ^2-119 \nu +77\right) \chi _S^2 +\delta \kapA (13-7 \nu )\nonumber\\
&\quad\qquad
+ \kapS \left(-2 \nu ^2-33 \nu +13\right)\Bigr]
\bigg\rbrace,
\end{align}
\end{subequations}
where we recall that $\Et \equiv - (E-Mc^2)/\mu$ and $h\equiv L/(GM\mu)$.

The time and phase eccentricities are related to $e_r$ via 
\begin{subequations}
\begin{align}
\frac{e_t}{e_r} &= 1 + \frac{(3 \nu -8) \tilde{E}}{c^2}
+\frac{\tilde{E} [h^2 \left(6 \nu ^2-19 \nu +36\right) \tilde{E}+14 \nu -8]+3 \sqrt{2} h (2 \nu -5) \tilde{E}^{3/2}}{c^4 h^2} \nn\\
&\quad
+   \frac{4\tilde{E}}{c^3 h} \left[\delta  \chi _A- (\nu -1) \chi _S\right]
+ \frac{2\tilde{E} }{c^5 h^3} \Bigr\lbrace
\delta  \chi _A \left[h^2 (7 \nu -11) \tilde{E}-4 \sqrt{2} h (\nu -3) \sqrt{\tilde{E}}-5 \nu +16\right] \nn\\
&\quad\qquad
+  \chi _S \left[h^2 \left(-7 \nu ^2+18 \nu -11\right) \tilde{E}+2 \sqrt{2} h \left(\nu ^2-8 \nu +6\right) \sqrt{\tilde{E}}+2 \nu ^2-21 \nu +16\right]
\Bigr\rbrace \nn\\
&\quad
+ \frac{ \tilde{E}}{c^4 h^2} \left[(4 \nu -1) \chi _A^2-2 \delta \chi _A \chi _S- \chi _S^2-\delta  \kapA+\kapS (2 \nu -1)\right] \nn\\
&\quad
+ \frac{\tilde{E}}{2 c^6 h^4} \Bigr\lbrace
\chi _A^2 \left[h^2 \left(36 \nu ^2-81 \nu +17\right) \tilde{E}-4 \sqrt{2} h \left(6 \nu ^2-46 \nu +11\right) \sqrt{\tilde{E}}-24 \nu ^2+350 \nu -84\right]\nn\\
&\quad\qquad
+2 \delta  \chi _A \chi _S \left[h^2 (17-21 \nu ) \tilde{E}+4 \sqrt{2} h (9 \nu -11) \sqrt{\tilde{E}}+70 \nu -84\right]\nn\\
&\quad\qquad
+\chi _S^2 \left[h^2 \left(16 \nu ^2-29 \nu +17\right) \tilde{E}-4 \sqrt{2} h \left(4 \nu ^2-16 \nu +11\right) \sqrt{\tilde{E}}-6 \left(8 \nu ^2-21 \nu +14\right)\right]\nn\\
&\quad\qquad
+\delta  \kapA \left[h^2 (17-5 \nu ) \tilde{E}+2 \sqrt{2} h (5 \nu -14) \sqrt{\tilde{E}}+10 \nu -36\right] \nn\\
&\quad\qquad
+\kapS \left[h^2 \left(18 \nu ^2-39 \nu +17\right) \tilde{E}-2 \sqrt{2} h \left(6 \nu ^2-33 \nu +14\right) \sqrt{\tilde{E}}-12 \nu ^2+82 \nu -36\right]
\Bigr\rbrace,\\
\frac{e_\phi}{e_r} &= 1 + \frac{\nu  \tilde{E}}{c^2}
+ \frac{\tilde{E}}{16 c^4 h^2} \left[\nu ^2 \left(22 h^2 \tilde{E}-15\right)+\nu  \left(357-2 h^2 \tilde{E}\right)+160\right] \nn\\
&\quad
-\frac{4 \nu  \tilde{E} \chi _S}{c^3 h}
+\frac{\tilde{E}}{4 c^5 h^3} \Bigr\lbrace
\delta \chi _A \left[\nu  \left(2 h^2 \tilde{E}-13\right)-128\right]
+ \chi _S \left[\nu ^2 \left(26-28 h^2 \tilde{E}\right)+\nu  \left(83-38 h^2 \tilde{E}\right)-128\right]
\Bigr\rbrace \nn\\
&\quad
+ \frac{\tilde{E}}{c^4 h^2}\left[\chi _A^2 (1-4 \nu )+2 \delta \chi _A \chi _S+\delta \kapA +\kapS (1-2 \nu  )+ \chi _S^2\right] \nn\\
&\quad
+ \frac{\tilde{E}}{8 c^6 h^4} \Bigr\lbrace
\chi _A^2 \left[-2 h^2 \left(36 \nu ^2-7 \nu +2\right) \tilde{E}+12 \nu ^2-1705 \nu +432\right]
+2 \delta \chi _A \chi _S \left[2 h^2 (31 \nu -2) \tilde{E}-169 \nu +432\right]\nn\\
&\quad\qquad
+ \chi _S^2 \left[2 h^2 \left(32 \nu ^2+63 \nu -2\right) \tilde{E}-64 \nu ^2-361 \nu +432\right]
+\delta  \kapA \left[2 h^2 (19 \nu -2) \tilde{E}-125 \nu +240\right] \nn\\
&\quad\qquad
+\kapS \left[-2 h^2 \left(18 \nu ^2-23 \nu +2\right) \tilde{E}+6 \nu ^2-605 \nu +240\right]
\Bigr\rbrace.
\end{align}
\end{subequations}

The quantities $f_{v-u}$ and $f_v$ that enter Eq.~\eqref{eq:QKl} for the mean anomaly read
\begin{subequations}
\begin{align}
f_{v-u} &= \frac{\tilde{E}^{3/2}}{h}\bigg\lbrace
\frac{3 \sqrt{2} (5-2 \nu) }{c^4}
+ \frac{2 \sqrt{2}}{c^5 h} \left[\delta  (4 \nu -12) \chi _A+\left(-2 \nu ^2+16 \nu -12\right) \chi _S\right] 
+ \frac{\sqrt{2} }{c^6 h^2} \Big[
\left(12 \nu ^2-92 \nu +22\right) \chi _A^2\nn\\
&\quad \qquad
+\left(8 \nu ^2-32 \nu +22\right) \chi _S^2
+\delta  (44-36 \nu ) \chi _A \chi _S
+\delta \kapA (14-5 \nu ) 
+\kapS \left(6 \nu ^2-33 \nu +14\right)
\Big] \bigg\rbrace, \\
f_v &= \frac{\tilde{E}^{3/2}}{h} \sqrt{2-4h^2 \tilde{E}} \, \bigg\lbrace
\frac{\nu (15-\nu)  }{4 c^4}
-\frac{1}{c^5 h} \left[\delta  (\nu +4) \chi _A+\left(-2 \nu ^2-3 \nu +4\right) \chi _S\right] 
+ \frac{1}{2 c^6 h^2} \Big[
\left(4 \nu ^2-31 \nu +8\right) \chi _A^2\nn\\
&\quad\qquad
+(8-15 \nu ) \chi _S^2
+\delta  (16-14 \nu ) \chi _A \chi _S
+\delta  \kapA (8-3 \nu ) 
+\kapS \left(2 \nu ^2-19 \nu +8\right)
\Big]
\bigg\rbrace,
\end{align}
\end{subequations}
while $g_{2v}$ and $g_{3v}$ that enter Eq.~\eqref{eq:QKphi} read
\begin{subequations}
\begin{align}
g_{2v} &= \frac{1-2 h^2 \tilde{E}}{h^4} \bigg\lbrace
\frac{\nu  (19-3 \nu )+1}{8 c^4}
+ \frac{\nu }{4 c^5 h} \left[(6 \nu -7) \chi _S-5 \delta  \chi _A\right]
+ \frac{1}{8 c^6 h^2} \Big[
\left(-12 \nu ^2-14 \nu +6\right) \chi _A^2 \nn\\
&\qquad\quad
+\delta  (24 \nu +12) \chi _A \chi _S+\delta  \kapA (6-7 \nu )+\kapS \left(-6 \nu ^2-19 \nu +6\right)+\left(-16 \nu ^2+14 \nu +6\right) \chi _S^2
\Big]
\bigg\rbrace,\\
g_{3v} &= \nu\frac{(1-2 h^2 \tilde{E})^{3/2}}{h^4} \bigg\lbrace
\frac{1-3 \nu }{32 c^4}
-\frac{\delta  \chi _A + (1-2 \nu)  \chi _S}{8 c^5 h}
+ \frac{(3-4 \nu ) \chi _A^2+3 \chi _S^2+6 \delta  \chi _A \chi _S-\delta  \kapA-\kapS (2 \nu +1)}{16 c^6 h^2}
\bigg\rbrace \,.
\end{align}
\end{subequations}

The gauge-independent mean motion $n$ and periastron advance $K$ are given by
\begin{subequations}
\begin{align}
n &= 2 \sqrt{2} \tilde{E}^{3/2} + \frac{(\nu -15) \tilde{E}^{5/2}}{\sqrt{2} c^2} + \frac{1}{c^4} \left[\frac{12 (2 \nu -5) \tilde{E}^3}{h}+\frac{(11 \nu^2 +30\nu+555) \tilde{E}^{7/2}}{8 \sqrt{2}}\right] \nn\\
&\quad
+ \frac{16 \tilde{E}^3}{c^5 h^2} \Big[(\nu^2 -8\nu +6) \chi _S-2 \delta  (\nu -3) \chi _A\Big] 
+ \frac{4 \tilde{E}^3}{c^6 h^3} \Big[
4 \delta  (9 \nu -11) \chi _A \chi _S
-2 \left(6 \nu ^2-46 \nu +11\right) \chi _A^2\nonumber\\
&\qquad\quad
-2 (4\nu^2 -16\nu +11) \chi _S^2
+\delta  \kapA (5 \nu -14)+\kapS \left(-6 \nu ^2+33 \nu -14\right)\Big],\\
K &=1+ \frac{3}{c^2 h^2} + \frac{1}{c^4} \left[\frac{\left(3 \nu -\frac{15}{2}\right) \tilde{E}}{h^2}-\frac{15 (2 \nu -7)}{4 h^4}\right]
+ \frac{2 (\nu -2) \chi _S-4 \delta  \chi _A}{c^3 h^3} 
+ \frac{1}{c^5}\bigg[
\frac{\frac{21}{2} \delta  (\nu -8) \chi _A+\left(-3 \nu ^2+\frac{147 \nu }{2}-84\right) \chi _S}{h^5} \nonumber\\
&\qquad
+\frac{\tilde{E} \left(4 \left(\nu ^2-8 \nu +6\right) \chi _S-8 \delta  (\nu -3) \chi _A\right)}{h^3}\bigg]
+ \frac{1}{c^4 h^4}\left[\left(\frac{3}{2}-6 \nu \right) \chi _A^2+3 \delta  \chi _A \chi _S+\frac{3 \delta  \kapA}{2}+\kapS \left(\frac{3}{2}-3 \nu \right)+\frac{3 \chi _S^2}{2}\right]\nonumber\\
&\quad
+ \frac{1}{c^6} \bigg\lbrace
\frac{1}{h^6}\bigg[\frac{15}{2} \left(2 \nu ^2-57 \nu +14\right) \chi _A^2+30 \delta  (7-4 \nu ) \chi _A \chi _S+\frac{15}{2} \left(4 \nu ^2-15 \nu +14\right) \chi _S^2+\frac{15}{4} \kapS \left(2 \nu ^2-27 \nu +12\right)\nonumber\\
&\qquad\quad
-\frac{45}{4} \delta  \kapA (\nu -4)\bigg] 
+\frac{1}{h^4}\bigg[\tilde{E} \left(-3 \left(6 \nu ^2-46 \nu +11\right) \chi _A^2+6 \delta  (9 \nu -11) \chi _A \chi _S+\left(-12 \nu ^2+48 \nu -33\right) \chi _S^2\right)\nonumber\\
&\qquad\quad
+\frac{3}{2} \delta \kapA (5 \nu -14)+\kapS \left(-9 \nu ^2+\frac{99 \nu }{2}-21\right)\bigg]
\bigg\rbrace \,,
\end{align}
\end{subequations}
which agree with the results of Refs.~\cite{Tessmer:2010hp,Tessmer:2012xr}.

The expressions for $(r,\rd,\phi,\fid)$ in terms of $(x,e_t,u)$, in addition to their eccentricity expansions in terms of $(x,e_t,l)$, are too lengthy to write here; we provide them in the Supplemental Material~\cite{ancMaterial}.

\bibliography{RefList_EccSpin.bib}

\end{document}